\documentclass[twocolumn,epjc3]{svjour3}  

\journalname{Eur. Phys. J. A}

\usepackage{lipsum}
\usepackage{xcolor}
\usepackage{amsmath,amssymb} 
\usepackage{widetext}
\usepackage{hhline}
\usepackage{bbm}
\usepackage{diagbox}
\usepackage{rotating}
\usepackage{float}
\restylefloat{table}
\usepackage{cite}
\usepackage{graphicx}

\usepackage{tikz}
\usepackage{lipsum} 
\usepackage[colorlinks=true, citecolor=blue]{hyperref}

\newcommand{\fet}[1]{\mbox{\boldmath $#1$}}

\newcommand{\La}{{\Lambda}}
\newcommand{\Si}{{\Sigma}}

\newcommand{\beq}{\begin{equation}}
\newcommand{\eeq}{\end{equation}}
\newcommand{\be}{\begin{eqnarray}}
\newcommand{\ee}{\end{eqnarray}}
\newcommand{\nn}{\nonumber \\ }

\newlength{\feynwidth} 
\newlength{\feynwidthbig} 

\usepackage{xcolor} 
\begin{document}
\title{Hyperon-nucleon interaction in chiral effective field theory at
next-to-next-to-leading order}
\author{Johann Haidenbauer\thanksref{addr1,e2}
\and Ulf-G. Mei{\ss}ner\thanksref{addr2,addr1,addr4,addr3,e3}
\and Andreas Nogga\thanksref{addr1,addr4,e4}
\and Hoai~Le\thanksref{addr1,e1}
}
\thankstext{e2}{e-mail: j.haidenbauer@fz-juelich.de}
\thankstext{e3}{e-mail: meissner@hiskp.uni-bonn.de}
\thankstext{e4}{e-mail: a.nogga@fz-juelich.de}
\thankstext{e1}{e-mail: h.le@fz-juelich.de}

\institute{IAS-4, IKP-3 and JHCP, Forschungszentrum J\"ulich, D-52428 J\"ulich, Germany \label{addr1}
           \and
           CASA, Forschungszentrum J\"ulich, D-52425 J\"ulich, Germany \label{addr4}
           \and
           HISKP and BCTP, Universit\"at Bonn, D-53115 Bonn, Germany \label{addr2}
           \and 
           Tbilisi State University, 0186 Tbilisi, Georgia \label{addr3}
}

\date{\today}

\maketitle

\begin{abstract}
A hyperon-nucleon potential for the strange\-ness $S=-1$ sector ($\Lambda N$, $\Sigma N$) 
up to third order in the chiral expansion is presented. 
SU(3) flavor symmetry is imposed for constructing the interaction, however, 
the explicit SU(3) symmetry breaking by the physical masses of the pseudoscalar 
mesons and in the leading-order contact terms is taken
into account. 
A novel regularization scheme is employed which has  already been successfully
used in studies of the nucleon-nucleon interaction within chiral
effective field theory up to high orders. 
An excellent description of the low-energy $\Lambda p$, $\Sigma^- p$ and
$\Sigma^+ p$ scattering data is achieved. New data from J-PARC on angular
distributions for the $\Sigma N$ channels are analyzed. Results for the
hypertriton and $A=4$ hyper-nuclear separation energies are presented. 
An uncertainty estimate for 
the chiral expansion is performed for selected hyperon-nucleon observables. 


\end{abstract}

%%%%%%%%%%%%%%%%%%%%%%%%%%%%%%%%%%%%%%%%%%%%%%%%%%%%%%%%%%%%%%
\section{Introduction} 
\label{sec:Intro} 

The hyperon-nucleon ($\La N$, $\Si N$) interaction has been under
scrutiny in various fields in recent times. Certainly most prominent
has been the discussion of its properties in an astrophysical
context. The discovery of neutron stars with masses around or 
in excess of twice the solar mass opened speculations about 
the role hyperons and specifically the $\La$ play in understanding
their characteristics. In particular, at densities realized in 
such compact objects, neutrons should be eventually converted to $\La$'s,
resulting in a softening of the equation-of-state (EoS) and a collapse of 
the conventional theoretical explanation of the observed mass radius
relation. This is the so-called hyperon puzzle, cf. the reviews
\cite{Chatterjee:2015pua,Schaffner-Bielich:2020bk,Tolos:2020aln,Weise:2022wuf}
and references therein. 
On a less spectacular (speculative) level, new measurements of $\La N$ and 
$\Si N$ scattering have been reported 
\cite{CLAS:2021gur,J-PARCE40:2021qxa,J-PARCE40:2021bgw,J-PARCE40:2022nvq},
including the first more extensive data on $\Si^+p$ and $\Si^-p$ differential 
cross sections away from the threshold. In addition, two-particle momentum
correlation functions involving strange baryons have been 
determined, in heavy-ion collision and in high-energy $pp$ collisions,
which allow access to the $YN$ interaction at very low momenta
\cite{STAR:2005rpl,HADES:2016dyd,ALICE:2021njx,ALICE:2019buq}.
Finally, there are ongoing efforts for a better determination of the 
binding energies of light $\La$ hypernuclei
\cite{Adam:2019phl,ALICE:2022rib,STAR:2022zrf}. 
On the theory side, lattice QCD simulations have matured to a stage where an 
evaluation of the $YN$ interaction for quark (pion) masses close to the physical 
point can be performed
\cite{Nemura:2017vjc,Nemura:2022wcs}.  Further,
{\em ab initio} methods like the no-core shell model (NCSM) have 
been pushed to a level where calculations of hypernuclei
up to $A=10$ and beyond can be performed, incorporating the full
complexity of the underlying elementary $YN$ interaction
\cite{Wirth:2014ko,Wirth:2018kh,Wirth:2019cpp,Le:2020zdu,Le:2021wwz,Le:2021gxa,Le:2022ikc}. 

Chiral effective field theory (EFT) for nuclear systems, formulated by Weinberg 
about 30 years ago \cite{Weinberg:1990bf,Weinberg:1991um}, constitutes
a rather powerful tool for studying the interaction between baryons. 
In this approach a potential is established via an expansion in terms of small
momenta and the pion mass, subject to an appropriate power counting, so that the results can be
improved systematically by going to higher orders,
while at the same time theoretical uncertainties can be estimated
\cite{Epelbaum:2009hy,Machleidt:2011gh}.
Furthermore, two- and three-baryon forces can be constructed in a consistent way.
The resulting interaction potentials can be readily employed in
standard two- and few-body calculations. They consist of contributions
from an increasing number of pseudoscalar-meson exchanges, determined
by the underlying chiral symmetry,
and of contact terms which encode the unresolved short-distance dynamics and whose
strengths are parameterized by a priori unknown low-energy constants (LECs).
Of course, there are further LECs related to higher order two-meson exchanges
which can in principle be fixed from meson-baryon scattering data.

While the description of the nucleon-nucleon ($NN$) interaction within chiral EFT 
has already progressed up to the fifth order and beyond \cite{Epelbaum:2014sza,Reinert:2018ip,Entem:2017hn}, 
corresponding applications 
of that framework to the $YN$ interaction are lagging far behind
\cite{Korpa:2001au,Polinder:2006zh,Li:2016paq,Ren:2019qow,Song:2021yab}. 
Here, NLO is presently the state-of-the-art \cite{Haidenbauer:2013oca,Haidenbauer:2019boi,Haidenbauer:2016gw,Haidenbauer:2018gvg}.
That status is primarily a consequence of the unsatisfactory situation with regard 
to the data base, practically only 
cross sections are available and primarily for energies near the thresholds. 
In particular, differential observables that would allow to 
fix the LECs in $P$- and/or higher partial waves, which emerge in the chiral 
expansion when going to higher orders, are rather scarce and of low statistics. 
Only within the last few years the overall circumstances became more promising,
thanks to the E40 experiment performed at the J-PARC facility. 
The measurements have already produced differential cross sections
for the $\Si^+ p$ and $\Si^- p$ channels for laboratory momenta from 
$440$ to $850$~MeV/c \cite{J-PARCE40:2021qxa,J-PARCE40:2021bgw,J-PARCE40:2022nvq} 
and corresponding studies for $\La p$,
including possibly even spin-dependent observables,
are in the stage of preparation \cite{Miwa:2022coz}. 

In this paper, we present a $YN$ potential up to next-to-next-to-leading
order (N$^2$LO), derived within SU(3) chiral EFT. 
The mentioned experimental development was one of the motivations 
to extend our study of the $\La N$-$\Si N$ interaction to the next order. 
However, there are also several theoretical aspects which make an
extension to N$^2$LO rather interesting. One of them is that in the
Weinberg counting three-baryon forces (3BFs) emerge at this order. Calculations of 
the four-body systems $^4_\La$H and $^4_\La$He for the NLO13 \cite{Haidenbauer:2013oca}
and NLO19 \cite{Haidenbauer:2019boi} potentials based on  
Faddeev-Yakubovsky equations
indicate that the experimental separation energies are underestimated
and dependent on the version of the YN interaction
\cite{Haidenbauer:2019boi}. Very likely this signals the need for including 
$\La NN$ and possibly also $\Si NN$ 3BFs \cite{Petschauer:2015elq}. 
Another appealing factor is (in view of the mentioned scarcity of
data) that no additional LECs appear at this order. At the same time,
results for $NN$ scattering indicate that there is some improvement in 
the energy dependence of the $S$-waves and, specifically, in several 
$P$-waves once the contributions involving the sub-leading $\pi N$ 
vertices that enter at N$^2$LO are taken into account. 

A further issue is the dependence on the regulator that has to be introduced
to remove high-momentum components when solving the scattering equations
\cite{Epelbaum:2004fk}. 
In general, a substantial reduction of the residual regulator dependence
can be achieved by going to high orders with a larger number of LECs, 
which then allow one to absorb those effects efficiently \cite{Epelbaum:2015vj}. 
Since our calculation is only up to N$^2$LO, we want to keep regulator 
artifacts as small as possible from the beginning. 
With regard to that, a novel regularization scheme 
proposed and applied in Ref.~\cite{Reinert:2018ip} seems to be rather promising. 
Here, a local regulator is applied to the pion-exchange contributions and only 
the contact terms, being non-local by themselves, are regularized with a 
non-local function. Accordingly, the resulting interactions are called 
``semilocal momentum-space regularized (SMS) chiral $NN$ potentials''
\cite{Reinert:2018ip}. 
In earlier works on the $NN$ interaction but also in our $YN$ 
studies, a non-local cutoff has been applied to the whole potential
\cite{Haidenbauer:2013oca,Haidenbauer:2019boi,Epelbaum:2004fk}.
A local regulator for pion-exchange contributions leads to a reduction of 
the distortion in the long-range part of the interaction and, thereby, 
facilitates a more rapid convergence already at 
low chiral orders.  
Of course, this effect cannot be directly quantified in case of $\La N$ 
and $\Si N$ because of the lack of more detailed empirical information,
specifically due to the absence of a proper partial-wave analysis. 
Nonetheless, given that we aim at comparing our results with the new 
J-PARC data at laboratory momenta around $500$~MeV/c, a reduction of 
regulator artifacts is definitely desirable.

The paper is structured in the following way: 
In Sect.~\ref{sec:Formalism}, we summarize the basics of the employed
formalism. More details are described in an appendix. 
Our results are presented in Sect.~\ref{sec:Results} where
we discuss in detail the scattering cross sections for 
the channels $\La p$, $\Si^+ p$ and $\Si^- p$. Predictions
for $S$- and $P$-wave phase shifts in the $\La N$ and 
$\Si N$ (isospin $I=3/2$) channels are also provided. 
Furthermore, 
results for the hypertriton and $A=4$ hyper-nuclear separation energies and for
the in-medium properties of the $\La$ and $\Si$ hyperons
are given. Finally, an uncertainty estimate of our
EFT calculations is presented. 
The paper closes with a brief summary and an outlook. 

%%%%%%%%%%%%%%%%%%%%%%%%%%%%%%%%%%%%%%%%
\section{Formalism} 
\label{sec:Formalism}

In this section and in \ref{sec:smspot}, we
provide a self-contained description of all the ingredients of
the new $YN$ interaction and its extension to N$^2$LO. However,
we refrain from repeating here the details of the derivation of the 
baryon-baryon interaction within SU(3) chiral EFT. This has been described 
and thoroughly discussed in Refs.~\cite{Polinder:2006zh,Haidenbauer:2013oca} 
and in the review \cite{Petschauer:2020urh}.
We refer the interested reader to those works.
Also, with regard to various aspects of the new regularization
scheme that forms the basis of the SMS potentials, we refer to 
Ref.~\cite{Reinert:2018ip} for details where this procedure was 
introduced and worked out. 
\vspace{1cm}
\subsection{One-boson exchange}
\label{sec:OBE} 

\begin{table*}[t]
%\begin{table}[h]
\caption{Isospin factors ${\mathcal I}$ for the various one--pseudoscalar-meson exchanges.}
\label{tab:Iso}
\vskip 0.1cm
\renewcommand{\arraystretch}{1.2}
\centering
\begin{tabular}{|c|c|c|r|r|r|}
%\begin{tabular}{rrrrr}
\hline
&Channel &Isospin &$\pi$ &$K$ &$\eta$\\
\hline
$S=0$ &$NN\rightarrow NN$ &$0$ &$-3$ &$0$ &$1$ \\
&                  &$1$ &$1$  &$0$ &$1$ \\
\hline
$S=-1$ &$\Lambda N\rightarrow \Lambda N$ &$\frac{1}{2}$ &$0$ &$1$ &$1$ \\
&$\Lambda N\rightarrow \Sigma N$ &$\frac{1}{2}$ &$-\sqrt{3}$ &$-\sqrt{3}$ &$0$ \\
&$\Sigma N\rightarrow \Sigma N$ &$\frac{1}{2}$ &$-2$ &$-1$ &$1$ \\
&                              &$\frac{3}{2}$ &$1$ &$2$ &$1$ \\
\hline
\end{tabular}
\renewcommand{\arraystretch}{1.0}
\end{table*}

Let us start with the one-boson-exchange
(OBE) contribution and with introducing the 
new regularization scheme. The formulae for the contributions from two-boson
exchanges which arise at NLO and N$^2$LO are given in \ref{sec:smspot}. 
The regularized potential for single-meson exchange $V_{P}$ ($P=\pi,\,K,\,\eta$) 
has the following form in momentum space:

\begin{widetext}
%\vskip 0.5cm 
\beq
\label{eq:OBE}
V^{\rm OBE}_{B_1 B_2\to B_3 B_4} ({\bf q} \, ) = - f_{B_1B_3P} f_{B_2B_4P} \,
  \bigg( \frac{{\bf \sigma_1} \cdot
  {\bf q} \, {\bf \sigma_2} \cdot {\bf q}}{{\bf q}^2 + M_P^2}    + C(M_P) \, 
 {\bf \sigma_1} \cdot {\bf \sigma_2} \bigg) \,
\exp\left(- \frac{{\bf q}^2 + M_P^2}{\Lambda^2}\right)\, {\cal I}_{B_1 B_2\to B_3 B_4}\,, 
\eeq
\end{widetext}

where the $f_{B_iB_jP}$ are baryon-baryon-meson coupling constants, $M_P$ is the mass 
of the exchanged pseudoscalar meson, and ${\cal I}_{B_1 B_2\to B_3 B_4}$
is the pertinent isospin factor. 
The transferred momentum ${\bf q}$ is defined in terms of
the final and initial center-of-mass (c.m.) momenta of the baryons, 
${\bf p}'$ and ${\bf p}$, as ${\bf q}={\bf p}'-{\bf p}$.
We adopt here the convention of Ref.~\cite{Reinert:2018ip} to include a leading-order
contact term in the one-boson exchange potential. It is chosen in such
a way that the (total) spin-spin part of the potential vanishes 
for $r\to 0$ in the configuration-space representation. The
expression of $C(M_P)$ which fulfills that requirement can be given 
in analytical form and amounts to \cite{Reinert:2018ip} 

\be
C (M_P) & = & -\Big[ \Lambda \left(\Lambda ^2-2 M_P^2 \right) \cr 
& & + 2 \sqrt{\pi } M_P^3 \exp\left(\frac{M_P^2}{\Lambda ^2}\right)
   \text{erfc}\left(\frac{M_P}{\Lambda }\right)\Big]/(3 \Lambda ^3) \,. \cr & & 
\ee
Here, $\text{erfc} (x)$ is the complementary error function
\beq
\text{erfc} (x) = \frac{2}{\sqrt{\pi}} \int_x^\infty dt \, e^{-t^2}\,.
\eeq

Under the assumption of strict SU(3) flavor symmetry, the various coupling constants 
$f_{B_iB_jP}$ are related to each other by
\cite{deSwart:1963pdg}

\begin{equation}
\begin{array}{rlrl}
f_{NN\pi}  = & f, 
& f_{NN\eta_8}  = & \frac{1}{\sqrt{3}}(4\alpha -1)f, \\
f_{\Lambda NK} = & -\frac{1}{\sqrt{3}}(1+2\alpha)f, 
& f_{\Xi\Xi\pi}  = & -(1-2\alpha)f, \\
f_{\Xi\Xi\eta_8}  = & -\frac{1}{\sqrt{3}}(1+2\alpha )f, 
& f_{\Xi\Lambda K} = & \frac{1}{\sqrt{3}}(4\alpha-1)f, \\
f_{\Lambda\Sigma\pi}  = & \frac{2}{\sqrt{3}}(1-\alpha)f, 
& f_{\Sigma\Sigma\eta_8}  = & \frac{2}{\sqrt{3}}(1-\alpha )f, \\
f_{\Sigma NK} = & (1-2\alpha)f, 
& f_{\Sigma\Sigma\pi}  = & 2\alpha f, \\ f_{\Lambda\Lambda\eta_8}  = & -\frac{2}{\sqrt{3}}(1-\alpha )f, & f_{\Xi\Sigma K} = & -f.
\end{array}
\label{eq:SU3}
\end{equation}
Accordingly, all coupling constants are given in terms of $f\equiv g_A/2f_0$ and 
the ratio $\alpha=F/(F+D)$. Here, $f_0$ is the Goldstone boson decay constant, 
$g_A$ is the axial-vector strength measured in neutron $\beta$-decay, and
$F+D = g_A$. Note that we will take the physical values of these various
parameters, though strictly speaking in the effective Lagrangian they appear
with their values in the chiral limit. This difference can be absorbed in higher order terms.
In the present calculation, deviations of the meson-baryon coupling constants 
from the SU(3) values are taken into account. Specifically, there is an explicit 
SU(3) symmetry breaking in the empirical values of the decay constants \cite{ParticleDataGroup:2012pjm},
\be
f_\pi & = &  92.4~{\rm MeV}, \cr 
f_K & = & (1.19\pm 0.01) f_\pi, \cr 
f_\eta & =& (1.30\pm 0.05) f_\pi \ .
\ee
The somewhat smaller SU(3) breaking in the axial-vector coupling constants,
see the pertinent discussion in Appendix~B of Ref.~\cite{Haidenbauer:2013oca}, 
is neglected in the present study. 
However, following the practice in chiral $NN$ potentials, 
we use $g_A = 1.29$, which is slightly larger than the experimental 
value, in order to account for the Goldberger-Treiman discrepancy. As
before in \cite{Haidenbauer:2013oca},
for the $F/(F+D)$ ratio, we adopt $\alpha = 0.4$ which is the SU(6) value. 
Further, the $\eta$ meson is identified with the octet-state $\eta_8$.
The isospin factors ${\cal I}_{B_1 B_2\to B_3 B_4}$ are summarized 
in Table~\ref{tab:Iso}. 

In the $NN$ case, where only pion exchanges are taken into account, cutoff values in 
the range $\Lambda = 350-550$~MeV were considered where $\Lambda = 450$~MeV yields 
the best results \cite{Reinert:2018ip}. 
The choice of the cutoff mass for the $YN$ interaction is more delicate. On the one hand, 
we want to preserve the principal features of the underlying approximate 
SU(3) flavor symmetry, in particular the explicit SU(3) breaking in the long-range 
part of the potential due 
to the mass splitting between the pseudoscalar mesons $\pi$, $K$, and $\eta$. 
Since the kaon mass is around $495$~MeV it seems appropriate to use cutoff
masses that are at least $500$~MeV, so that the essential role of 
the $K$ meson for the $YN$ dynamics can be incorporated. 
At the other end, large values, say $650$~MeV 
or beyond, lead to highly non-perturbative
potentials and bear the risk of the appearance 
of spurious bound states, according to the experience from $NN$
studies~\cite{Epelbaum:2002ji}. 
Considering that aspect implicates that two-meson 
exchange contributions involving a $K$ and/or 
$\eta$ ($\pi K$, $K K$, etc.), where then the combined masses
exceed the cutoff value, will be strongly suppressed. 
Therefore, there is no point to include them explicitly. 
Rather their effect should be subsumed 
into the contact terms. Thus, contrary to our earlier work \cite{Haidenbauer:2013oca,Haidenbauer:2019boi}, 
we expect and allow for SU(3) symmetry breaking of the LECs in the 
$\La N$ and $\Si N$ systems. In this context, it should be mentioned that also the
counterterms in Eq.~(\ref{eq:OBE}) constitute effectively an SU(3) symmetry 
breaking contact interaction. 

In the present work we consider the cutoff values
$\La = 500$, $550$, and $600$~MeV. Clearly, for the lowest value 
$\eta$ exchange will be already strongly suppressed
and, in fact, we neglected its contribution in this
case. The highest value is well above the masses of
the $K$- and $\eta$ mesons so that the effect of 
the SU(3) symmetry breaking in the masses of the 
pseudoscalar mesons on the $YN$ interaction is well accounted for.
In the discussion below we focus predominantly on the 
results for $\La = 550$~MeV. However, some results, 
notably the $\chi^2$, the effective range parameters
and the hypertriton and $A=4$ separation energies will be given
for all three cutoffs in order to provide an overview
of the quality of our chiral $YN$ interactions. 
Anticipating the results, we stress that an equally good description 
of the considered $YN$, $YNN$ and $A=4$ hyper-nuclear observables can be achieved for 
all three cutoffs. 

At leading order (LO), only a very basic description of the 
$YN$ interaction can be obtained \cite{Polinder:2006zh}.
In particular, unrealistic small scattering lengths
emerge from fits to the low-energy data under the 
prerequisite that the lightest $\La$ hypernuclei are not 
too strongly bound. 
Nonetheless, we construct also a LO interaction in the
present study because we want to perform an uncertainty
estimate of our chiral $YN$ potentials following the
procedure proposed in Ref.~\cite{Epelbaum:2015vj}.
It turns out that under the assumption of SU(3) symmetry
(note that SU(3) breaking contact terms arise first at NLO \cite{Petschauer:2013uua})
a LO fit with 
a decent $\chi^2$ is only possible for a cutoff of $\Lambda = 700$~MeV and
without subtraction. For smaller cutoffs, the $\chi^2$
increases dramatically. We use that potential for
the uncertainty estimate below but do not discuss its result in
detail. Anyway, the LO results 
($\chi^2\approx 30$, $a^{\La N}_s = -2.1$~fm, $a^{\La N}_t = -1.2$~fm)  
%($\chi^2$, scattering lengths, etc.) 
are very similar to those based on a non-local 
cutoff reported in Ref.~\cite{Polinder:2006zh}.

\begin{table*}[ht]
\caption{SU(3) relations for the interactions in different $B_1B_2\to B_3B_4$
channels, with isospin $I$ and strangeness $S$.
$C^{27}_{\xi}$ etc. refers to the corresponding irreducible SU(3) representation
for a particular partial wave ${\xi}$ \cite{Polinder:2006zh,Haidenbauer:2013oca}.
}
\label{tab:SU3}
\vskip 0.1cm
\renewcommand{\arraystretch}{1.4}
\centering
\begin{tabular}{|l|c|c|l|l|l|}
\hline
&Channel &I &\multicolumn{3}{c|}{$V({\xi})$} \\
\hline
&        &  &$\xi= \, ^1S_0, \, ^3P_0, \, ^3P_1, \, ^3P_2 $
& $\xi = \, ^3S_1, \, ^3S_1$-$^3D_1, \, ^1P_1$ & $\xi = \, ^1P_1$-$^3P_1$ \\
\hline
${S=\phantom{-}0}$&$NN\rightarrow NN$ &$0$ & \ \ -- & $C^{10^*}_{\xi}$ & \ \ -- \\
                       &$NN\rightarrow NN$ &$1$ & $C^{27}_{\xi}$ & \ \ -- & \ \ -- \\
\hline
${S=-1}$&$\La N \rightarrow \La N$ &$\frac{1}{2}$ &$\frac{1}{10}\left(9C^{27}_{\xi}+C^{8_s}_{\xi}\right)$
& $\frac{1}{2}\left(C^{8_a}_{\xi}+C^{10^*}_{\xi}\right)$ & $\frac{-1}{\sqrt{20}} C^{8_s8_a}_{\xi}$\\
&$\La N \rightarrow \Si N$ &$\frac{1}{2}$        &$\frac{3}{10}\left(-C^{27}_{\xi}+C^{8_s}_{\xi}\right)$
& $\frac{1}{2}\left(-C^{8_a}_{\xi}+C^{10^*}_{\xi}\right)$ & $\frac{-3}{\sqrt{20}} C^{8_s8_a}_{\xi}$\\
&$\Si N \rightarrow \La N$ & & & & $\frac{1}{\sqrt{20}} C^{8_s8_a}_{\xi}$\\
&$\Si N \rightarrow \Si N$  &$\frac{1}{2}$        &$\frac{1}{10}\left(C^{27}_{\xi}+9C^{8_s}_{\xi}\right)$
& $\frac{1}{2}\left(C^{8_a}_{\xi}+C^{10^*}_{\xi}\right)$ & $\frac{3}{\sqrt{20}} C^{8_s8_a}_{\xi} $\\
&$\Si N \rightarrow \Si N$  &$\frac{3}{2}$        &$C^{27}_{\xi}$
& $C^{10}_{\xi}$ & \ \ -- \\
\hline
\end{tabular}
\renewcommand{\arraystretch}{1.0}
\end{table*}

\subsection{Contact terms} 
\label{sec:2CT}

The spin dependence of the potentials due to the LO contact terms is given by
\cite{Weinberg:1990bf}
\begin{eqnarray}
V^{(0)}_{BB\to BB}&=& C_{S} + C_{T}\,
\mbox{\boldmath $\sigma$}_1\cdot\mbox{\boldmath $\sigma$}_2\,,
\label{V0}
\end{eqnarray}
where the parameters $C_{S}$ and $C_{T}$ are low-energy
constants (LECs) depending on the considered baryon-bary\-on channel.
These need to be determined by a fit to data.
At NLO the spin- and momentum-dependence of the contact terms reads \hfill
%\begin{widetext}
\begin{eqnarray}
&&V^{(2)}_{BB\to BB}= C_1 {\bf q}^{\,2}+ C_2 {\bf k}^{\,2} + (C_3 {\bf q}^{\,2}+ C_4 {\bf k}^{\,2})
\,\mbox{\boldmath $\sigma$}_1\cdot\mbox{\boldmath $\sigma$}_2
 \nonumber \\
&& \quad \quad + \frac{i}{2} C_5 (\mbox{\boldmath $\sigma$}_1+\mbox{\boldmath $\sigma$}_2)\cdot
 ({\bf q} \times {\bf k})
 + C_6 ({\bf q} \cdot \mbox{\boldmath $\sigma$}_1) ({\bf q} \cdot \mbox{\boldmath $\sigma$}_2)
 \nonumber \\ 
&&\quad \quad + C_7 ({\bf k} \cdot \mbox{\boldmath $\sigma$}_1) ({\bf k} \cdot \mbox{\boldmath $\sigma$}_2)
 + \frac{i}{2} C_8 (\mbox{\boldmath $\sigma$}_1-\mbox{\boldmath $\sigma$}_2)\cdot ({\bf q} \times {\bf k}) \ ,
  \nonumber \\
\label{V1}
\end{eqnarray}
%\end{widetext}
where the $C_i$ ($i=1,\dots,8$) are additional LECs and ${\bf k}$  is
the average momentum defined by ${\bf k}=({\bf p}'+{\bf p})/2$.
When performing a partial-wave projection, these terms contribute to the two $S$--wave
($^1S_0$, $^3S_1$) potentials, the four $P$--wave
($^1P_1$, $^3P_0$, $^3P_1$, $^3P_2$) potentials, and the $^3S_1$-$^3D_1$ and $^1P_1$-$^3P_1$
transition potentials in the following way \cite{Epelbaum:2004fk} (note that due to the
absence of the Pauli principle, there is one more term than in the $NN$ case):
\begin{eqnarray}
\label{VC0}
V(^1S_0) 
&=& \tilde{C}_{^1S_0} + {C}_{^1S_0} ({p}^2+{p}'^2)~, \label{C1S0}\\
V(^3S_1) 
&=& \tilde{C}_{^3S_1} + {C}_{^3S_1} ({p}^2+{p}'^2)~, \\
V(^3D_1 -\, ^3S_1) &=& {C}_{^3SD_1}\, {p'}^2~,\\
V(^3S_1 -\, ^3D_1) &=& {C}_{^3SD_1}\, {p}^2~, \label{C3S1} \\
V(^3P_0) &=& {C}_{^3P_0}\, {p}\, {p}'~,\\
V(^1P_1) &=& {C}_{^1P_1}\, {p}\, {p}'~,\\
V(^3P_1) &=& {C}_{^3P_1}\, {p}\, {p}'~, \\ 
\label{VC1}
V(^3P_1 - {^1P_1}) &=& {C}_{^3P_1-^1P_1}\, {p}\, {p}'~,\\
\label{VC2}
V(^1P_1 - {^3P_1}) &=& {C}_{^1P_1-^3P_1}\, {p}\, {p}'~,\\
V(^3P_2) &=& {C}_{^3P_2}\,  {p}\, {p}'~,
\label{VC}
\end{eqnarray}
with $p = |{\bf p}\,|$ and ${p}' = |{\bf p}\,'|$.
$\tilde C_\alpha$ and $C_\alpha$ are appropriate combinations of the
$C_i$'s appearing in Eqs.~(\ref{V0}) and (\ref{V1}), see Ref.~\cite{Haidenbauer:2013oca}. 

Assuming only isospin symmetry, the LECs for each spin-isospin state of 
the $BB\to BB$ potentials are independent. 
When imposing ${\rm SU(3)}$ flavor symmetry one obtains 
relations between the LECs in the strangeness $S=0$ and $S=-1$ systems, 
see Table~\ref{tab:SU3}, 
so that the total number of independent terms is noticeably
reduced \cite{Haidenbauer:2013oca}. Specifically, for the partial waves 
relevant at low energies, 
$^1S_0$ and $^3S_1$, within SU(3) symmetry there are only $10$ independent LECs 
($5$ at LO and $5$ at NLO) altogether, whereas with isospin symmetry alone 
there would be $16$. 
Like in our previous studies \cite{Haidenbauer:2013oca,Haidenbauer:2019boi}, we impose SU(3)
constraints on the LECs. However, for the reasons discussed above, those constraints 
are relaxed in the course of the fitting procedure whenever required for 
improving the description 
of the $\La p$ and $\Sigma N$ low-energy data. In practice, a departure from SU(3) symmetry
is only necessary for the LO $S$-wave LECs, which is anyway in line with the 
employed power counting, see Refs.~\cite{Haidenbauer:2013oca} (Appendix~B) and 
\cite{Petschauer:2013uua}. 
 
Note that we do not consider the possible $^1P_1$-$^3P_1$ transition 
at the present stage. 
In principle, one could fix the pertinent LECs which correspond to an antisymmetric 
$\La N$-$\Si N$ spin-orbit force, c.f. the term involving $C_8$ in Eq.~(\ref{V1}), 
by considering the Scheerbaum factor \cite{Scheerbaum:1976zz} in nuclear matter as 
done by us in Refs.~\cite{Haidenbauer:2014uua,Petschauer:2016jqa}. 
However, we intend to extend our calculations of $\La$-hypernuclei within the
NCSM approach \cite{Le:2020zdu,Le:2022ikc} up to $A=9$ systems in the future. Then we 
can  directly use the empirical information on the level splitting of the
$^9_{\La}$Be hypernucleus \cite{Akikawa:2002tm} to investigate the
strength needed for the elementary antisymmetric spin-orbit force.

\begin{table*}
\caption{Comparison between the 36 $YN$ data and the theoretical results
for the various cutoffs in terms of the achieved $\chi^2$. The last two 
columns are results for the NLO13 \cite{Haidenbauer:2013oca} and 
NLO19 \cite{Haidenbauer:2019boi} $YN$ potentials. 
}
\renewcommand{\arraystretch}{1.3}
\label{tab:R1}
\vspace{0.2cm}
\centering
\begin{tabular}{|c|c||rrr|rrr||r|r|}
\hline
& data & \multicolumn{3}{c|}{SMS NLO} & \multicolumn{3}{c||}{SMS N$^2$LO} & NLO13 & NLO19 \\
\hline
\multicolumn{2}{|l||}{$\Lambda$ (MeV)} 
&$500$ & $550$ & $600$  & $500$& $550$ & $600$  & $600$ & $600$ \\
\hline
\hline
$\La p \to \La p$          &Sechi-Zorn \cite{SechiZorn:1969hk} 
&$1.8$  &$1.6$  &$1.5$  &$1.9$ &$1.9$ &$1.8$  & $1.4$ &$1.9$\\
                           &Alexander \cite{Alexander:1969cx}
&$2.2$  &$2.5$  &$2.7$  &$2.0$ &$2.1$ &$2.2$  & $3.0$ &$1.6$\\
$\Sigma^- p\to \La n$      &Engelmann \cite{Engelmann:1966pl} 
&$3.6$  &$3.8$  &$4.0$  &$3.6$ &$4.0$ &$3.6$  & $4.1$ &$4.0$\\
$\Sigma^- p\to \Sigma^0 n$ &Engelmann \cite{Engelmann:1966pl} 
&$5.9$  &$5.8$  &$5.8$  &$5.9$ &$5.9$ &$5.9$  & $5.8$ &$6.0$\\
$\Sigma^- p\to \Sigma^- p$ &Eisele \cite{Eisele:1971mk}    
&$1.9$  &$1.8$  &$1.8$  &$2.0$ &$1.9$ &$1.9$  & $1.9$ &$2.2$\\
$\Sigma^+ p\to \Sigma^+ p$ &Eisele \cite{Eisele:1971mk}    
&$0.1$  &$0.3$  &$0.4$  &$0.2$ &$0.2$ &$0.3$  & $0.5$ &$0.4$\\
\hline
$r_R$                      & \cite{Hepp:1968zza,StephenPhD:1970wd}    
&$0.1$  &$0.0$  &$0.0$  &$0.3$ &$0.1$ &$0.1$  & $0.1$ &$0.1$\\
\hline
total $\chi^2$             &                       
&$15.52$ &$15.67$ &$16.15$  &$15.78$ &$15.56$ &$15.74$  &$16.82$ &$16.29$\\
\hline
\end{tabular}
\renewcommand{\arraystretch}{1.0}
\end{table*}
\subsection{Scattering equation} 
Once the $YN$ potential is established, a partial-wave projection is
performed \cite{Polinder:2006zh} and the ($\La N$ or $\Si N$) 
reaction amplitudes are obtained from the solution of a coupled-channel 
Lippmann-Schwinger (LS) equation,
\begin{widetext}
\begin{eqnarray}
T^{\ell''\ell',J}_{\nu''\nu'}(p'',p';\sqrt{s})&=&V^{\ell''\ell',J}_{\nu''\nu'}(p'',p')
\nonumber\\&+&
\sum_{\ell,\nu}\int_0^\infty \frac{dpp^2}{(2\pi)^3} \, V^{\ell''\ell,J}_{\nu''\nu}(p'',p)
\frac{2\mu_{\nu}}{k_{\nu}^2-p^2+i\eta}T^{\ell\ell',J}_{\nu\nu',J}(p,p';\sqrt{s})\ . 
\label{LS} 
\end{eqnarray}
\end{widetext}%
The label $\nu$ indicates the channels and the label $\ell$ the partial 
wave. $\mu_\nu$ is the pertinent reduced mass. The on-shell momentum in the 
intermediate state, $k_{\nu}$, is
defined by $\sqrt{s}=\sqrt{m^2_{B_{1,\nu}}+k_{\nu}^2}+\sqrt{m^2_{B_{2,\nu}}+k_{\nu}^2}$.
Relativistic kinematics is used for relating the laboratory momentum 
$p_{{\rm lab}}$ of the hyperons to the c.m. momentum.
For evaluating phase shifts, the LS equation is solved in the isospin basis.
For observables, all calculations are performed in the particle basis, so that 
the correct physical thresholds can be incorporated. 
The Coulomb interaction (in the $\Si^-p$ and $\Si^+p$ channels) is taken 
into account appropriately via the Vincent-Phatak method \cite{Vincent:1974zz}.

%%%%%%%%%%%%%%%%%%%%%%%%%%%%%%%%%%%%%%%
\section{Results} 
\label{sec:Results} 

In fitting to the $YN$ data we proceed as before \cite{Haidenbauer:2013oca,Haidenbauer:2019boi}, i.e. 
we consider the set of $36$ data for $\La p$, $\Si^-p$ and $\Si^+p$ 
scattering at low energies \cite{SechiZorn:1969hk,Alexander:1969cx,Engelmann:1966pl,Eisele:1971mk,Hepp:1968zza,StephenPhD:1970wd}
for determining the LECs in the $S$-waves. And, like before, as additional
constraint, we require the hypertriton to be bound, which enables us to 
fix the relative strength of the singlet- and triplet S-waves in the 
$\La p$ channel. 
SU(3) symmetry is imposed for the contact terms at the
initial stage but eventually relaxed for the LO LECs, 
$\tilde C_{^1S_0}$ and $\tilde C_{^3S_1}$ in Eqs.~(\ref{C1S0},\ref{C3S1}), 
in line with the 
power counting where SU(3) breaking terms arise from mass insertions in 
the chiral Lagrangian at the NLO level \cite{Petschauer:2013uua}. 
Anyway, as said, we do expect some SU(3) breaking in the contacts terms in view of 
the fact that two-meson exchange contributions from $\pi K$, $\pi \eta$, etc. 
are not explicitly included. The achieved $\chi^2$ is comparable to the one found 
for our NLO interactions \cite{Haidenbauer:2013oca,Haidenbauer:2019boi}, and typically 
around $16$ for the $36$ data points, see Table~\ref{tab:R1}.  
An overview of the scattering lengths and effective
ranges for the various $YN$ channels is provided in
Table~\ref{tab:ereF}. 
Preliminary results have been reported in Ref.~\cite{Haidenbauer:2022esw}. 

In the detailed discussion of the results, we focus on the ones for the
cutoff $550$~MeV. Those for the other considered cutoffs, $500$ and $600$~MeV, 
are very similar as one can conjecture from the $\chi^2$ values. 
Also, we start with the $\Si N$ channels where new data from the J-PARC 
E40 experiment have become available \cite{J-PARCE40:2021qxa,J-PARCE40:2021bgw,J-PARCE40:2022nvq}.  
Here, $\Si^+ p$ scattering is of particular interest for theory since 
it is a pure isospin $I=3/2$ system. Thus, there is no 
coupling to the $\La N$ channel which simplifies the dynamics.  
Moreover, there are, in principle, rather restrictive constraints from SU(3)
symmetry. Specifically, the space-spin antisymmetric states ($^1S_0$,
$^3P_{0,1,2}$, ...) belong all to the $\{27\}$ irrepresentation (irrep) of SU(3) 
(cf. Table~\ref{tab:SU3}) 
\cite{Haidenbauer:2013oca,Haidenbauer:2019boi} and thus
the corresponding interactions would be identical to those in the $NN$ 
system provided that SU(3) symmetry is exactly fulfilled. While there is a 
sizable SU(3) symmetry breaking in case of the $^1S_0$
partial wave \cite{Haidenbauer:2014rna}, the amplitudes in the $P$- and higher 
partial waves could be much closer to those found for $NN$ scattering.

\begin{table*}[t]
\caption{Scattering lengths ($a$) and effective ranges ($r$) 
for singlet (s) and triplet (t) $S$-waves (in fm), 
for $\Lambda N$, $\Si N$ with isospin $I=1/2,\, 3/2$, and for 
$\Si^+ p$ with inclusion of the Coulomb interaction. 
  }
 \label{tab:ereF}
\vskip 0.1cm
\renewcommand{\arraystretch}{1.4}
\begin{center}
\begin{tabular}{|l||rrr|rrr||r|r|}
\hline
\hline
& \multicolumn{3}{c|}{SMS NLO} & \multicolumn{3}{c||}{SMS N$^2$LO} & NLO13 & NLO19  \\
\hline
${\Lambda}$ [MeV] 
& 500     & 550     & 600     & 500     & 550     & 600     & 600     & 600  \\
\hline
\hline
$a^{\La N}_s$ 
& $-2.80$ & $-2.79$ & $-2.79$ & $-2.80$ & $-2.79$ & $-2.80$  & $-2.91$ & $-2.91$  \\
$r^{\La N}_s$ 
& $ 2.87$ & $ 2.72$ & $ 2.63$ & $ 2.82$ & $ 2.89$ & $ 2.68$  & $ 2.78$ & $ 2.78$  \\
\hline
$a^{\La N}_t$ 
& $-1.59$ & $-1.57$ & $-1.56$ & $-1.56$ & $-1.58$ & $-1.56$  & $-1.54$ & $-1.41$ \\
$r^{\La N}_t$ 
& $ 3.10$ & $ 2.99$ & $ 3.00$ & $ 3.16$ & $ 3.09$ & $ 3.17$  & $ 2.72$  & $ 2.53$ \\
\hline
\hline
Re\,$a^{\Si N \ (I=1/2)}_{s}$
& $ 1.14$ & $ 1.15$ & $ 1.10$ & $ 1.03$ & $ 1.12$ & $ 1.06$  & $ 0.90$ & $ 0.90$ \\
Im\,$a^{\Si N}_{s}$ 
& $ 0.00$ & $ 0.00$ & $ 0.00$ & $ 0.00$ & $ 0.00$ & $ 0.00$  & $ 0.00$ & $ 0.00$ \\
\hline
Re\,$a^{\Si N \ (I=1/2)}_{t}$
& $ 2.58$ & $ 2.42$ & $ 2.31$ & $ 2.60$ & $ 2.38$ & $ 2.53$  & $ 2.27$ & $ 2.29$  \\
Im\,$a^{\Si N}_{t}$ 
& $-2.60$ & $-2.95$ & $-3.09$ & $-2.56$ & $-3.26$ & $-2.64$  & $-3.29$ & $-3.39$  \\
\hline
$a^{\Si N \ (I=3/2)}_s$
& $-4.21$ & $-4.05$ & $-4.11$ & $-4.37$ & $-4.19$ & $-4.03$  & $-4.45$ & $-4.55$  \\
$r^{\Si N}_s$ 
& $ 3.93$ & $ 3.89$ & $ 3.75$ & $ 3.73$ & $ 3.89$ & $ 3.74$  & $ 3.68$ & $ 3.65$  \\
\hline
$a^{\Si N \ (I=3/2)}_t$
& $ 0.46$ & $ 0.47$ & $ 0.47$ & $ 0.38$ & $ 0.44$ & $ 0.41$  & $ 0.44$ & $ 0.43$  \\
$r^{\Si N}_t$ 
& $-5.08$ & $-4.74$ & $-4.82$ & $-5.70$ & $-4.96$ & $-5.72$  & $-4.59$ & $-5.27$  \\
\hline
\hline
$a^{\Si^+ p}_s$ 
& $-3.41$ & $-3.30$ & $-3.44$ & $-3.47$ & $-3.39$ & $-3.25$  & $-3.56$ & $-3.62$  \\
$r^{\Si^+ p}_s$ 
& $ 3.75$ & $ 3.73$ & $ 3.59$ & $ 3.61$ & $ 3.73$ & $ 3.65$  & $ 3.54$ & $ 3.50$  \\
\hline
$a^{\Si^+ p}_t$ 
& $ 0.51$ & $ 0.52$ & $ 0.52$ & $ 0.41$ & $ 0.48$ & $ 0.45$  & $ 0.49$ & $ 0.47$  \\
$r^{\Si^+ p}_t$ 
& $-5.46$ & $-5.12$ & $-5.19$ & $-6.74$ & $-5.50$ & $-6.41$  & $-5.08$ & $-5.77$  \\
\hline
\hline
\end{tabular}
\end{center}
\renewcommand{\arraystretch}{1.0}
\end{table*}

Note that the cross sections in the $\Sigma^+ p\to \Sigma^+ p$ 
and $\Sigma^- p\to \Sigma^- p$ channels in past studies
were obtained from experiments with an incomplete angular coverage 
by defining \cite{Eisele:1971mk} 
\begin{eqnarray}
\label{eq:sigtot}
\sigma&=&\frac{2}{\cos \theta_{{\rm max}}-\cos \theta_{{\rm min}}}
\int_{\cos \theta_{{\rm min}}}^{\cos \theta_{{\rm max}}}\frac{d\sigma(\theta)}{d\cos \theta}d\cos \theta \ .
\end{eqnarray}
We use the same prescription, and specifically $\cos \theta_{{\rm min}}=-0.5$ and
$\cos \theta_{{\rm max}}=0.5$,
for obtaining ``integrated'' $\Sigma^+ p$ and $\Sigma^- p$ cross sections.

%%%%%%%%%%%%%%%%%%%%%%%%%%%%%%%%%%%%%%%%%%%%%%%
\subsection{The $\Si^+ p$ channel} 
\label{sec:Spp}

\begin{figure*}[t]
\centering
\includegraphics[width=0.41\textwidth]{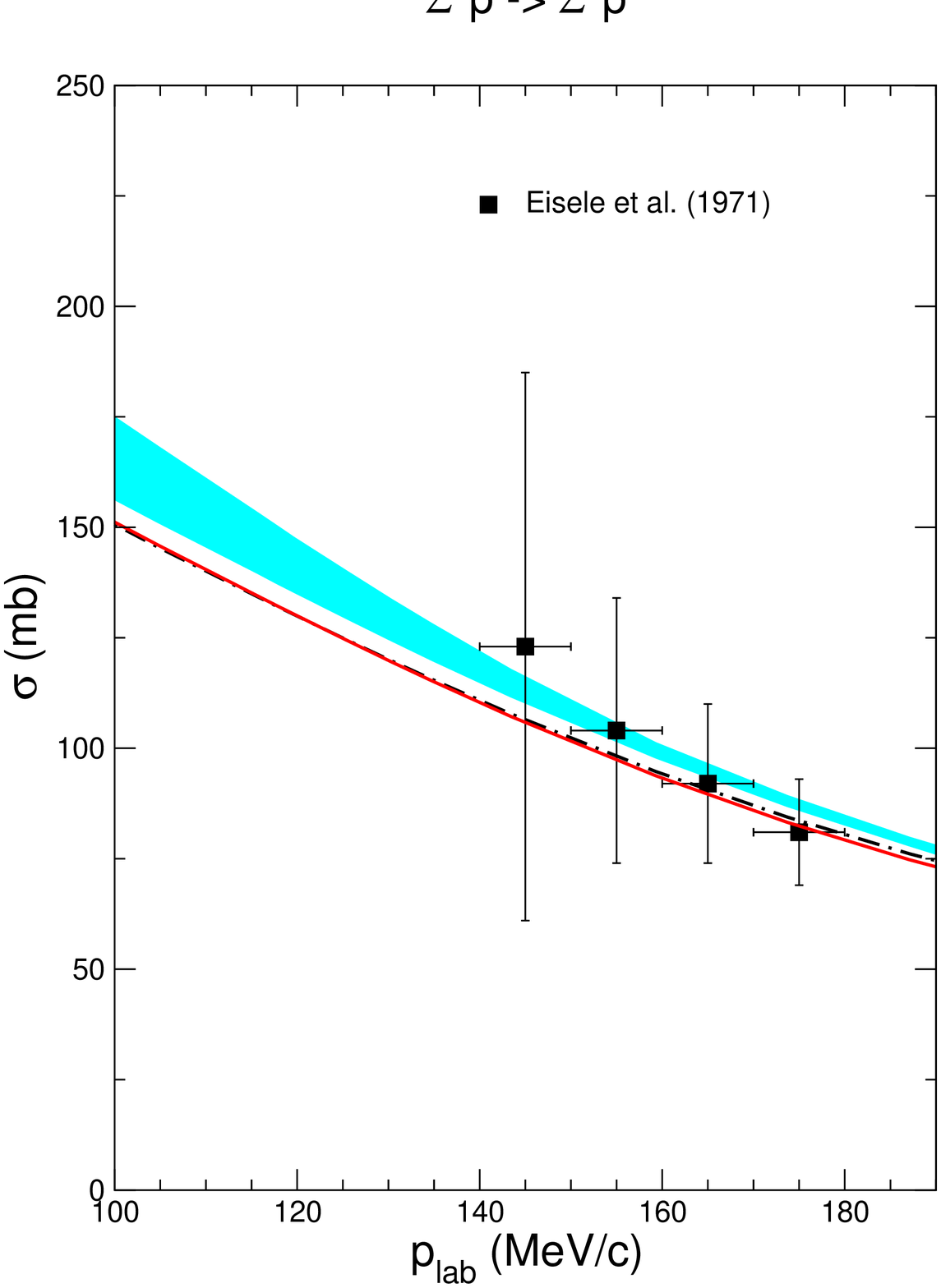}\includegraphics[width=0.41\textwidth]{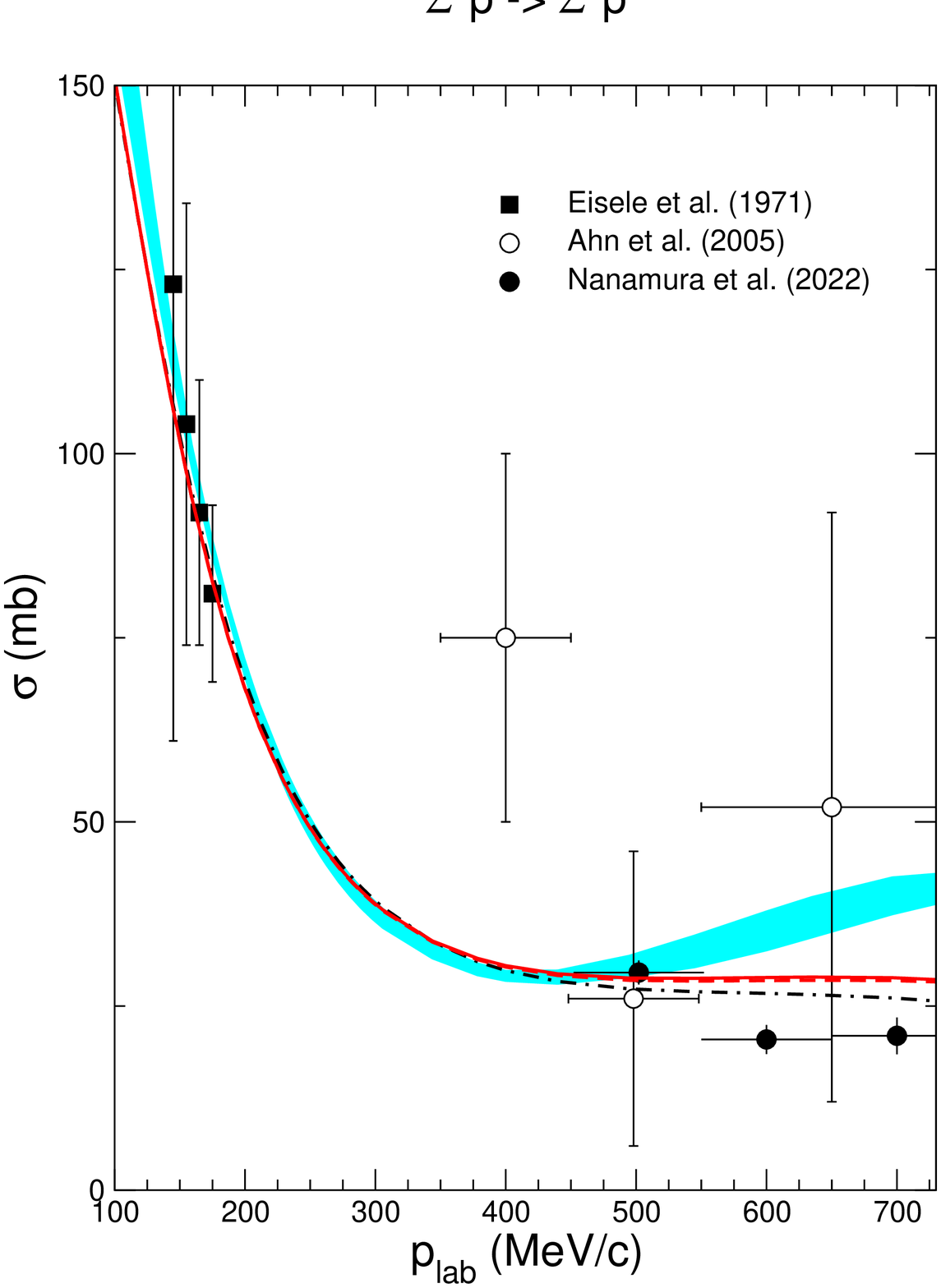}

\vspace*{-0.3cm}
\includegraphics[width=0.33\textwidth]{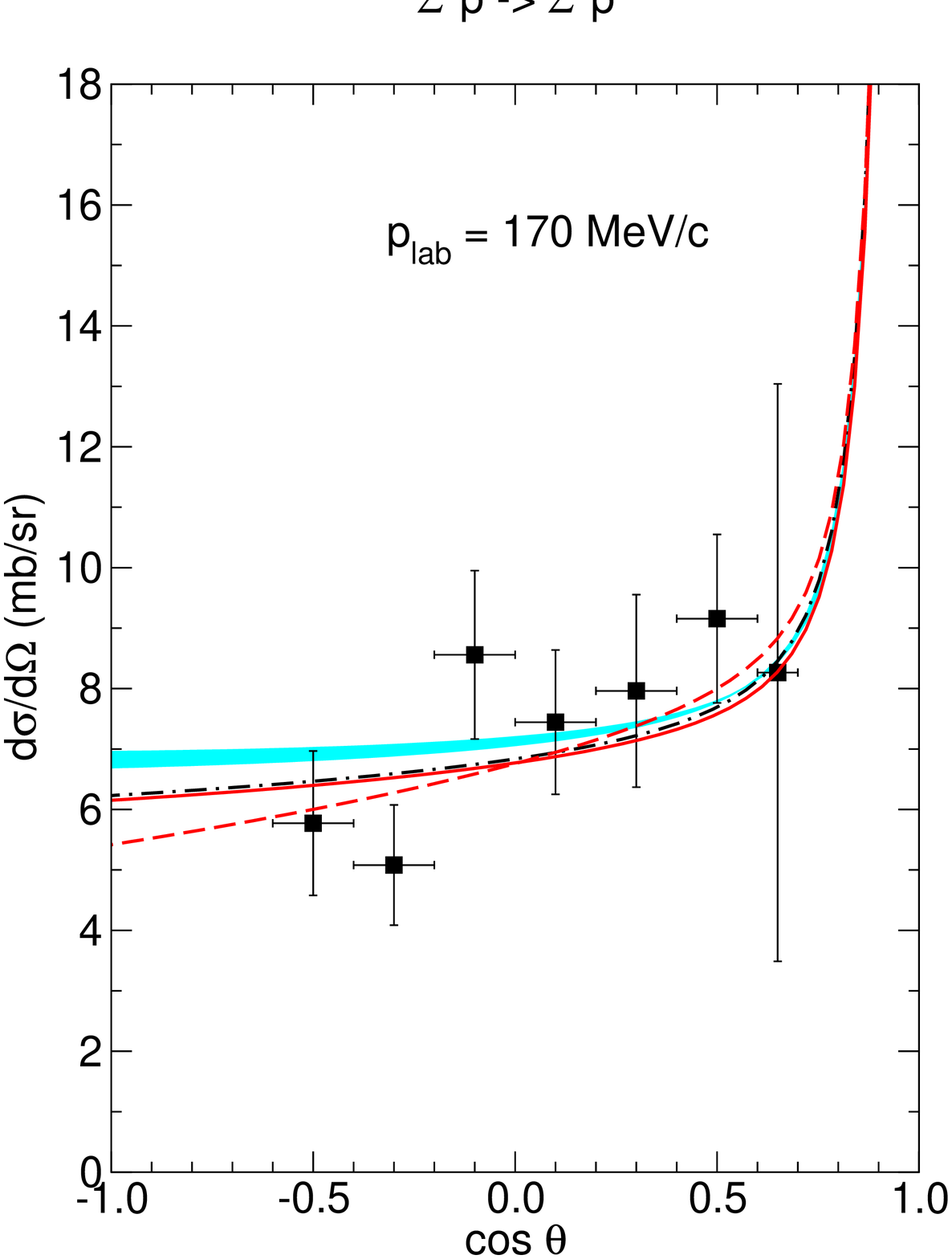}\includegraphics[width=0.33\textwidth]{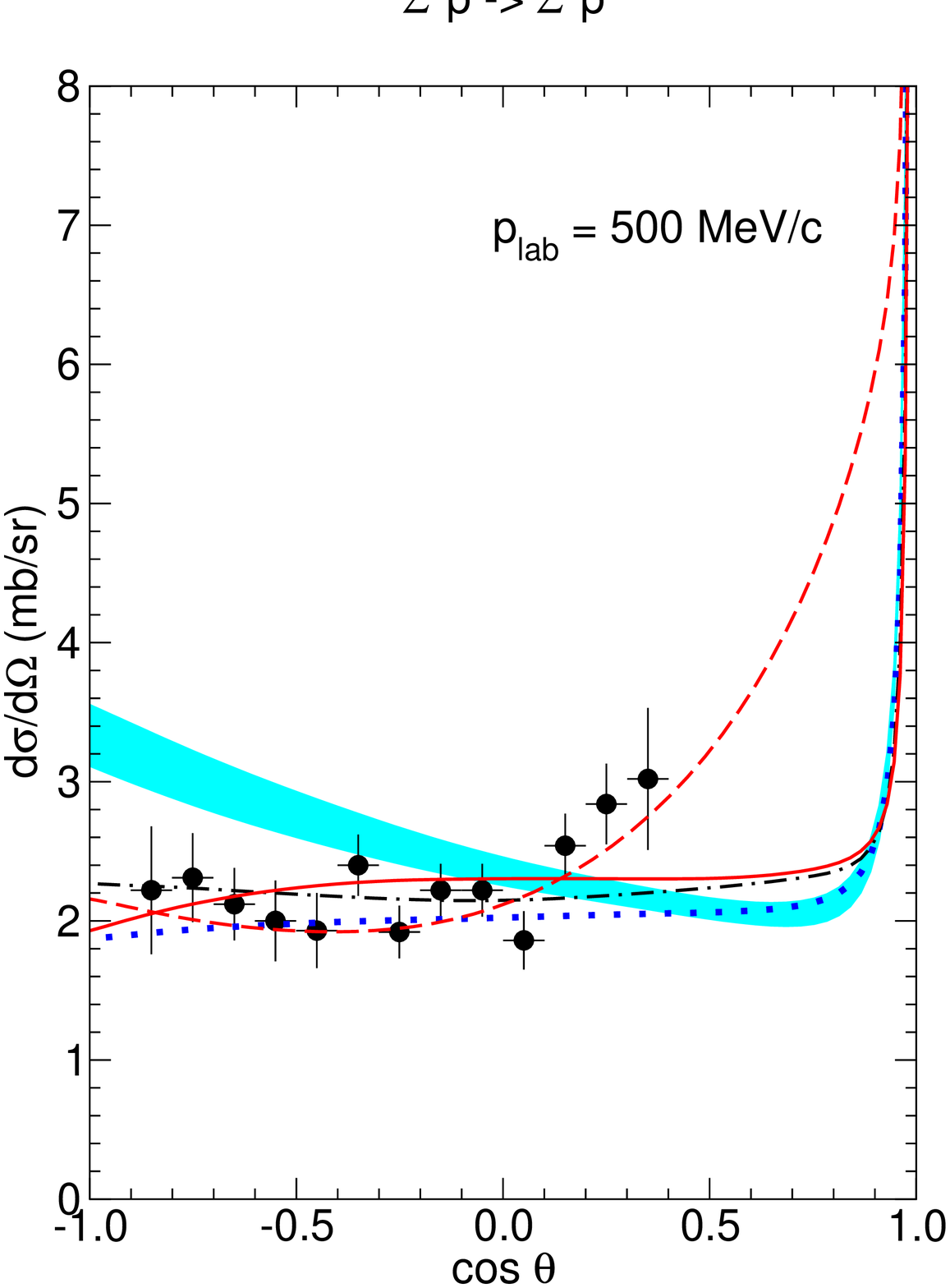}
\includegraphics[width=0.33\textwidth]{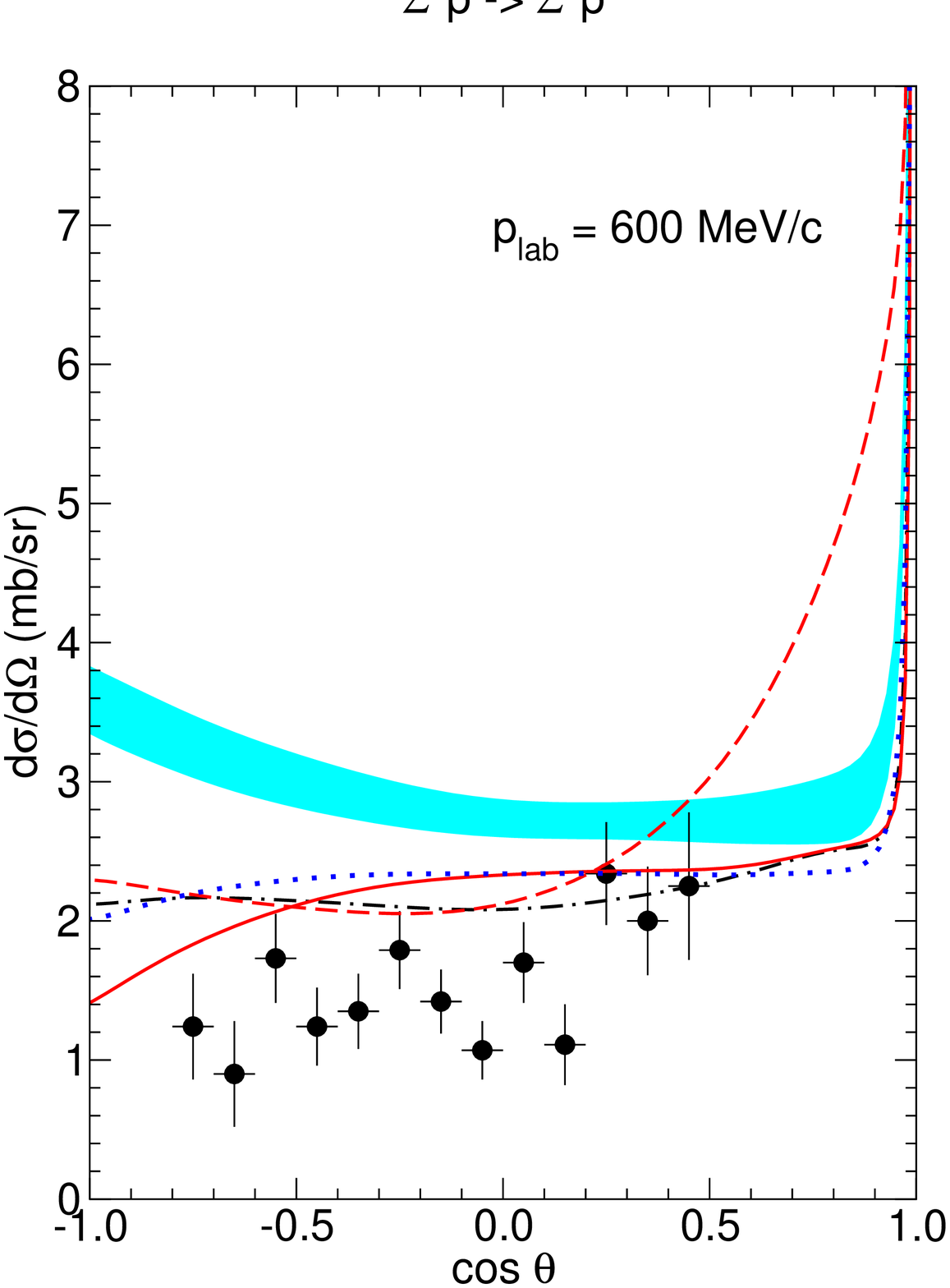}
\vspace*{-0.8cm}
\caption{Cross section for $\Si^+ p$ scattering as a function of $p_{\rm lab}$. Results 
are shown for the SMS NLO (dash-dotted) and N$^2$LO (solid) $YN$ potentials
with cutoff $550$~MeV. 
The dashed line corresponds to an alternative fit at N$^2$LO, see text. 
The cyan band is the result for NLO19 \cite{Haidenbauer:2019boi}. The dotted line
is the result for NLO19(600) with readjusted $C_{^3SD_1}$, see text. 
Data are from the E40 experiment \cite{J-PARCE40:2022nvq} for the momentum
regions $440-550$ and $550-650$~Mev/c, respectively, 
and from Refs.~\cite{Eisele:1971mk,KEK-PSE289:2005nsj}.  
%\vspace*{-0.7cm}
}
\label{fig:cs5}
\end{figure*}

$\Si^+ p$ scattering cross sections for the SMS $YN$ interactions are 
presented in Fig.~\ref{fig:cs5}, and compared with data and with
the results obtained from the NLO19 potential. The latter are shown as 
bands, representing the cutoff dependence \cite{Haidenbauer:2019boi}. 
On the upper left side the cross section at low energies is displayed. 
This is the region with the data of Eisele et al.~\cite{Eisele:1971mk}, which are 
included in the fitting procedure for the $S$-wave LECs. One can see that
the results for the SMS potentials are slightly below those of NLO19. 
The main reason for that is that we no longer impose strict SU(3) constraints
on the $S$-wave contact terms.

\begin{figure*}[t]
\centering
\includegraphics[width=0.62\textwidth]{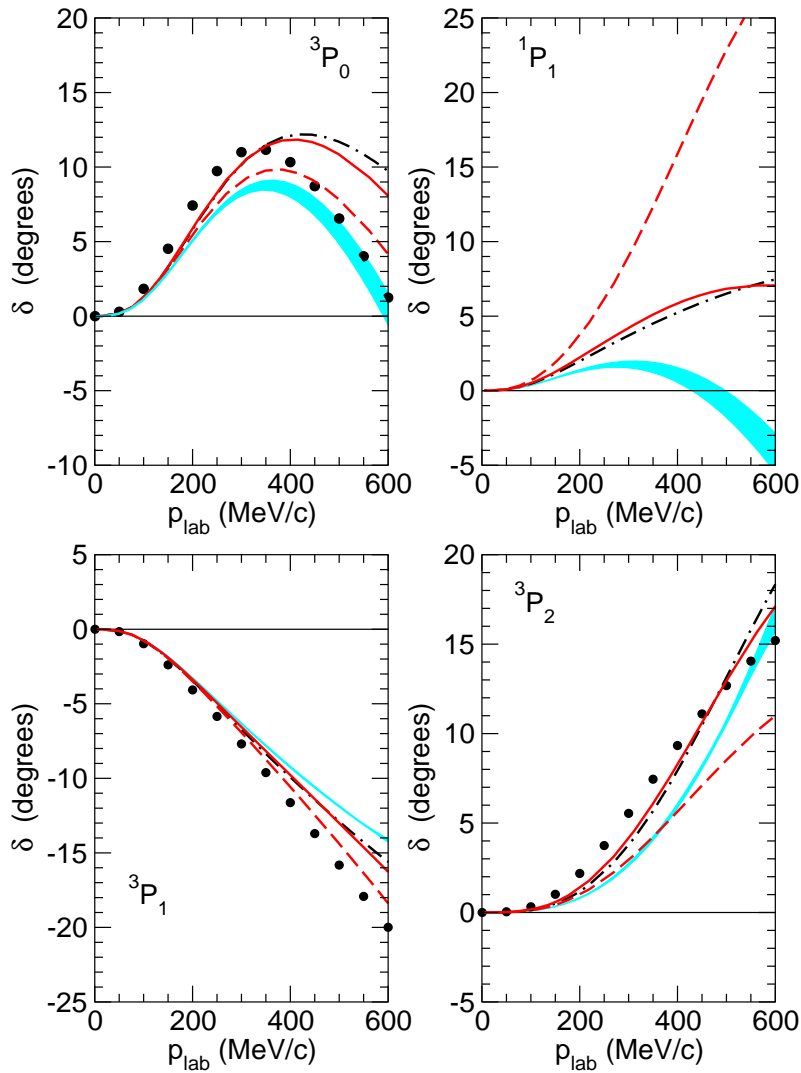}
\caption{$\Si N$ $I=3/2$ phase shifts: $P$-waves.
Same description of the curves as in Fig.~\ref{fig:cs5}. For illustrating
the extent of SU(3) symmetry breaking, $NN$ phase shifts \cite{Arndt:1994br,SAID:webpage} 
for partial waves in the pertinent $\{27\}$ irrep are indicated 
by circles. 
}
\label{fig:Sphp}
\end{figure*}

\begin{figure*}[t]
\centering
\includegraphics[width=0.62\textwidth]{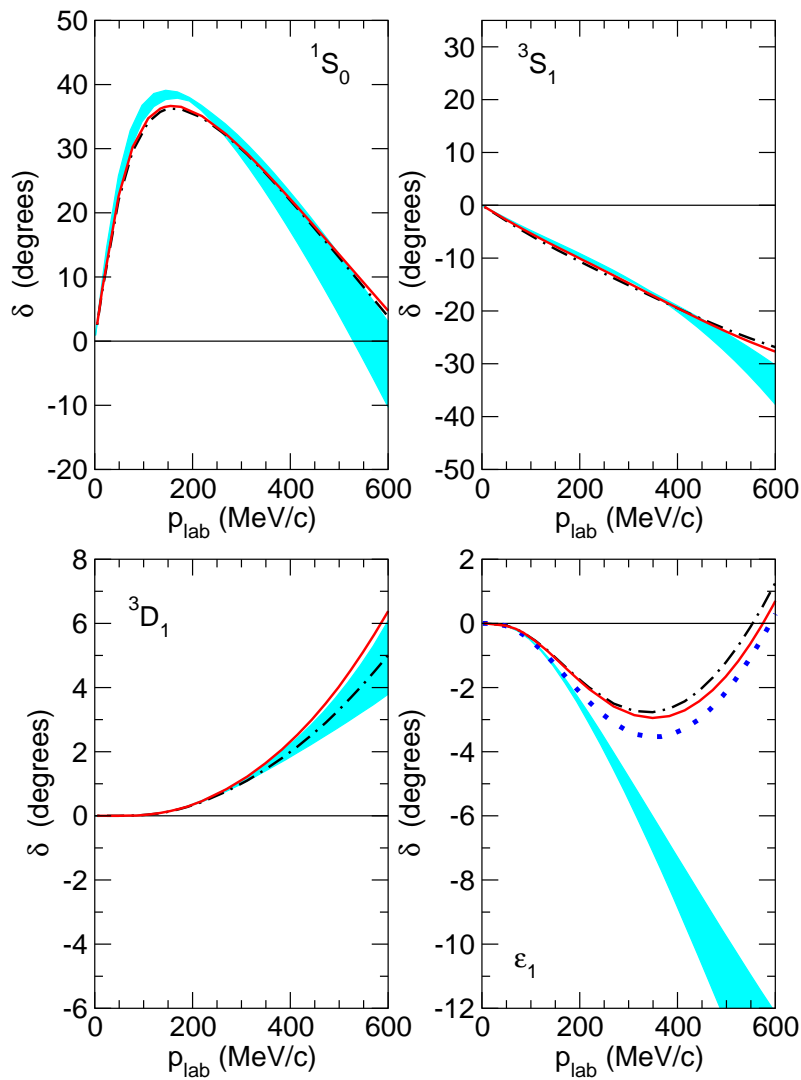}
\caption{$\Si N$ $I=3/2$ phase shifts: $^1S_0$ and $^3S_1$-$^3D_1$.
Same description of the curves as in Fig.~\ref{fig:cs5}. 
}
\label{fig:Sphs}
\end{figure*}

\begin{figure*}[t]
\centering
\includegraphics[width=0.41\textwidth]{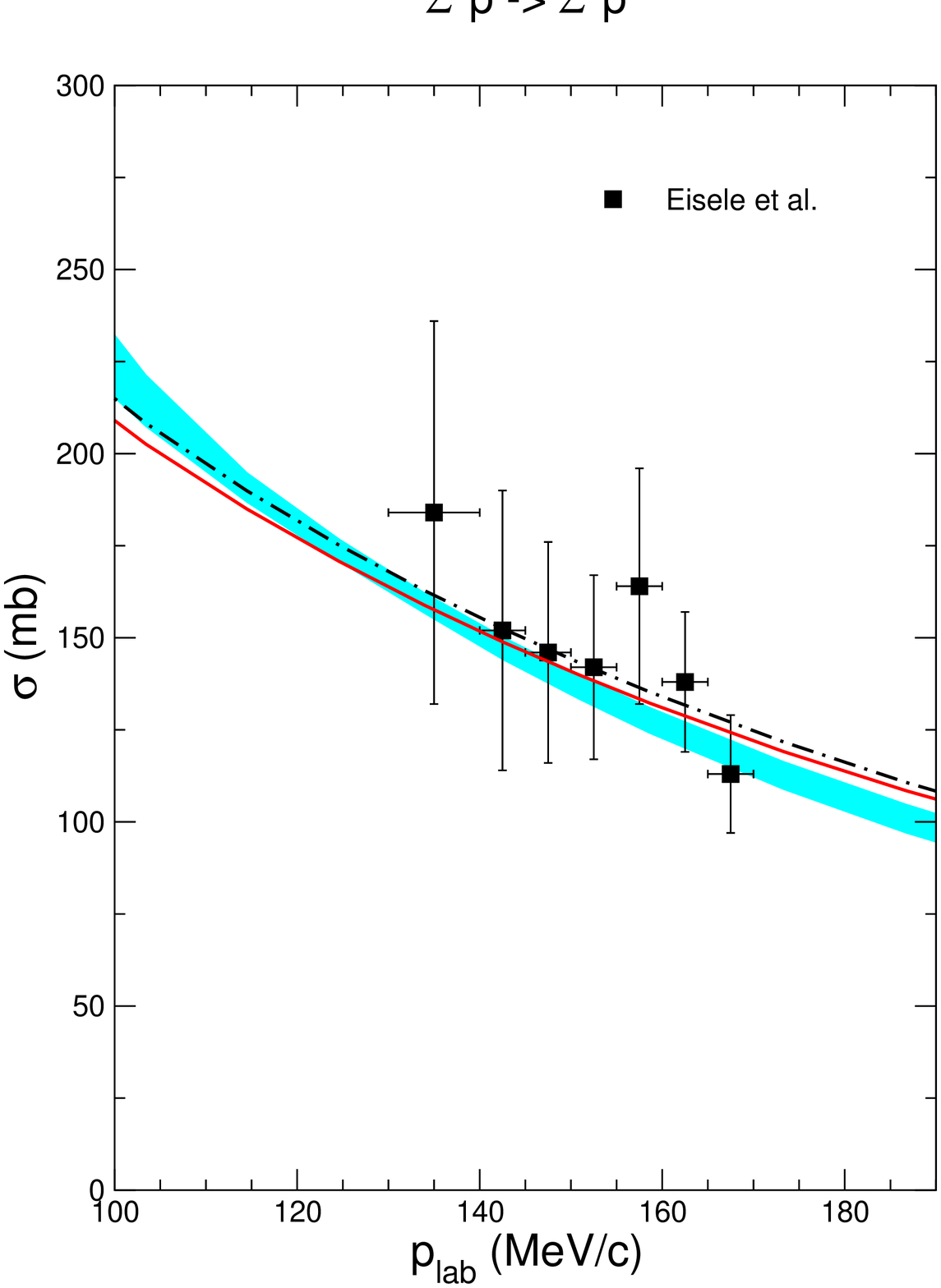}\includegraphics[width=0.41\textwidth]{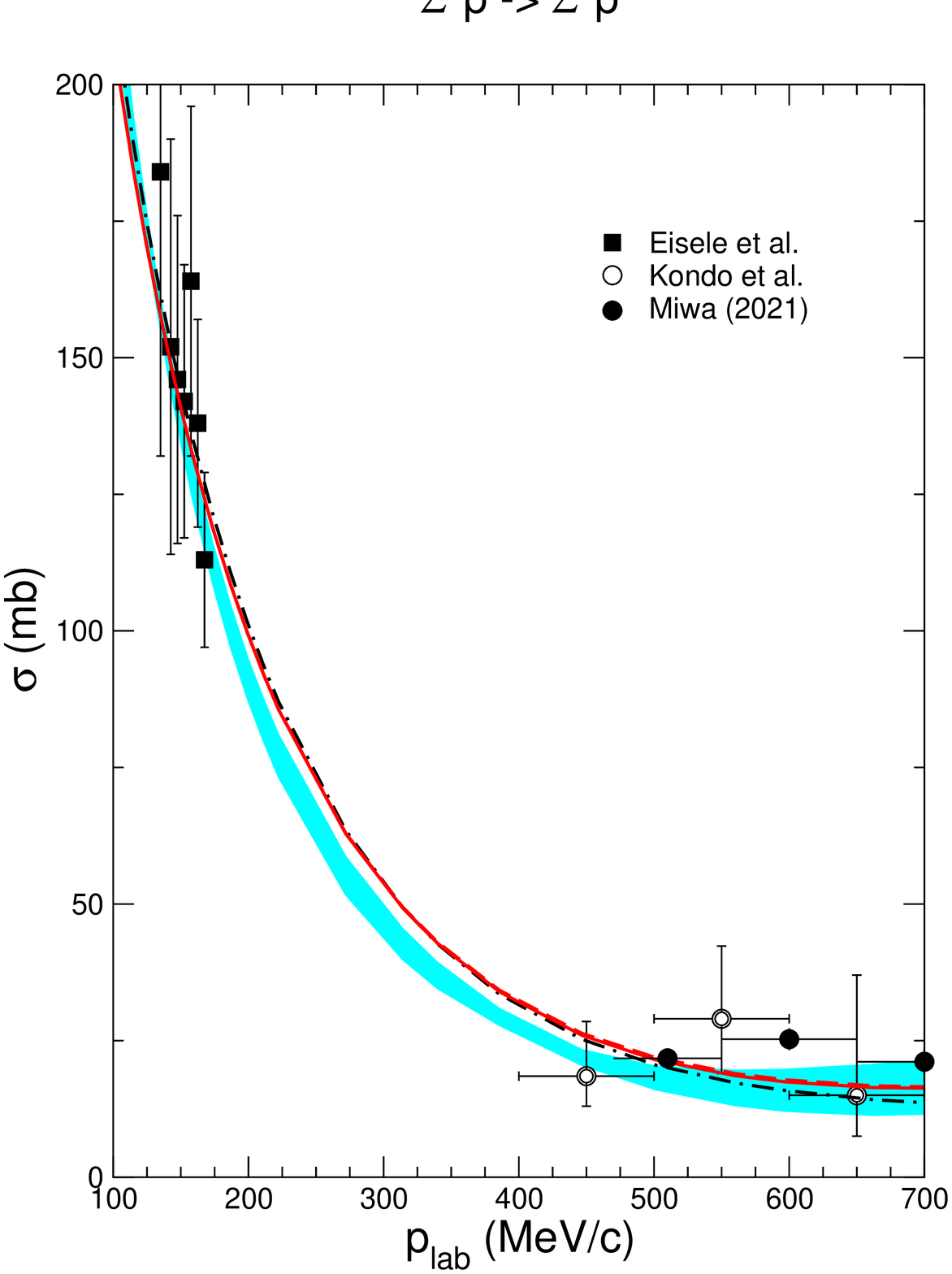}

 \includegraphics[width=0.33\textwidth]{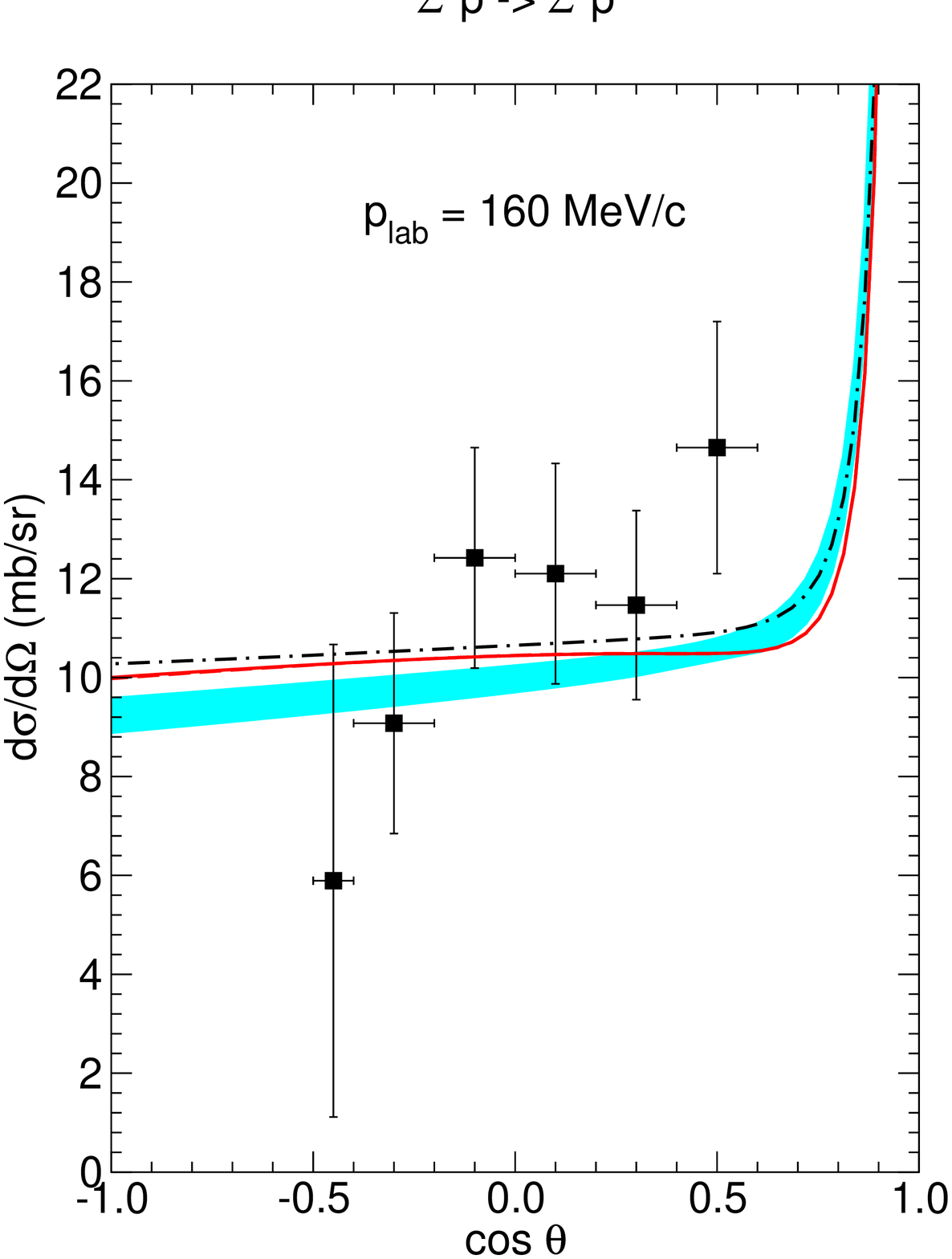}\includegraphics[width=0.33\textwidth]{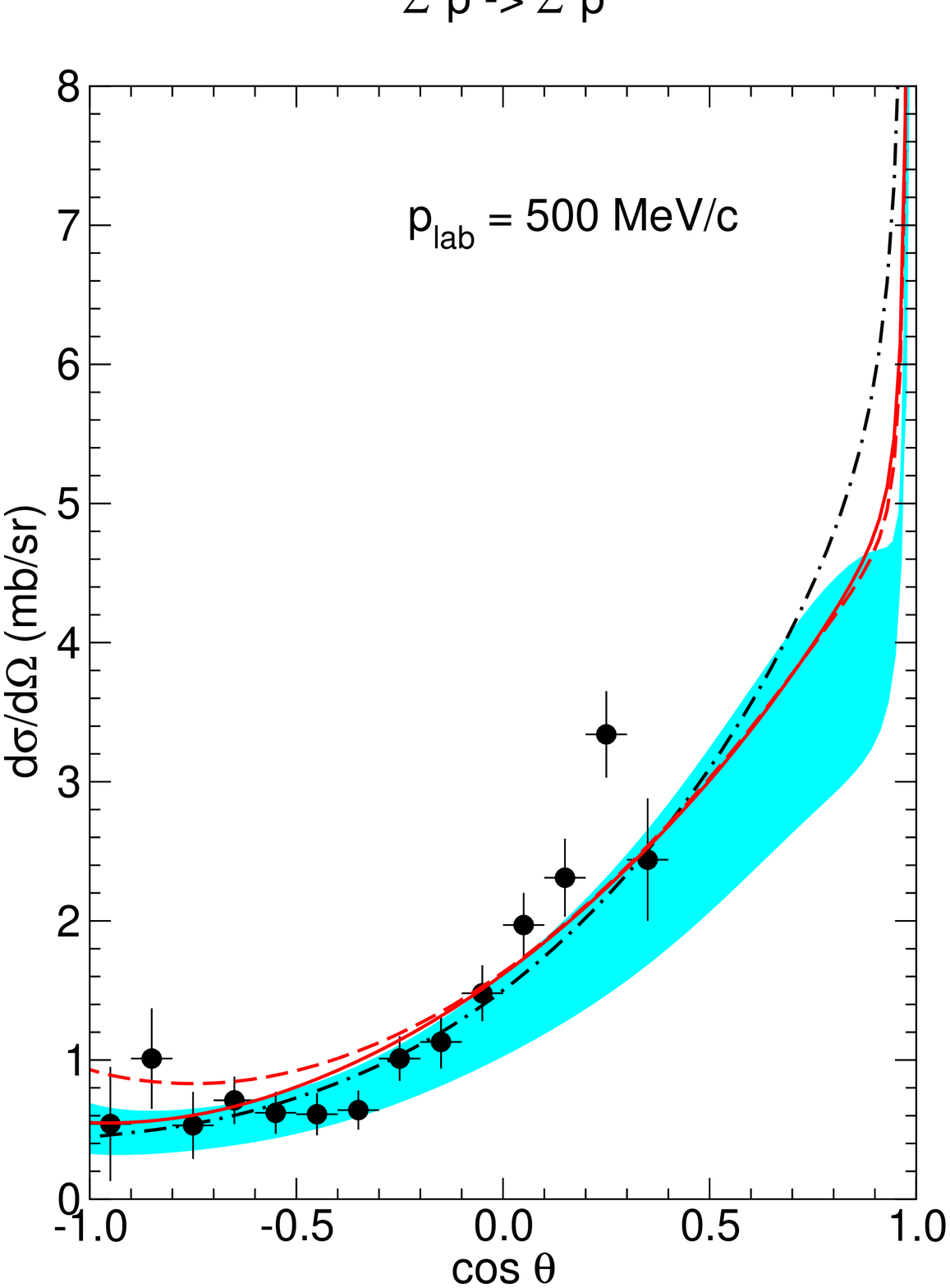}
 \includegraphics[width=0.33\textwidth]{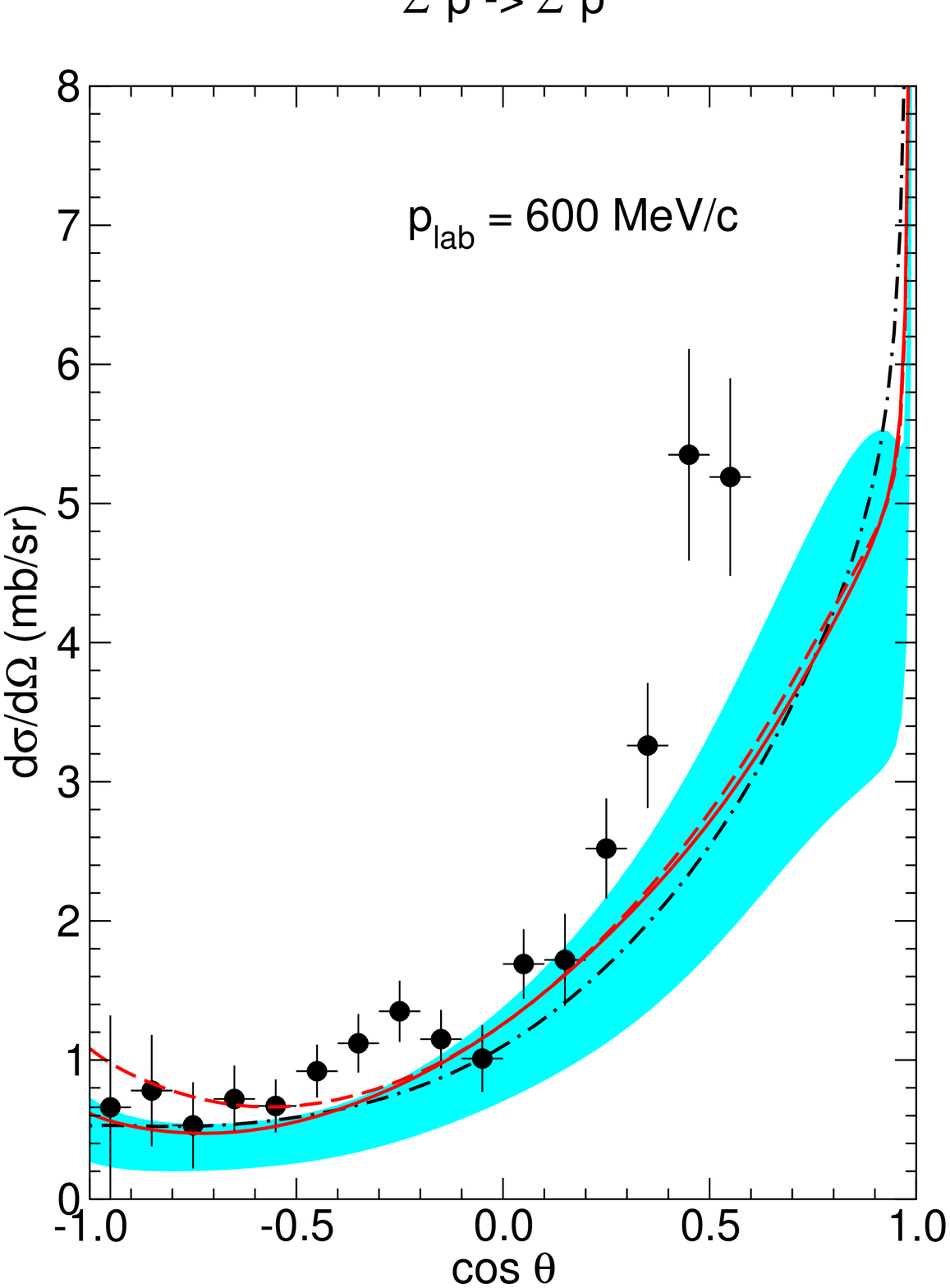}
 \vspace*{-0.6cm}
\caption{Cross section for $\Si^- p$ scattering as a function of $p_{\rm lab}$. 
Same description of the curves as in Fig.~\ref{fig:cs5}. 
Data are from the E40 Collaboration \cite{J-PARCE40:2021qxa} for the momentum
regions $470-550$ and $550-650$~MeV/c, respectively, 
and from Refs.~\cite{Eisele:1971mk,KEK-PS-E289:2000ytt}.  
%\vspace*{-0.7cm}
}
\label{fig:cs4}
\end{figure*}

\begin{figure*}[t]
\centering
\includegraphics[width=0.41\textwidth]{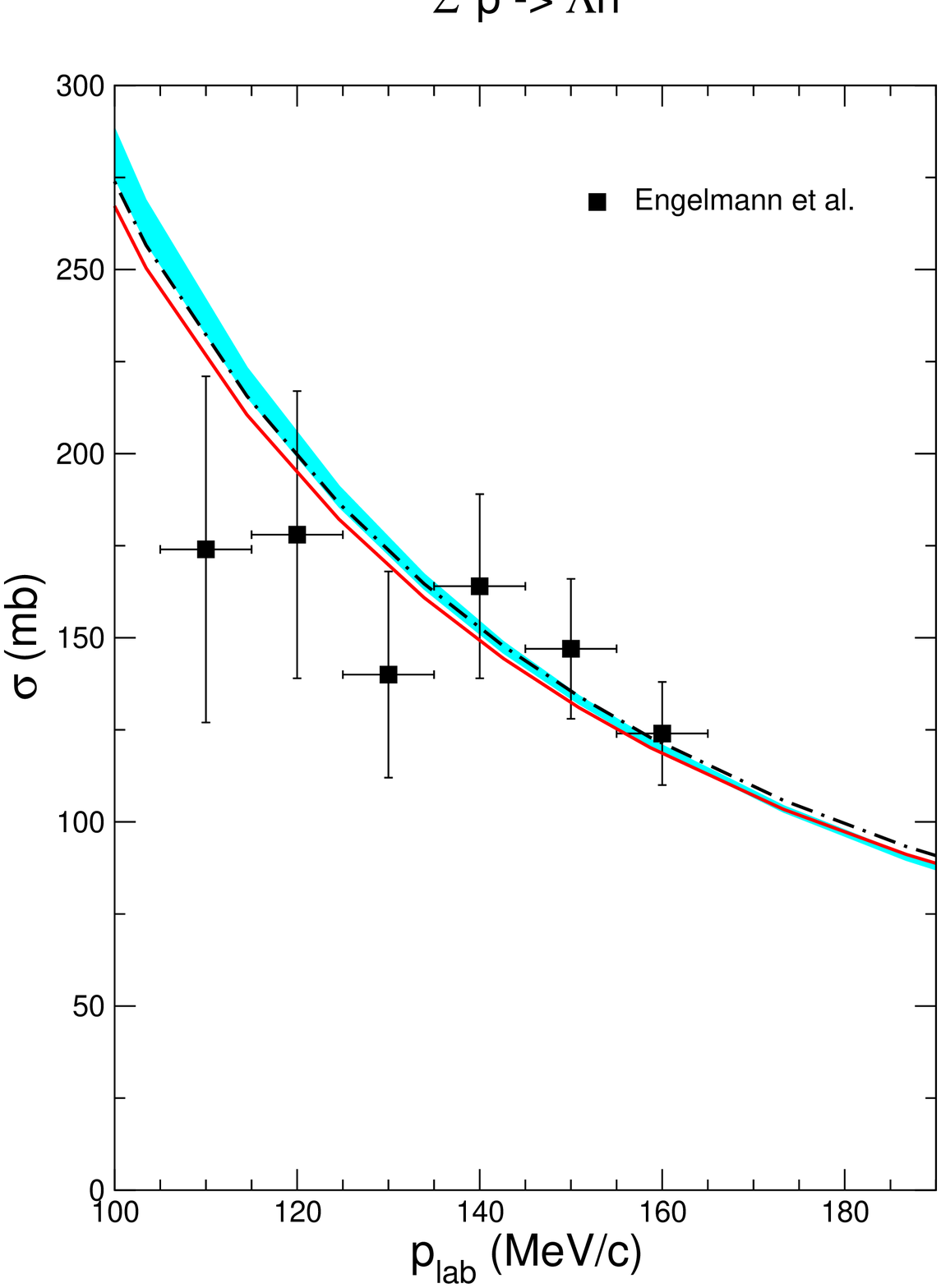}\includegraphics[width=0.41\textwidth]{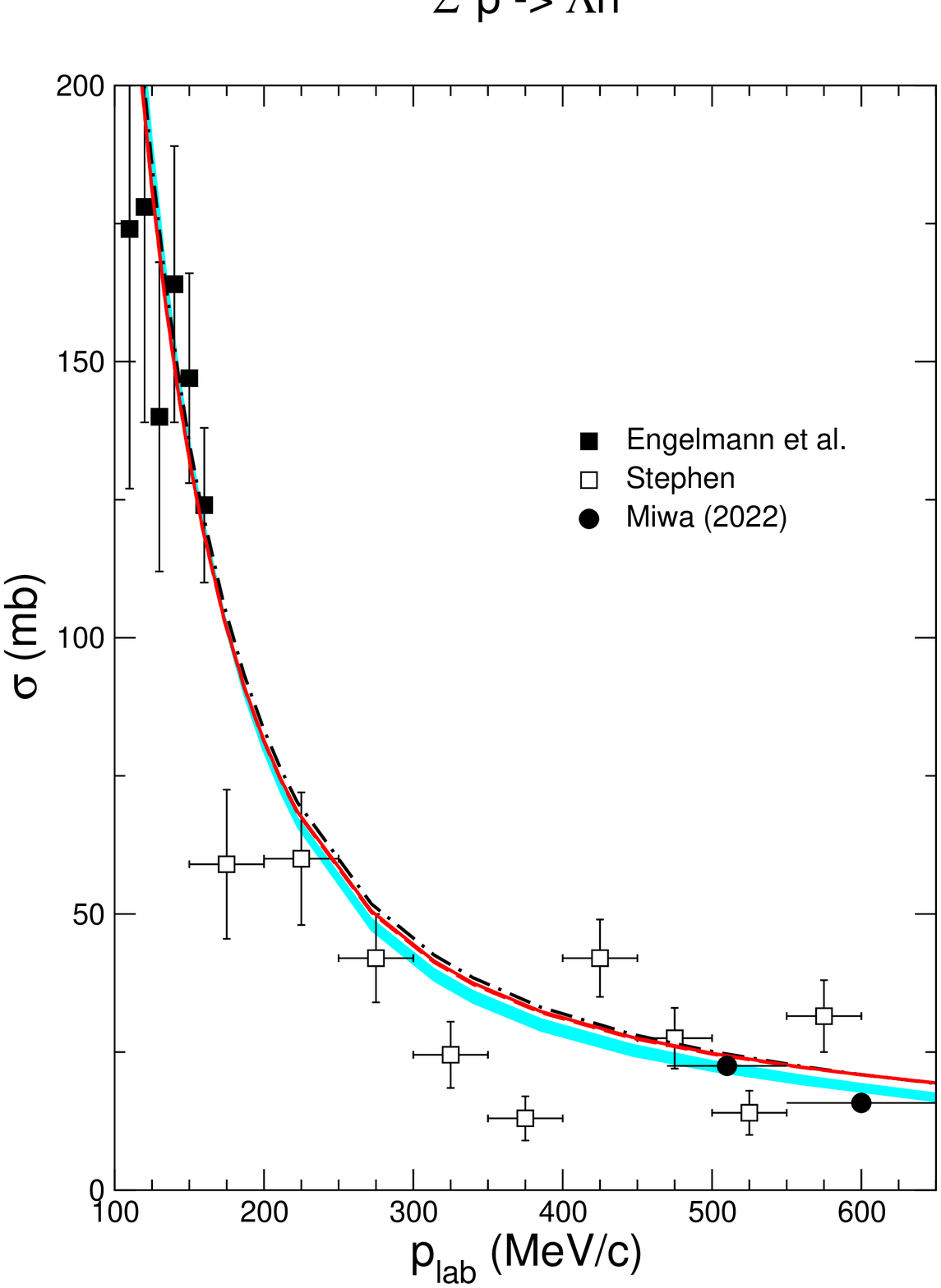}

\includegraphics[width=0.33\textwidth]{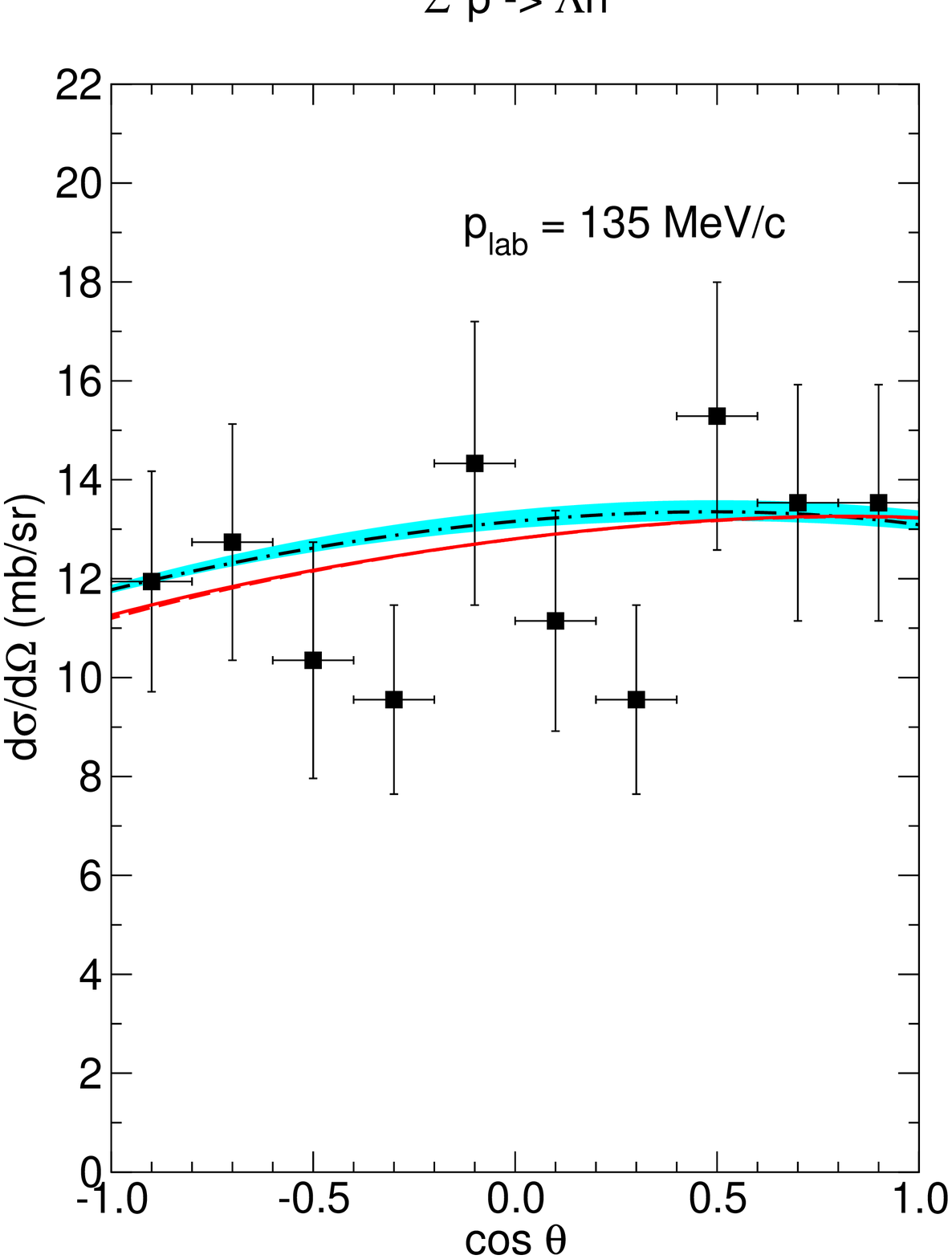}\includegraphics[width=0.33\textwidth]{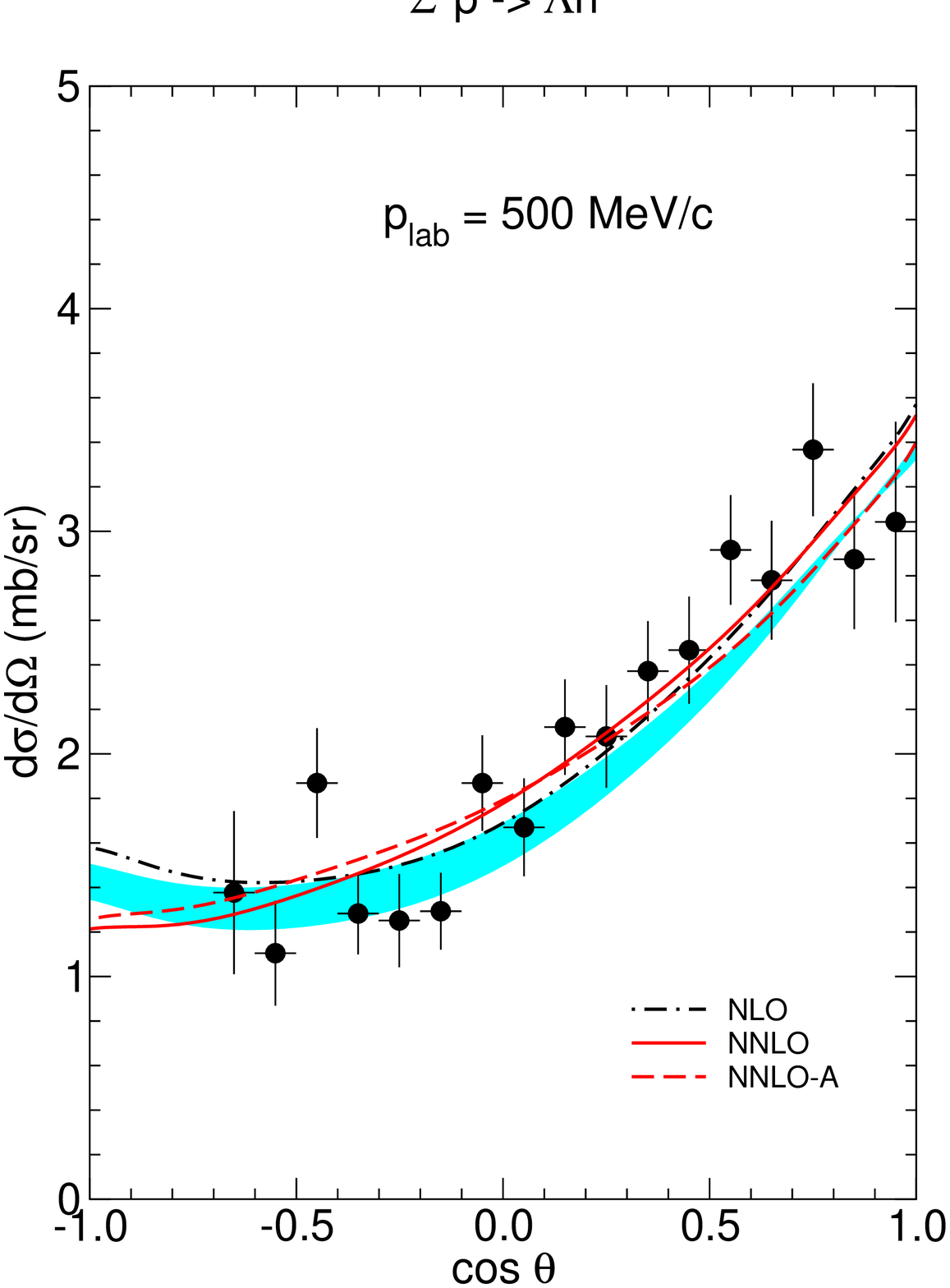}
\includegraphics[width=0.33\textwidth]{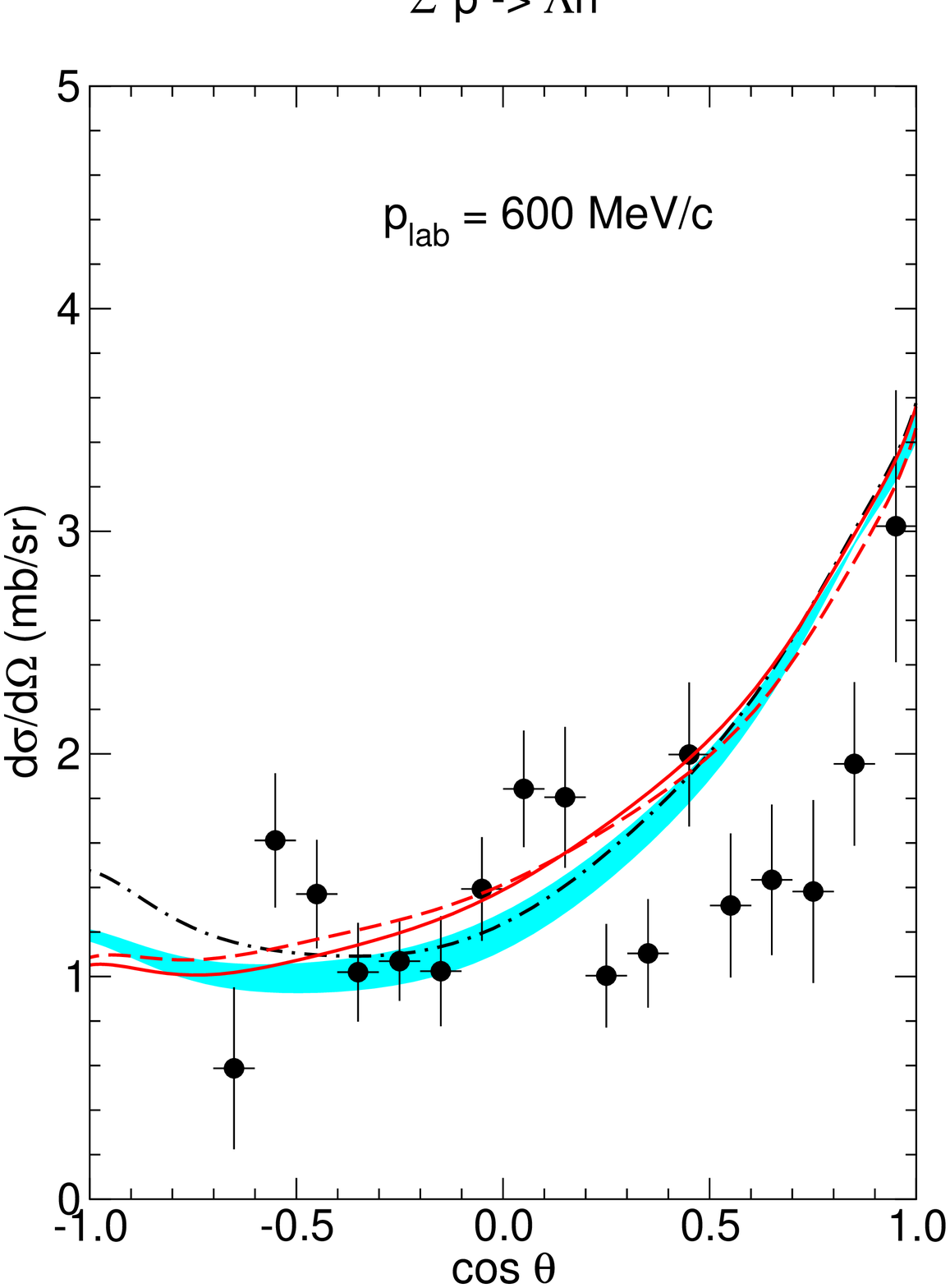}
 \vspace*{-0.6cm}
\caption{Cross section for $\Si^- p \to \La n$ as a function of $p_{\rm lab}$. 
Same description of the curves as in Fig.~\ref{fig:cs5}. 
Data are from the E40 Collaboration \cite{J-PARCE40:2021bgw} for the momentum
regions $470-550$ and $550-650$~MeV/c, respectively, 
and from Refs.~\cite{Engelmann:1966pl,StephenPhD:1970wd}.  
%\vspace*{-0.7cm}
}
\label{fig:cs2}
\end{figure*}
 
\begin{figure*}[t]
\centering
\includegraphics[width=0.41\textwidth]{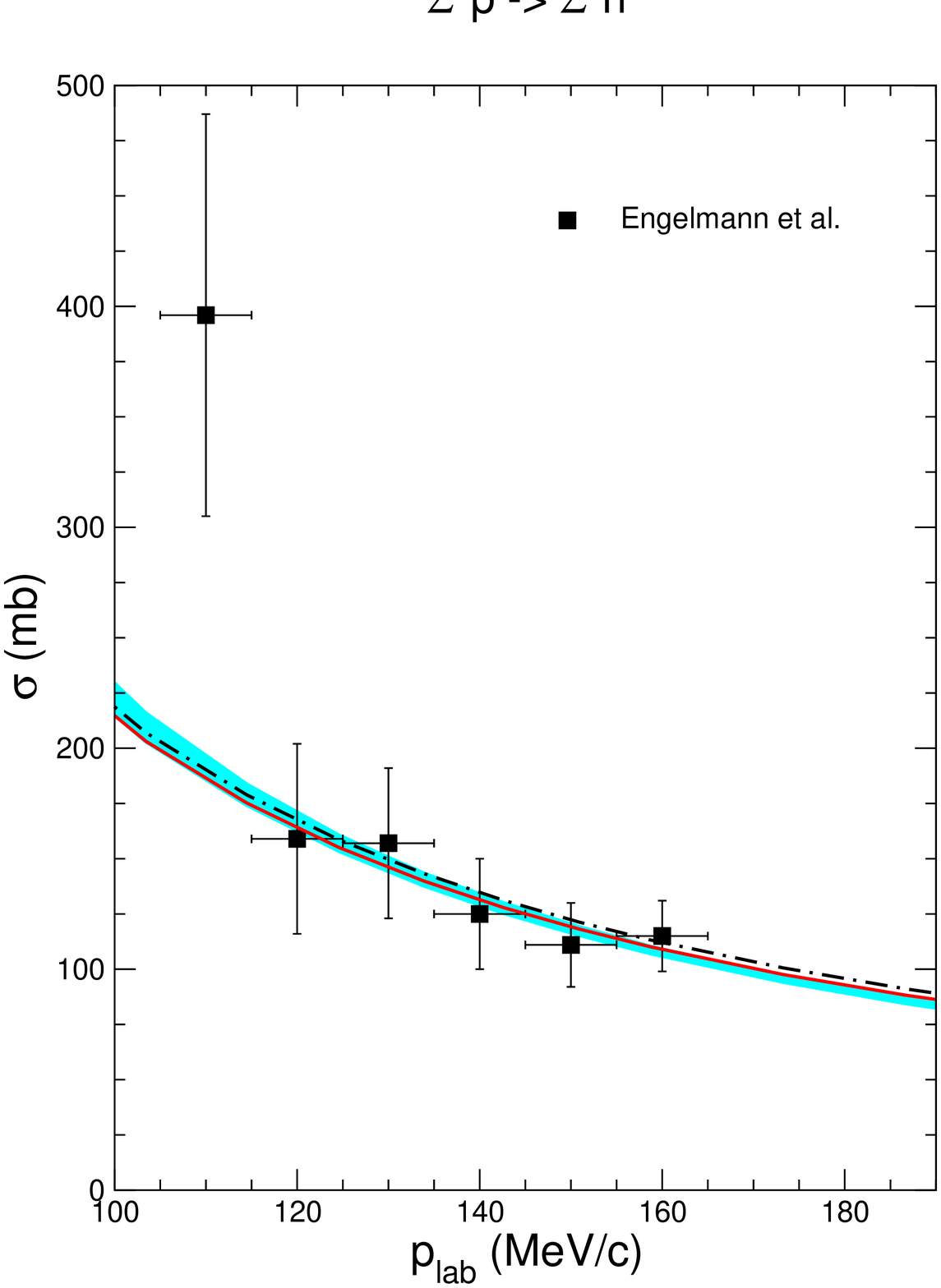}\includegraphics[width=0.41\textwidth]{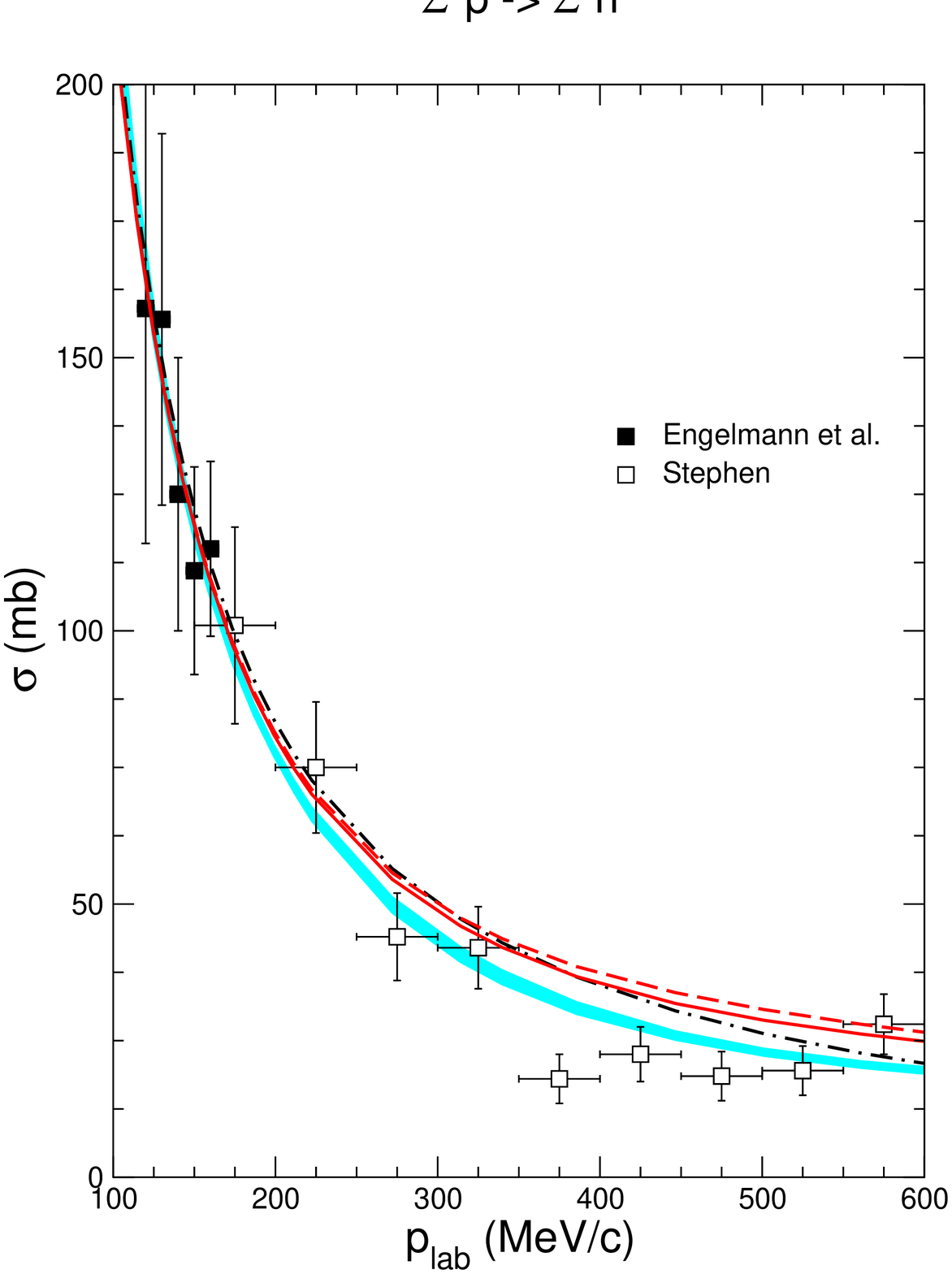}
 \vspace*{-0.6cm}
\caption{Cross section for $\Si^- p \to \Si^0 n$ as a function of $p_{\rm lab}$. 
Same description of the curves as in Fig.~\ref{fig:cs5}. 
Data are from Refs.~\cite{Engelmann:1966pl,StephenPhD:1970wd}.  
%\vspace*{-0.7cm}
}
\label{fig:cs3}
\end{figure*}
 
\begin{figure*}[t]
\centering
\includegraphics[width=0.41\textwidth]{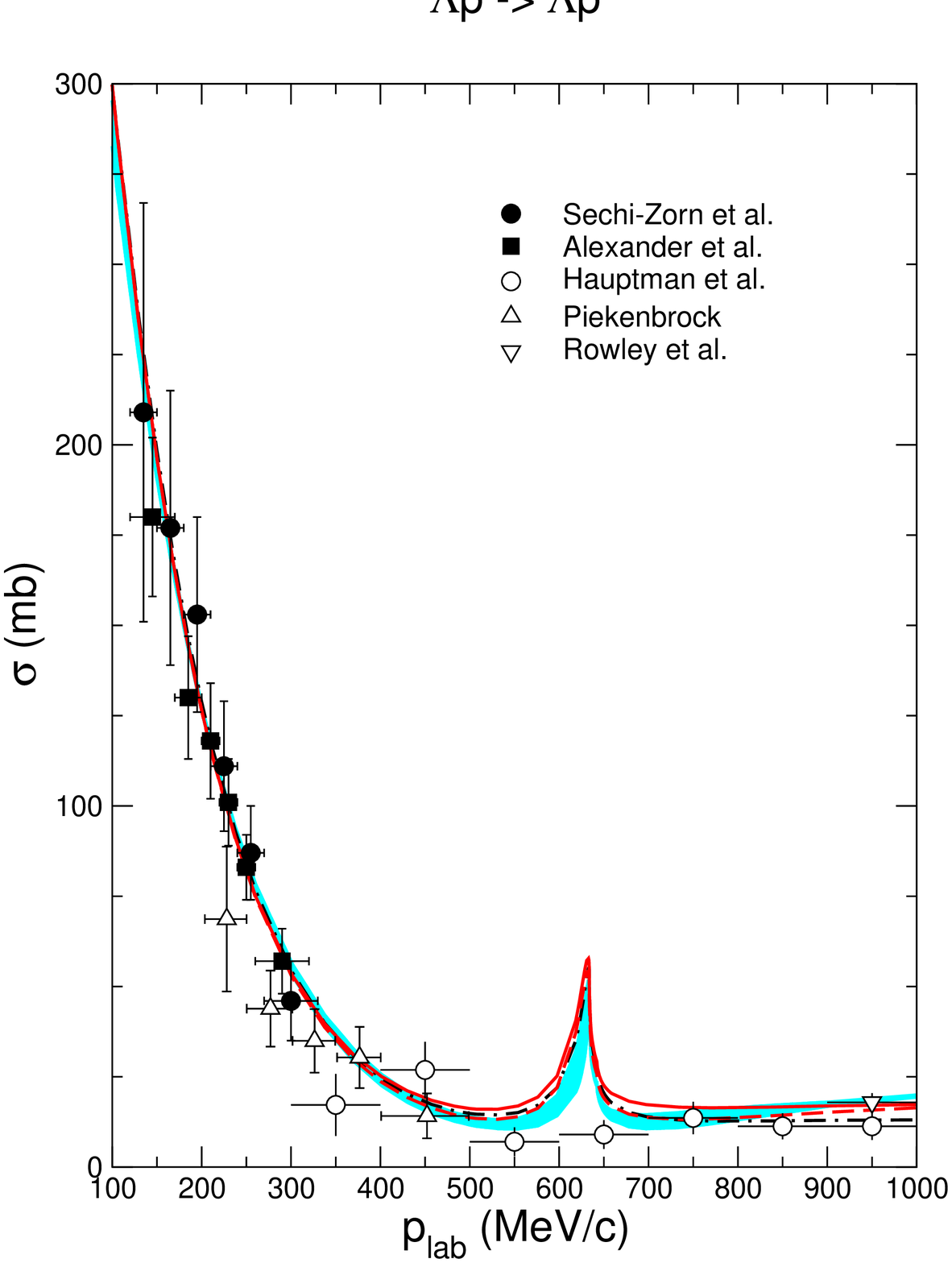}\includegraphics[width=0.41\textwidth]{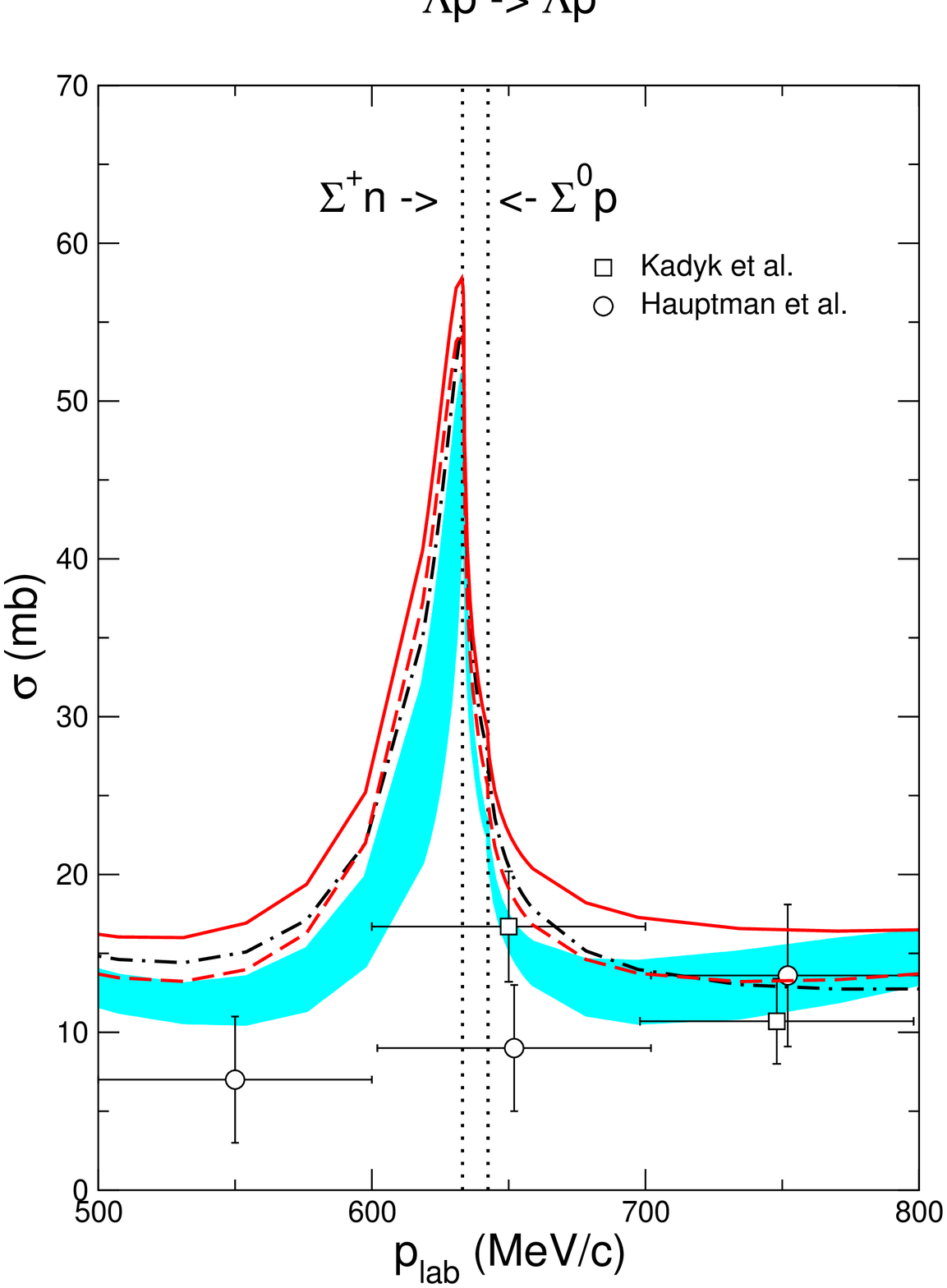}

\includegraphics[width=0.33\textwidth]{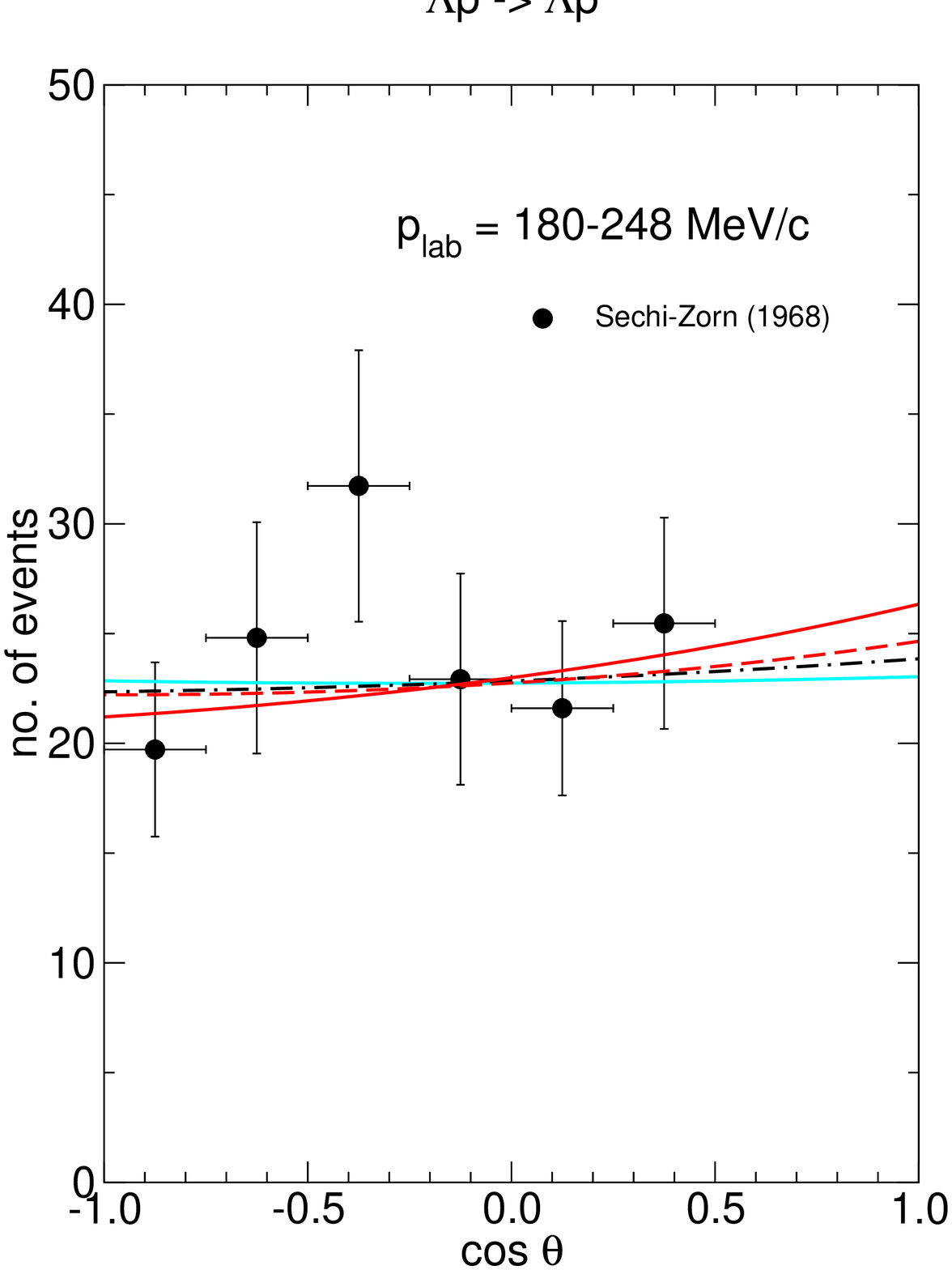}\includegraphics[width=0.33\textwidth]{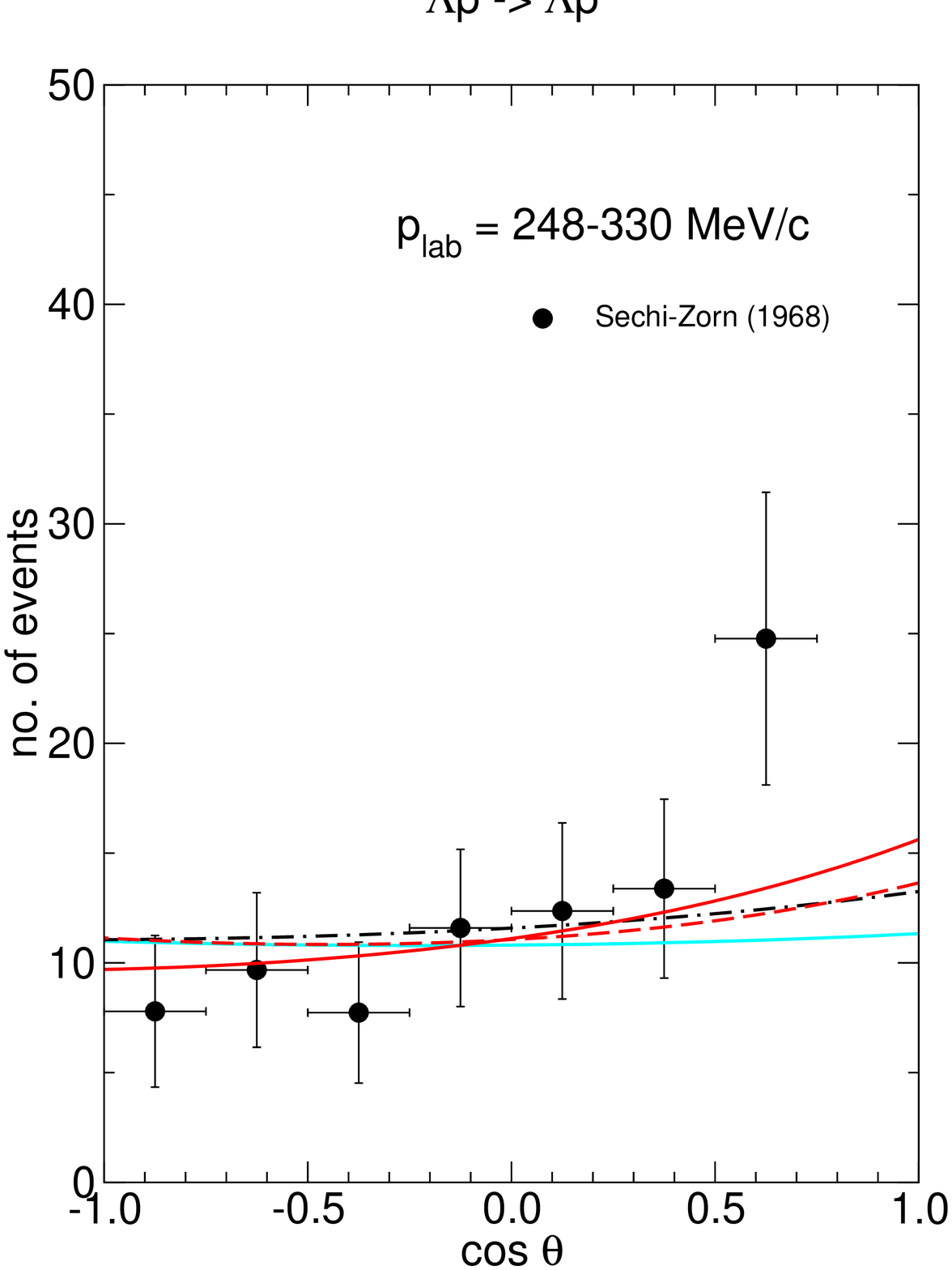}
\includegraphics[width=0.33\textwidth]{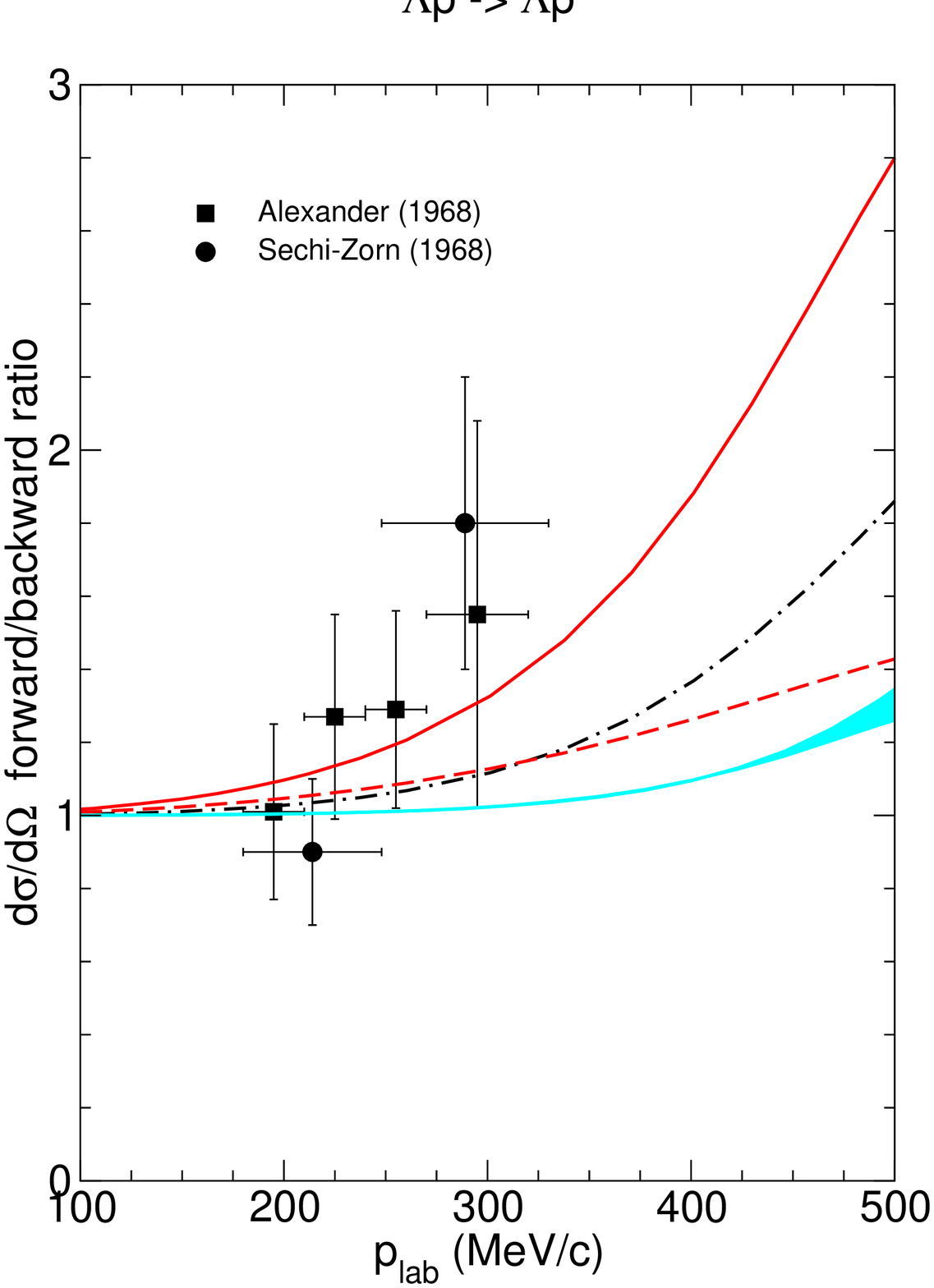}
 \vspace*{-0.5cm}
\caption{Cross section for $\La p$ as a function of $p_{\rm lab}$. 
Same description of the curves as in Fig.~\ref{fig:cs5}. 
Data are from Refs.~\cite{SechiZorn:1969hk} (filled circles), 
\cite{Alexander:1969cx} (filled squares),
\cite{Piekenbrock:unpub,Herndon:1967zz} (open triangles),
\cite{Kadyk:1971tc} (open squares), \cite{Hauptman:1977hr} (open circles)
and \cite{CLAS:2021gur} (inverted triangles).
%\vspace*{-0.7cm}
}
\label{fig:cs1}
\end{figure*}
 
Once the $S$-wave LECs are fixed from a combined fit to the $\La p$ and $\Si N$
cross sections, the differential cross sections established in the 
E40 experiment are analyzed. 
Interestingly, in the NLO case taking over the LECs from the corresponding $NN$ 
potential by Reinert et al.~\cite{Reinert:2018ip} for the $^3P_{0,1,2}$ partial 
waves, in accordance with SU(3) symmetry, and assuming 
the LEC in the $^1P_1$ to be zero, yields already a good description of the 
E40 data in the region $440$-$550$~MeV/c, cf. Fig.~\ref{fig:cs5} 
(center of the lower panel). 
For the N$^2$LO interaction all $P$-wave LECs are adjusted to the data. 
Actually, here we explore two scenarios (denoted by the superscripts $a$ and $b$ 
in the tables below so that one can distinguish them), 
one where the resulting angular 
distribution is similar to that obtained for NLO (solid line) and one which 
produces an overall more pronounced angular dependence (dashed line).
The latter is clearly 
preferred by the available data in that momentum range. However, a view on
the situation in the next momentum region, $550$-$650$~MeV/c, see Fig.~\ref{fig:cs5} 
(lower right), tells us that one has to be careful with conclusions. Here
the experiment suggest an overall somewhat different angular dependence, 
which seems to be more in line with a flat behavior or a very moderate increase
in forward direction. 
In any case, note that the alternative fit provides an at least visually 
slightly better description of the old low-energy data (lower left). 
Indeed, those data from the momentum region $160$-$180$~MeV/c~\cite{Eisele:1971mk}
($T_{\rm lab}\approx 12$~MeV) seem to exhibit a more pronounced angular dependence
than the E40 data at much higher momenta. Thus, it would be very interesting to 
explore the energy region in between by experiments. Such data could also 
help to pin down the $P$-wave contributions more reliably since higher 
partial waves should be much less important. For completeness, let us mention
that the fitting ranges considered for establishing the SMS $NN$ potential 
are $p_{\rm lab} \lesssim 480$~MeV/c at NLO and $p_{\rm lab} \lesssim 540$~MeV/c 
at N$^2$LO \cite{Reinert:2018ip}. 

The predictions by NLO19 are definitely at odds with the E40 experiment. 
However, it should be said that the pronounced rise of the cross section for
backward angles, excluded by the data, is mainly due to an accidental choice 
of the LEC $C_{^3SD_1}$ in the $\Si N$ $I=3/2$ contact interaction in 
\cite{Haidenbauer:2013oca,Haidenbauer:2019boi}. 
Its value can be easily re-adjusted,
without any change in the overall quality of those $YN$ potentials.
Pertinent results, for NLO19(600) as example, are indicated by dotted lines 
in Fig.~\ref{fig:cs5}. 

The integrated $\Si^+ p$ cross section over a larger energy range is shown
in Fig.~\ref{fig:cs5} (upper right). Note that again the angular
averaging according to Eq.~(\ref{eq:sigtot}) is applied to the theory
results. It is likewise done to obtain the indicated E40 data points 
because only differential cross sections in a limited angular range
are available \cite{J-PARCE40:2022nvq}.
Once more the NLO19 potential does not reproduce the trend of the data. 
Specifically, contrary to the experiment, there is a rise of the cross 
section for 
larger $p_{\rm lab}$ which we observed also for NLO13 and which seems to be present
also in results by the so-called covariant chiral EFT \cite{Li:2016paq,Song:2021yab}. 
This behavior could be simply an artifact of the employed regulator. 
Anyway, since $p_{\rm lab}=600$~MeV/c corresponds
to a laboratory energy of $T_{\rm lab}\approx 150$~MeV, one is certainly
in a region where NLO and possibly even N$^2$LO cannot be expected to be 
still quantitatively reliable. In this context, one should keep in mind
that the $\La N\pi$ channel opens around that energy which clearly marks 
the formal limit for the applicability of any effective two-body potential.
However, whether the noticeable drop in the experimental cross section,
which can not be reproduced by theory, has something to do with the opening
of that channel or not, remains unclear at present. 
 
The authors of Ref.~\cite{J-PARCE40:2022nvq} have attempted to perform
a phase-shift analysis, including 
partial waves up to the total angular momentum of $J=2$, with the aim to
determine the phase in the $^3S_1$ channel. 
For that different scenarios have been considered where the phase shifts in
the partial waves in the $\{27\}$ irrep of SU(3), 
cf. Table~\ref{tab:SU3}, 
were fixed either from $NN$ results (exploiting SU(3) symmetry) 
or from predictions of $YN$ models. Earlier efforts for establishing
the $\Si N$ $I=3/2$ phase shifts,
based on the differential cross section of Eisele et al. 
(lower-left of Fig.~\ref{fig:cs5}), can be found in 
Refs.~\cite{Nagels:1973is,Nagata:2002vj}. 
Our predictions for the phase shifts are displayed in
Figs.~\ref{fig:Sphp} and \ref{fig:Sphs}. For illustration we include the
$NN$ phase shifts in the $^3P_{0,1,2}$ partial waves (circles) which,
as said, would be identical to the ones for $\Si N$ with $I=3/2$ 
under strict validity of SU(3) symmetry. 
It is interesting to see that the difference is indeed fairly small. 
In comparison, the predictions of the chiral potentials for 
$^1P_1$, not constrained 
by SU(3), vary sizably. 
The results for the $^1S_0$ and $^3S_1$ partial waves
shown in Fig.~\ref{fig:Sphs} are, of course, strongly constrained by the 
available low-energy cross section data. The behavior of the $^1S_0$ 
is qualitatively similar to that in the $NN$ case~\cite{Reinert:2018ip}, 
as expected from the approximate SU(3) symmetry. 
One can observe a large difference in the results for the mixing 
angle $\epsilon_1$ between the SMS $YN$ potentials and NLO19. 
As discussed above, its large value is the reason for the rise of the
cross section at backward angles, cf. Fig.~\ref{fig:cs5}.
At the time when NLO19 and NLO13 were established, the existing data 
did not
allow to fix the relevant LEC ($C_{^3SD_1}$) reliably. However, it can be
re-adjusted (see the dotted line) without changing the overall $\chi^2$ and then
the pertinent results can be brought in line with the E40 measurement.

%%%%%%%%%%%%%%%%%%%%%%%%%%%%%%%%%%%%%%%%%%%%%%%%%%%%
\subsection{The $\Si^- p$ channel} 

Results for $\Si^-p$ elastic scattering are presented in Fig.~\ref{fig:cs4}. 
The SMS $YN$ potentials produce a slightly weaker energy dependence 
of the 
integrated cross section than NLO19. In the momentum region of 
the new E40 data \cite{J-PARCE40:2021qxa}, $p_{\rm lab} = 500-700$~MeV/c,
the predictions of all our $YN$ potentials are similar and in agreement 
with the experiment. Also the differential cross sections agree with 
the experiment, cf. the lower panel of Fig.~\ref{fig:cs4}. It should be
said, however, that the proper behavior in forward direction remains 
somewhat unclear since the experimental information is too sparse in 
that angular region. Nonetheless, 
the data points available for the momentum region $550-650$~MeV/c could
point to a somewhat steeper rise for small angles.   
The predictions based on NLO19 exhibit a sizable cutoff 
dependence. It is due to the fact that
the hadronic amplitude is overall attractive for some cutoffs and repulsive
for others so that there is either a destructive or constructive interference 
with the attractive Coulomb interaction. In case of a destructive interference
there is a small dip in the differential cross section at very forward angles.
Data with high resolution would be needed in order to resolve
that issue. 

Results for the transition $\Si^-p \to \La n$ are presented 
in Fig.~\ref{fig:cs2}. Also in this case the predictions of the SMS
$YN$ potentials and those of NLO19 are rather similar. Specifically, 
all interactions yield a reaction cross section in line with the
E40 data \cite{J-PARCE40:2021bgw}. The angular distributions are likewise
reproduced, cf. Fig.~\ref{fig:cs2} (center and left of the lower panel). 
One should keep in mind that in case of NLO19 no actual fitting of 
the $P$-wave LECs was performed. The ones belonging to the $\{27\}$ 
and $\{10^*\}$ irreps were taken over from fits to $NN$ $P$-waves, exploiting SU(3) symmetry 
constraints, whereas the others were fixed qualitatively by 
requiring that
the contribution of each $P$-wave to the $\La p$ cross section 
for momenta above the $\Si N$ threshold remains small
\cite{Haidenbauer:2013oca}. 
We note that for $\Si^- p \to \La n$ partial waves up to $J=8$ are 
needed to achieve converged results for the differential cross section 
at $600$~MeV/c.

In the context of the inelastic $\Si^- p$ data by Engelmann et al.~\cite{Engelmann:1966pl},
we would like to point to a footnote in that paper which emphasizes
the role of the $\Si^-$ lifetime in their determination of the
cross sections. The fact that the present value is almost $10$\,\% smaller
\cite{ParticleDataGroup:2022pth} suggests that the actual cross sections could be smaller, too.

There are no new data for the charge-exchange reaction 
$\Si^-p \to \Si^0 n$. 
The predictions of chiral EFT are in agreement with the existing
experimental evidence, as one can see in Fig.~\ref{fig:cs3}.

\begin{figure*}[t]
\centering
\includegraphics[width=0.62\textwidth]{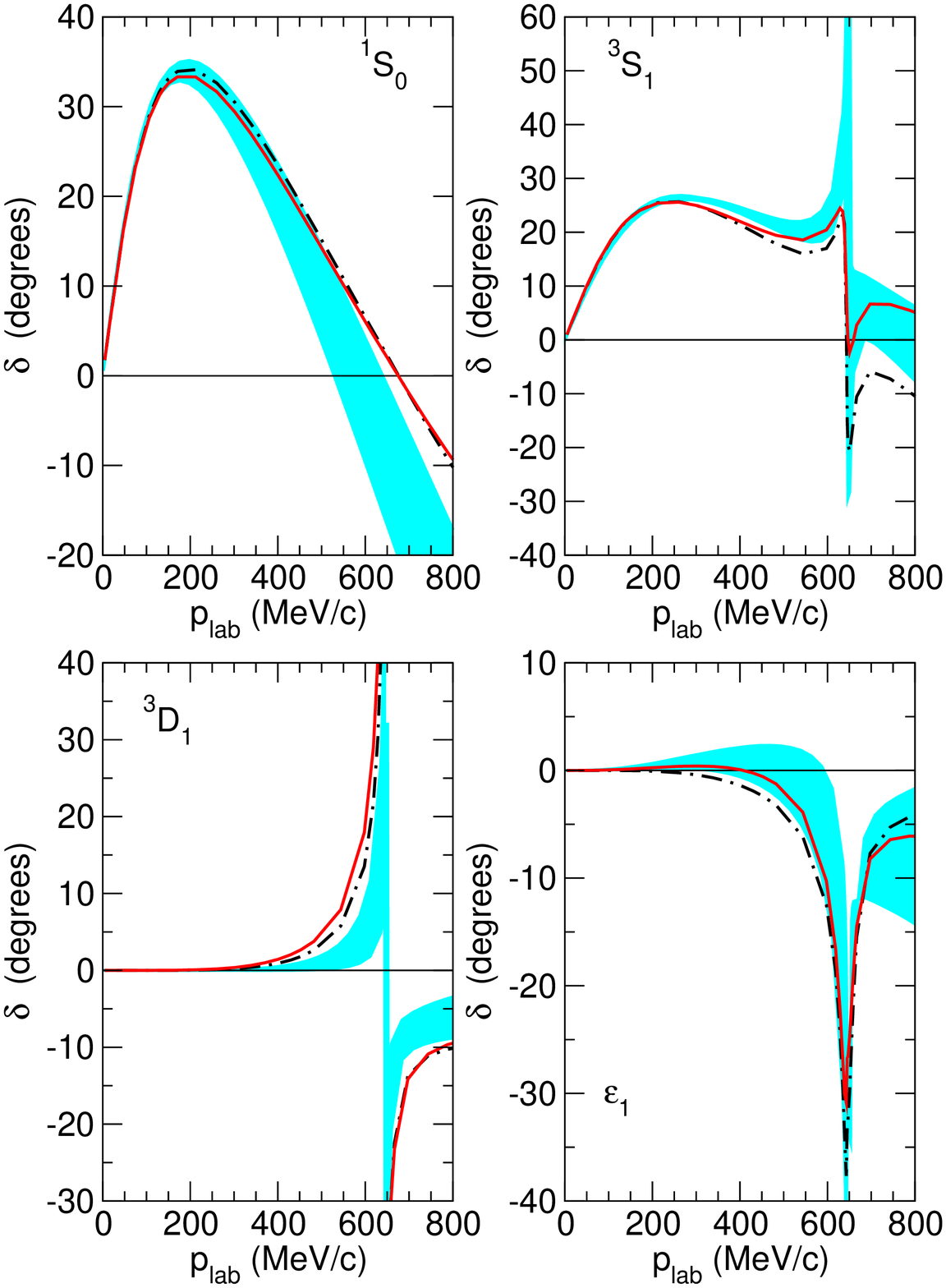}
\caption{$\La N$ phase shifts: $^1S_0$ and $^3S_1$-$^3D_1$.
Same description of the curves as in Fig.~\ref{fig:cs5}. 
The results for the $^3S_1$ and $^3D_1$ phases are shown 
modulo $180^0$.
}
\label{fig:Lphs}
\end{figure*}

\begin{figure*}[t!]
\centering
\includegraphics[width=0.62\textwidth]{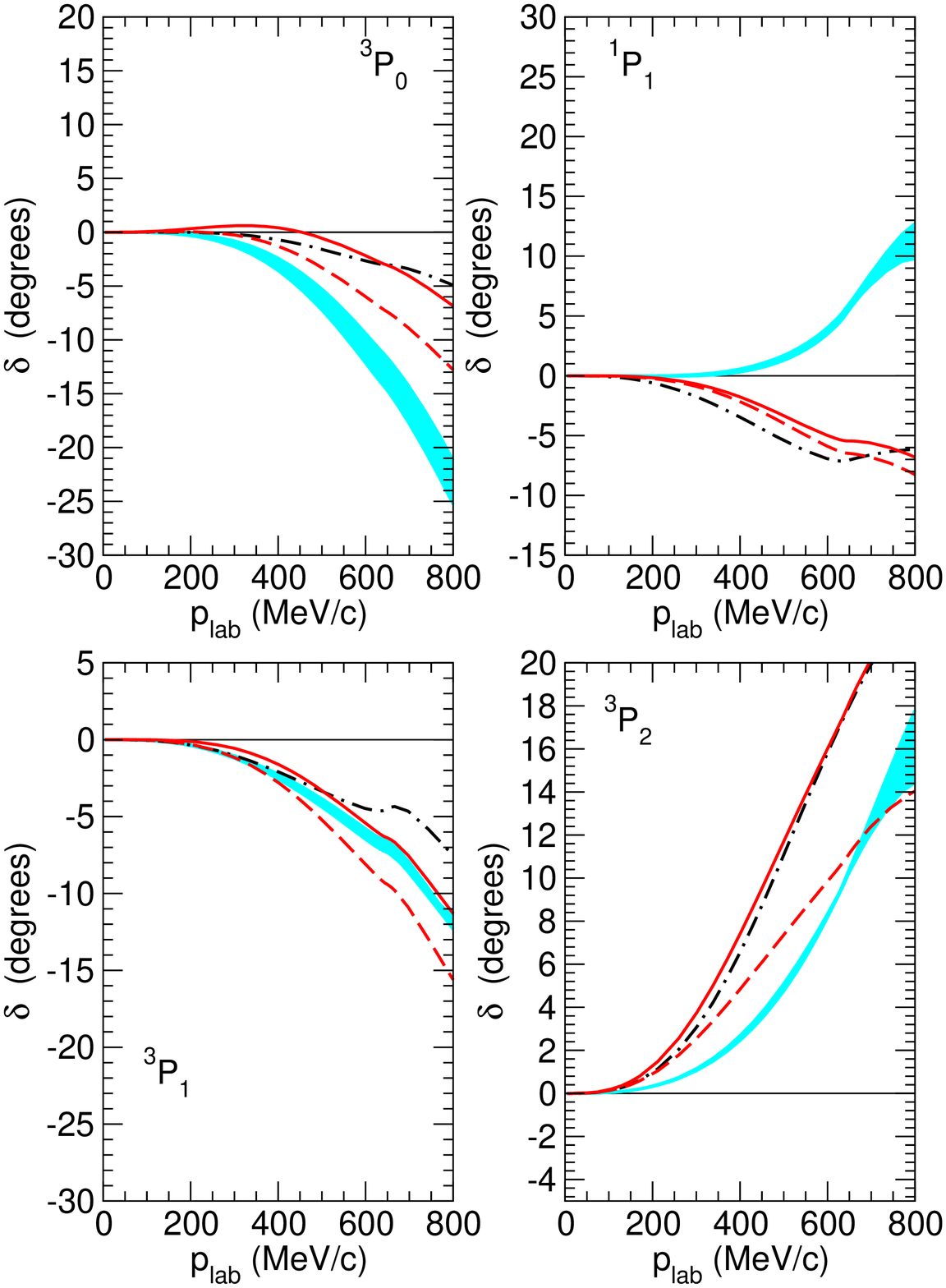}
\caption{$\La N$ phase shifts: $P$-waves.
Same description of the curves as in Fig.~\ref{fig:cs5}. 
}
\label{fig:Lphp}
\end{figure*}

%%%%%%%%%%%%%%%%%%%%%%%%%%%%%%%%%%%%%
\subsection{The $\La p$ channel} 

Results for $\La p$ scattering are presented in Fig.~\ref{fig:cs1}. 
So far there are no data from J-PARC for this channel. 
The new $\La p$ data from CLAS/Jlab \cite{CLAS:2021gur} are at 
fairly high momenta ($p_{\rm lab} \ge 900$~MeV/c) so that a quantitative 
comparison with our NLO and N$^2$LO predictions is not really 
sensible. Nonetheless, we display the momentum region up to their lowest data point 
(inverted triangle) so that one can see that the trend of our predictions is 
well in line with that measurement. 
Anyway, the low-energy data are reproduced with similar
quality by all chiral potentials, as expected in view of the excellent and 
low $\chi^2$ achieved in all fits. It is interesting
though that even the predicted cusp at the $\Si N$ threshold is practically
identical, cf. Fig.~\ref{fig:cs1} (upper right). This testifies that
the actual shape of the cusp is to a large extent determined by the 
$\Si N$ low-energy data \cite{Haidenbauer:2021smk} which, of course, 
are all described well by the
considered $YN$ potentials as discussed above.  

There are no genuine differential cross sections available for 
$\La p$ scattering.
However, some data on the angular distribution and the forward/backward
ratio can be found in Refs.~\cite{SechiZorn:1969hk,Alexander:1969cx}. Those are shown in the
lower panel of Fig.~\ref{fig:cs1} and compared with predictions normalized
to the number of events. Evidently, all our chiral potentials predict the
trend of the data that indicate a rise of the cross section in forward
direction. The present data would favor a more pronounced angular dependence
as produced by one of the N$^2$LO interactions (solid line). 
But for a quantitative conclusion more accurate data 
are needed. Also measurements for somewhat higher momenta, closer to the $\Si N$
thresholds, would be quite instructive \cite{Haidenbauer:2019boi}. 
Such data are expected to be provided by the future E86 
experiment at J-PARC \cite{Miwa:2022coz}. 

Results for $\La N$ phase shift in the $S$- and $P$-waves are shown in
Figs.~\ref{fig:Lphs} and \ref{fig:Lphp}. Like in case of $\Si N$
discussed above, the predictions for the $^1S_0$ and $^3S_1$ 
partial waves are strongly constrained by fitting the cross section data. 
And, as already mentioned, like in our previous works 
\cite{Haidenbauer:2013oca,Haidenbauer:2019boi,Haidenbauer:2005hx} 
the empirical binding energy of the hypertriton $^3_\La \rm H$
is used as a further constraint. Thereby we can exploit the fact that
the spin-singlet and triplet amplitudes contribute with different
weights to the $\La p$ cross section and to the $^3_\La \rm H$
binding energy, see Eq.~(9) in \cite{Haidenbauer:2019boi}.
Without that feature it would not be possible to fix the relative 
strength of the spin-singlet and spin-triplet $S$-wave components 
of the $\Lambda p$ interaction. 
A more detailed discussion on the
hypertriton will be provided in the next subsection. However, we want to
mention already here that we fixed the strength in the spin-singlet
interaction based on some exploratory calculations with SMS NLO (550).
The resulting scattering length, $a_s \approx -2.80$~fm, was then
used to calibrate all other NLO and N$^2$LO interactions. This
value is slightly smaller in magnitude that what has been found and
used for the NLO13 and NLO19 interactions with non-local cutoff, see 
Table~\ref{tab:ereF}. Nonetheless, the chiral $YN$ interactions with
the new regularization scheme tend to be overall slightly more
attractive. This is best seen in Fig.~\ref{fig:Lphs} from the 
$^1S_0$ phase shifts, where the predictions by the SMS potentials 
drop off more slowly with increasing momentum as compared to those 
of our former $YN$ interactions. 

As discussed in Ref.~\cite{Haidenbauer:2021smk}, most of 
the $YN$ potentials, that include the $\La N$-$\Si N$ coupling
and provide a quantitative description of the data,
predict an unstable $\Si N$ bound state near the $\Si N$ threshold.
This is reflected in the behavior of the $^3S_1$-$^3D_1$ phase shifts, where 
either the $^3S_1$ or $^3D_1$ phase pass through $90^\circ$ 
\cite{Polinder:2006zh,Haidenbauer:2013oca}.  
In case of the SMS potentials this happens in the $^3D_1$ state. 
Note that for convenience, and to keep the scales of the figures commensurable, 
we show the results in Fig.~\ref{fig:Lphs} modulo $180^\circ$. 

The results for the $P$-waves are qualitatively rather similar, 
except for the $^1P_1$ where the NLO19 prediction is of opposite sign. 
Certainly, the $^3P$ states are all dominated by the \{27\} irrep of 
SU(3) (Table \ref{tab:SU3}) and, thus, strongly constrained by 
fixing the pertinent LECs in a fit to the $NN$ phases (in case of 
NLO19 and SMS NLO) and to the new $\Si^+ p$
data (in the SMS N$^2$LO $YN$ potentials). It will be interesting to see
whether those predictions are consistent with $\La p$ differential
cross sections, once such data become available from J-PARC \cite{Miwa:2022coz}.

Recently, the $\La p$ two-particle momentum correlation function has
been measured with high precision by the ALICE Collaboration in 
$pp$ collisions at $13$~TeV \cite{ALICE:2021njx}. An exploratory analysis 
of those data suggests that the $\La p$ interaction could be slightly 
less attractive than what follows from the low-energy $\La p$ cross
section data \cite{SechiZorn:1969hk,Alexander:1969cx}. 
However, since additional ingredients (and additional uncertainties) 
arise in an in-depth analysis \cite{Haidenbauer:2019if}, those data 
cannot be included straightforwardly into our fitting procedure. 
Therefore, we refrain from taking into account constraints provided by 
such correlation functions at the present stage.

Finally, we want to mention that there are data for the $\La$
polarization, $\alpha \bar P(\theta^*_{\La})$, 
for forward and backward angles, see Table~II of Ref.~\cite{Alexander:1969cx}.
$\alpha$ is the weak decay parameter of the $\Lambda$ \cite{ParticleDataGroup:2022pth}.
These suggest that the polarization is practically consistent 
with zero for $p_{\rm lab} \le 320$~MeV/c. Since the experimental
uncertainties are rather large, we do not display these data here. 
However, we want to mention that the results of the SMS potentials 
for $\alpha P$ in that momentum region are all smaller than $0.1$.  
Also, we would like to point to Ref.~\cite{Miwa:2022coz}
where results of NLO13 and NLO19 for the $\La p$ analyzing
power are shown, and where one can see that those predictions 
are likewise rather small at low momenta. 

%%%%%%%%%%%%%%%%%%%%%%%%%%%%%%%%%%%%%%%%%%%
\subsection{$A=3$ and $A=4$ $\La$ hypernuclei} 

The binding energy of the hypertriton is obtained by solving 
Faddeev equations in momentum space. This method is well suited for
the chiral $YN$ and $NN$ potentials which involve local as well as
non-local components. A detailed description of the formalism can
be found in \cite{Miyagawa:1993rd,Nogga:2001ef}. 
In the discussion, we focus on the separation energy which is the
difference between the hypertriton binding energy and that of the
core nucleus, i.e. that of the deuteron. 
As shown by us in Ref.~\cite{Haidenbauer:2019boi}, the $\La$
separation energies of light hypernuclei are not very
sensitive to the employed $NN$ interaction. Therefore,
we use in all calculations the same state-of-the-art chiral 
$NN$ interaction, namely the SMS $NN$ potential of Ref.~\cite{Reinert:2018ip} 
at order N$^4$LO$^+$ with cutoff $\La = 450$~MeV.
The variation of the separation energy with the cutoff of the 
chiral $NN$ potentials is only in the order of $10$~keV, see Table~3
of Ref.~\cite{Haidenbauer:2019boi}. Note that some recent studies suggest a larger dependence on the $NN$ potential \cite{Gazda:2022fte,Wirth:2018ho}. This is in part due to using lower order $NN$ interactions but also because the dependence on the $NN$ interaction seems to be larger for the $LO$ YN interactions. We are currently investigating the $NN$ force dependence in more detail \cite{Nogga:inprep}. Our preliminary results confirm the small $NN$ force dependence of the order of $10$~keV for the NLO and N$^2$LO calculations presented here. The dependence is certainly much smaller than the 
experimental uncertainty of $\pm 40$~keV. 

As already mentioned, we require the hypertriton to be bound as an additional 
constraint for our $YN$ interaction. 
However, we do not include the $^3_\La$H separation energy in the 
actual fitting procedure because of its large experimental uncertainty.
While for a long time the value given by Juri{\v c} et al.
\cite{Juric:1973cz}, $B_\La = 0.13\pm 0.05$~MeV, has been accepted 
as the standard, recent measurements
reported by the STAR and ALICE Collaborations indicate that the
separation energy could be either significantly larger 
($0.41 \pm 0.12 \pm 0.11$~MeV \cite{Adam:2019phl}) or somewhat smaller
($0.072 \pm 0.063 \pm 0.036$~MeV \cite{ALICE:2022rib}). 
The latest average from the Mainz Group is
$0.148\pm 0.040$~MeV \cite{Eckert:2022srr}. 
New high-precision experiments to determine the hypertriton
binding energy are planned at the Mainz Microtron (MAMI)~\cite{Eckert:2022srr} 
and at JLab~\cite{Gogami:2022zha} and will hopefully resolve
those discrepancies. 

Given these variations, as a guideline of the present work,  
we aimed at achieving a $^3_\La$H separation energy in the order 
of $150$~keV with our chiral $YN$ interactions.
An arbitrary fine-tuning to one or the other value is not really
meaningful at the present stage. It would be also questionable 
in view of the fact that there should be a contribution from chiral 
three-body forces (3BF) \cite{Petschauer:2015elq}. Those could contribute up 
to $50$~keV to the binding, as argued in Ref.~\cite{Haidenbauer:2019boi}.
Incidentally, since the present experimental uncertainties exceed that 
estimation, there is no way of fixing the pertinent
3BF LECs from the hypertriton and, therefore, we refrain from 
including 3BFs in the present work. A possible and viable way
to fix the 3BFs is, in our opinion, via studies of the 
$^4_\La$H/$^4_\La$He and $^5_\La$He systems and we intend 
to explore that option in the future. 

\begin{table*}[t]
\renewcommand{\arraystretch}{1.5}
\caption{Overview of results for the hypertriton up to N$^2$LO
and for the $\La$ and $\Si$ single-particle potentials in
symmetric nuclear matter at saturation density. The superscripts
$a$ and $b$ denote the two variants introduced in Sect.~\ref{sec:Spp}
with different $P$-wave interactions. For the $NN$ interactions SMS N$^4$LO$^+$(450)
is used \cite{Reinert:2018ip}. The NLO13 and NLO19 results 
are from \cite{Haidenbauer:2019boi}.
 }
\label{tab:BE1}
\vspace{0.5cm}
\centering
\begin{tabular}{|l||ccc||rr|}
\hline
$YN$ potential  &  $B_\La$ [MeV] &  E [MeV] &  $P_\Si$ [\%] 
  & $U_\La (0)$ & $U_\Si (0)$ \\
\hline
SMS LO(700) & 0.135 &  $-$2.359 &  0.20 & $-$37.8 & 10.2 \\
           \hline
SMS NLO(500) & 0.127 &  $-$2.350  & 0.28 & $-$30.1 & 0.2 \\
SMS NLO(550) & 0.124 &  $-$2.347  & 0.23 & $-$32.1 & $-$1.6\\
SMS NLO(600) & 0.122 &  $-$2.345  & 0.32 & $-$29.7 & $-$3.1\\
\hline
SMS N$^2$LO(500) & 0.147 &  $-$2.371  & 0.25 & $-$33.1 & 6.4 \\
SMS N$^2$LO(550)$^a$ & 0.139 &  $-$2.362  & 0.25 & $-$38.5 & 2.5 \\
SMS N$^2$LO(550)$^b$ &0.125& $-$2.348  & 0.24 & $-$35.9 & 2.5 \\
SMS N$^2$LO(600) & 0.172 &  $-$2.395  & 0.22 & $-$37.8 & 0.1 \\
\hline
 NLO13(600)     & 0.090 &  $-$2.335  & 0.25 & $-$21.6 & 17.1 \\
 NLO19(600)     & 0.091 &  $-$2.336  & 0.21 & $-$32.6 & 16.9 \\
\hline
\end{tabular}
\end{table*}

Results of the SMS $YN$ potentials for the hypertriton separation 
energy are summarized in Table~\ref{tab:BE1}. It is interesting to
see that the predicted values lie fairly close together, keeping in mind,
of course, that the NLO and N$^2$LO potentials have been all tuned to the 
same $\La N$ scattering length in the $^1S_0$ partial wave.
Evidently, the separation energies are well in
line with the experimental values by Juri{\v c} et al. and agree also
with the new ALICE measurement within the uncertainty. 
Compared to the previous chiral $YN$ interactions NLO13 and NLO19, the 
separation energies are slightly larger indicating that the new 
interactions are more attractive than the previous ones. 

\begin{table*}[t]
\renewcommand{\arraystretch}{1.5}
\caption{Overview of results for the $^4_\Lambda$He
separation energy up to N$^2$LO. 
The superscripts $a$ and $b$ denote the two variants 
introduced in Sect.~\ref{sec:Spp} with different $P$-wave interactions.
For the $NN$ interactions SMS N$^4$LO$^+$(450) \cite{Reinert:2018ip} is used. 
For our new SMS results, we also apply the properly adjusted 
three-nucleon interaction (see \cite{LENPIC:2022cyu}). The NLO13 and NLO19 results 
are from \cite{Haidenbauer:2019boi}
}
\label{tab:BEA4}
\vspace{0.5cm}
\centering
\begin{tabular}{|l||rr|rr|}
\hline
      &  \multicolumn{4}{c|}{$^4_\Lambda$He}   \\
      &  \multicolumn{2}{c}{$J^\pi=0^+$} &  \multicolumn{2}{c|}{$J^\pi=1^+$}  \\ 
\hline      
$YN$ potential & $B_\La$ [MeV] & $P_\Si$ [\%]  & $B_\La$ [MeV]  & $P_\Si$ [\%]  \\
\hline
   SMS LO(700)   &       3.088 &        1.36 &       2.275 &        1.72        \\
\hline   
  SMS NLO(500)   &       2.009 &        2.32 &       1.041 &        2.05        \\
  SMS NLO(550)   &       2.102 &        2.13 &       1.102 &        1.96  \\
  SMS NLO(600)   &       2.021 &        2.34 &       0.927 &        1.69      \\
\hline  
SMS N$^2$LO(500) &       2.001 &        2.01 &       1.002 &        2.07       \\
SMS N$^2$LO(550)$^a$ &   2.024 &        1.81 &       1.251 &        2.01        \\
SMS N$^2$LO(550)$^b$ &   1.969 &        1.82 &       1.188 &        1.99        \\
SMS N$^2$LO(600) &       2.263 &        1.79 &       1.181 &        1.81        \\
\hline 
NLO13(600)       &       1.477 &        2.02 &     0.580  &   1.51    \\         
NLO19(600)       &       1.461 &        1.37 &     1.055  &   1.68    \\         
\hline
\end{tabular}
\end{table*}

It is now interesting to apply the same interactions to a more densely 
bound system, namely $^4_\Lambda$He. For this hypernucleus, charge symmetry breaking (CSB) 
is expected to contribute of the order of $100$~keV to the separation 
energies \cite{Haidenbauer:2021wld,Gazda:2016ir}. We do not include 
CSB terms here and likewise no
$YNN$ interactions since for now we are only interested in a
first comparison with our previous calculations for NLO13 and NLO19.
Without CSB interactions, the mirror hypernuclei $^4_\Lambda$He and
$^4_\Lambda$H have very 
similar separation energies. Therefore, we only present results for $^4_\Lambda$He. 

The binding energy for $A=4$ hypernuclei are obtained by solving 
Yakubovsky equations in momentum space \cite{Nogga:2001ef}. 
Such calculations 
require a large number of partial wave states for being converged. 
We have used here all partial waves with orbital angular momenta up to 
$l=6$ and a sum of the three orbital angular momenta related to the three 
relative momenta necessary up $8$. With this restriction of partial waves, 
our accuracy is of the order of $50$~keV for the separation energies. 

The results are summarized in Table~\ref{tab:BEA4}. It can be seen 
that the trend to larger separation energies applies also for $A=4$. 
For the $J^\pi=0^+$ ground 
state, the energies are now significantly closer to the experiment, 
where the current average value is $2.377\pm0.036$~MeV \cite{Mainz:webpage}. 
The same observation holds for the  $J^\pi=1^+$ excited state for which the 
experimental average is $0.942\pm 0.036$~MeV. This indicates 
that the contribution of $\La NN$ (and/or $\Si NN$) three-body forces  
is probably significantly smaller for 
the new series of interactions compared to NLO19 and NLO13. Surprisingly, at the same time, the SMS interactions lead to larger $\Sigma$ probabilities 
than NLO19. In past calculations it was observed that such larger
contributions of $\Sigma$'s to the hypernuclear states usually lead 
to smaller binding energies, c.f. the comparison 
of NLO13 and NLO19.

%%%%%%%%%%%%%%%%%
\subsection{$\La$ and $\Si$ in nuclear matter} 
For completing the picture, we include results for the in-medium 
properties of the $\La$ and $\Si$ based on the new $YN$ interactions. 
Specifically, we provide the predictions for the single-particle potentials 
$U_Y(p_Y)$ at nuclear matter saturation density ($k_F=1.35$~fm$^{-1}$), 
evaluated self-consistently within a conventional $G$-matrix calculation, 
utilizing the formalism described in detail in
Refs.~\cite{Haidenbauer:2019boi,Haidenbauer:2014uua}. 
As one can see from Table~\ref{tab:BE1}, $U_\La(p_\La=0)$ for the SMS $YN$ 
potentials is around $-30$ to $-38$~MeV, while $U_\Si (p_\Si=0)$ 
is around $-3$ to $+6$~MeV. 

The predicted value for $U_\La (0)$ is comparable to the result for NLO19 and 
also well in line with the usually cited empirical value of 
$U_\La = -27 \sim -30$~MeV \cite{Gal:2016boi}. Thus, the conclusions drawn 
in Refs.~\cite{Haidenbauer:2016vfq,Gerstung:2020ktv}
on the properties of neutron stars and a possible solution of the hyperon puzzle
based on the NLO13 and NLO19 potentials remain unchanged. In that works it 
was argued that the combined repulsive effects of the two-body interaction 
and a chiral $\La NN$ three-body force could be sufficiently strong to prevent 
the appearance of $\La$ hyperons in neutron stars. 
We want to emphasize that the somewhat larger result
for N$^2$LO (550)$^a$ is mainly due to the $P$-wave contributions. The 
alternative fit (550)$^b$ considered in the discussion of the $\Si^+ p$ 
cross section in Sect.~\ref{sec:Spp}, where only the $P$-waves were readjusted, 
yields $U_\La = -35.9$~MeV.

By contrast, $U_\Si$ is definitely less repulsive than what
was found for NLO13 and NLO19 and also below the range of
$10-50$~MeV advocated in Ref.~\cite{Gal:2016boi}. A detailed comparison
reveals that the more strongly repulsive $U_\Si$ of NLO13 and NLO19
is primarily due to the $^3S_1$ interaction in the $I=3/2$ channel
which is more repulsive at large momenta for those potentials.
However, the latter feature is precisely the reason why for NLO19 
the scattering results are in conflict with the J-PARC data on $\Si^+p$ 
(cf. Fig.~\ref{fig:cs5}), as we have discussed in Sect.~\ref{sec:Spp}.
Specifically, the artificial rise of the cross section at large
momenta is a direct result of the increasingly negative values for
the $^3S_1$ phase shift (Fig.~\ref{fig:Sphs}).                 
The same conflicting situation occurs for NLO13 and our LO interactions.
Indeed, as far as we can see, also phenomenological
$YN$ potentials that predict a more strongly repulsive $U_\Si$,
like those of Fujiwara et al.~\cite{Fujiwara:2006yh} based on 
the constituent-quark model, overestimate the $\Si^+p$ cross section  
at large momenta, see Fig.~24 in \cite{J-PARCE40:2022nvq}. 

At the moment, it remains unclear to us whether one can reconcile 
the constraints provided by the J-PARC data for the $\Si^+ p$ interaction with the request for 
a strongly repulsive $U_\Si$. Clearly, with regard to the $\Si$ single-particle
potential, the situation could be more complicated because of the overall  
spin-isospin structure of the $\Si N$ interaction where some of the 
relevant $S$-waves are attractive and others repulsive so that there
are possible cancellations in the evaluation of $U_\Si$. 
That being said, and may be more importantly,
one should keep in mind that the $\La$ single-particle potential 
follows from the rich spectrum of bound $\La$ hypernuclei and can 
be considered as well established.
Evidence for the $\Si$ single-particle potential comes only from the 
analysis of level shifts and widths of $\Sigma^-$ atoms and from 
measurements of inclusive $(\pi^-,K^+)$
spectra related to $\Sigma^-$-formation in heavy nuclei \cite{Gal:2016boi}.
It is worth mentioning that a conflicting situation has been likewise 
observed for the $\Xi$ single-particle potential. 
Also in that case the results from 
Brueckner calculations, using $YN$ interactions either constrained by
available data \cite{Haidenbauer:2018gvg} or from lattice QCD simulations
\cite{Inoue:2018axd} differ noticeably from phenomenological
results deduced again mainly from atomic states and inclusive 
($K^-$, $K^+$) spectra \cite{Gal:2016boi,Friedman:2021rhu}.

%%%%%%%%%%%%%%%%%%%%%%%%%%%%%%%%%%
\subsection{Uncertainty estimate}
Since the range and the strength of the $YN$ interaction is comparable
to that in the $NN$ system, considering the approximate validity
of SU(3) flavor symmetry, we expect overall a very similar convergence 
pattern with increasing order in the chiral expansion
as that found in the $NN$ studies in Refs.~\cite{Reinert:2018ip,Epelbaum:2015vj}. Anyway, 
to corroborate this expectation, we adopt here the tools proposed in Ref.~\cite{Epelbaum:2015vj} for 
an uncertainty estimate and present some selected results below. For simplicity
reasons, we focus on the elastic channels, namely $\La p$ and $\Si^+ p$.
One can see from the $NN$ results~\cite{Reinert:2018ip} that
$S$- and, in general, also $P$-waves are already well described
at the N$^2$LO level, say for laboratory energies up to $150$~MeV. The 
situation is different for $D$- and higher partial waves because, in this 
case, contact terms appear only at N$^3$LO or even higher order. 

The concrete expression used to calculate an uncertainty 
$\Delta X^{\rm N^2LO} (k)$ to the N$^2$LO prediction
$X^{\rm N^2LO}(k)$ of a given observable $X(k)$ is \cite{Epelbaum:2015vj}
\begin{eqnarray}
\label{Error}
\Delta X^{\rm N^2LO} (k) &=& \max  \bigg( Q^4 \times \Big| X^{\rm
    LO}(k) \Big|, \cr
  && Q^2 \times \Big|
  X^{\rm LO}(k) -   X^{\rm NLO}(k) \Big|, \;\; \;  \nonumber \\
  && Q \times \Big| X^{\rm NLO}(k) -   X^{\rm N^2LO}(k) \Big| \bigg)\,,
\end{eqnarray}
where the expansion parameter $Q$ is defined by
\begin{equation}
\label{expansion}
Q =\max \left( \frac{k}{\Lambda_b}, \; \frac{M_\pi}{\Lambda_b} \right)\,,
\end{equation}
with $k$ the on-shell center-of-mass momentum corresponding to the 
considered laboratory energy/momen\-tum, and $\Lambda_b$ the breakdown scale
of the chiral EFT expansion.
For the latter, we take over the value established in 
Ref.~\cite{Epelbaum:2015vj}, i.e. $\Lambda_b \sim 600$~MeV.
Analogous definitions are used for calculating the uncertainty up to NLO.
Note that the quantity $X(k)$ represents not only a ``true'' observable such 
as a cross section or an analyzing power, but also a phase shift.

\begin{figure*}[t]
\centering
\includegraphics[width=0.33\textwidth]{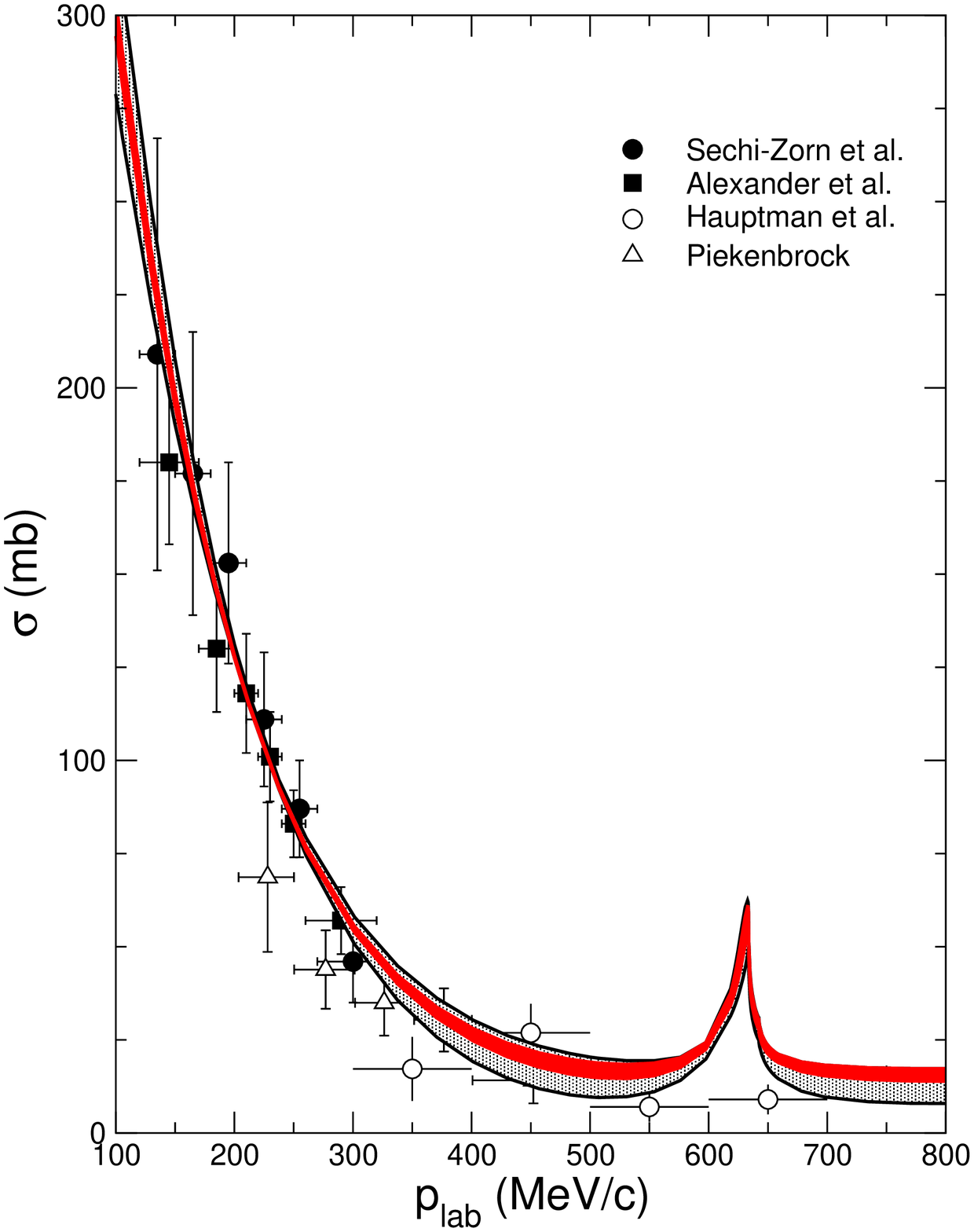}\includegraphics[width=0.33\textwidth]{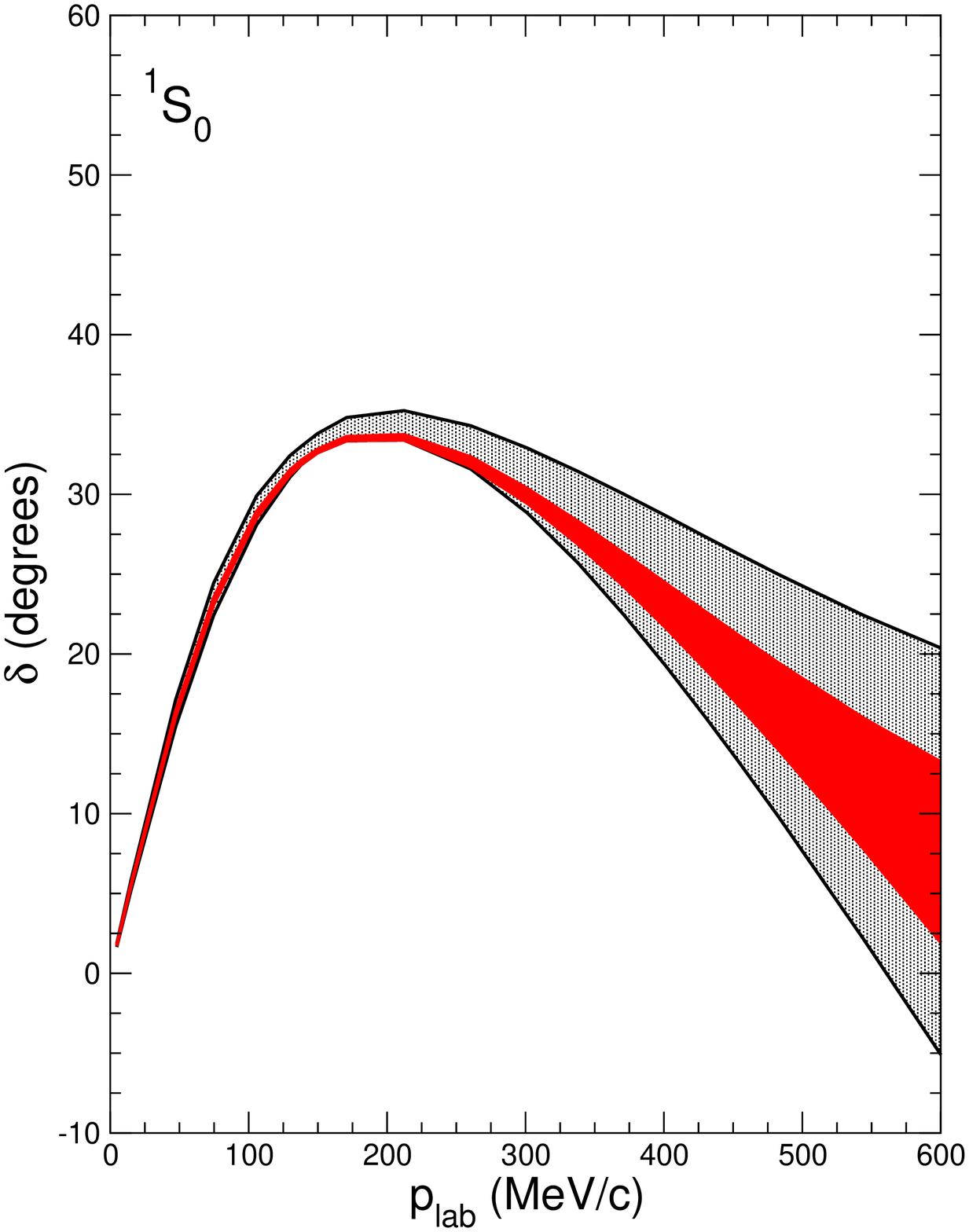}
\includegraphics[width=0.33\textwidth]{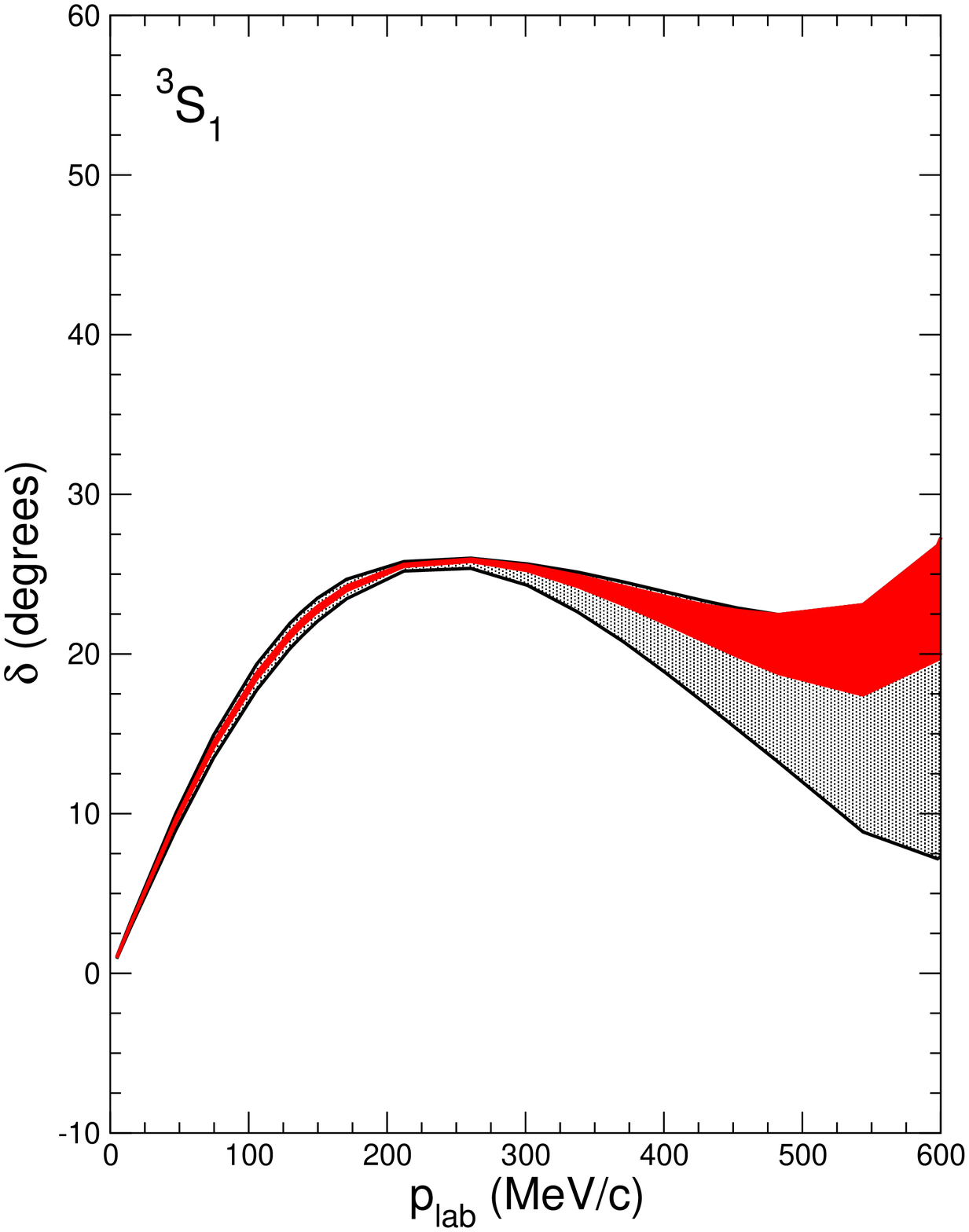}
\vspace*{-0.8cm}
\caption{Uncertainty estimate for the $YN$ interaction in the $\La p$ 
channel, employing the method suggested in Ref.~\cite{Epelbaum:2015vj}.   
As basis the LO(700), and the NLO(550) and N$^2$LO(550) results are used.
The grey (light) band corresponds to $\Delta X^{\rm NLO}$, the red (dark)
band to $\Delta X^{\rm N^2LO}$.
%\vspace*{-0.7cm}
}
\label{fig:uln}
\end{figure*}

\begin{figure*}[t]
\centering
\includegraphics[width=0.33\textwidth]{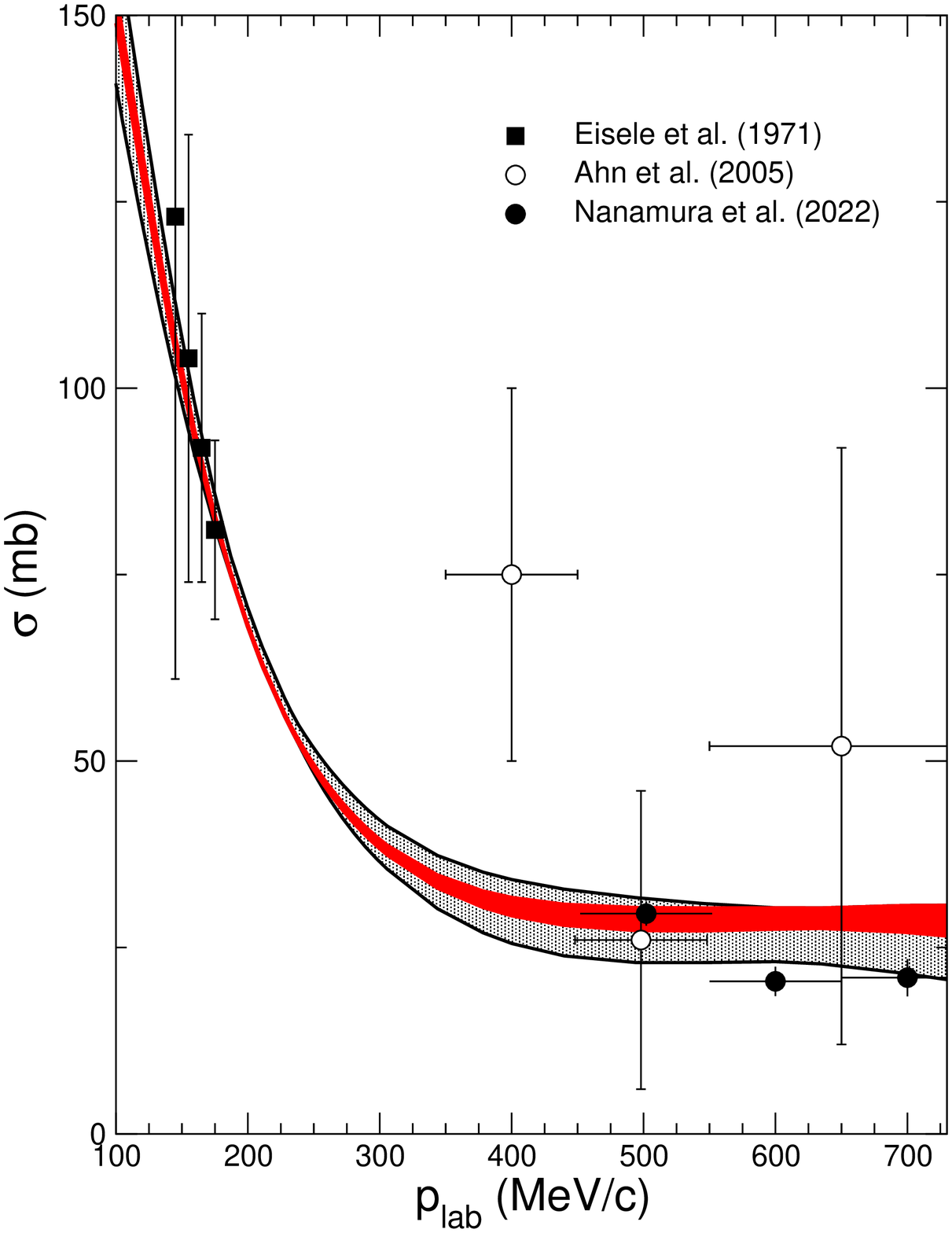}\includegraphics[width=0.33\textwidth]{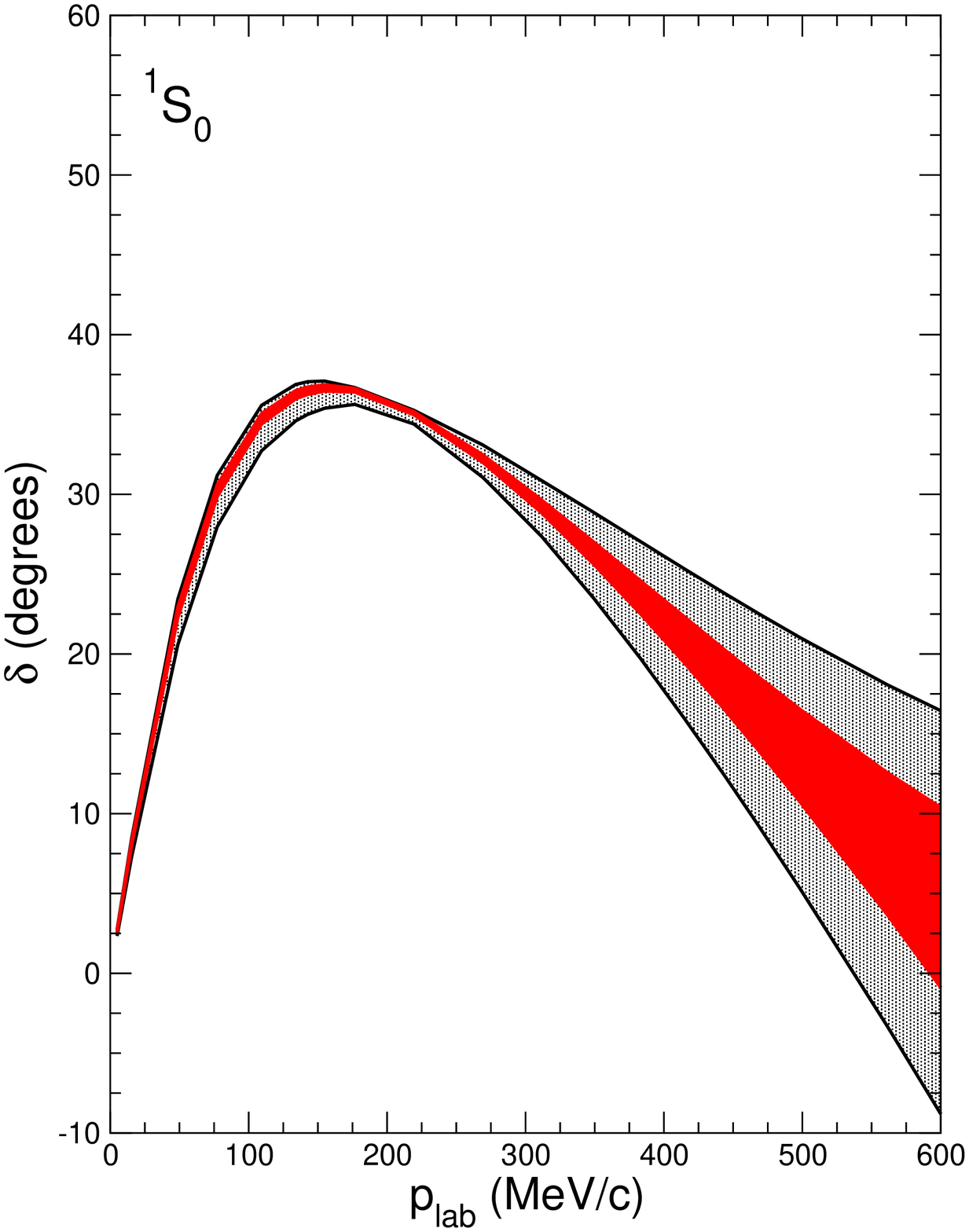}
\includegraphics[width=0.33\textwidth]{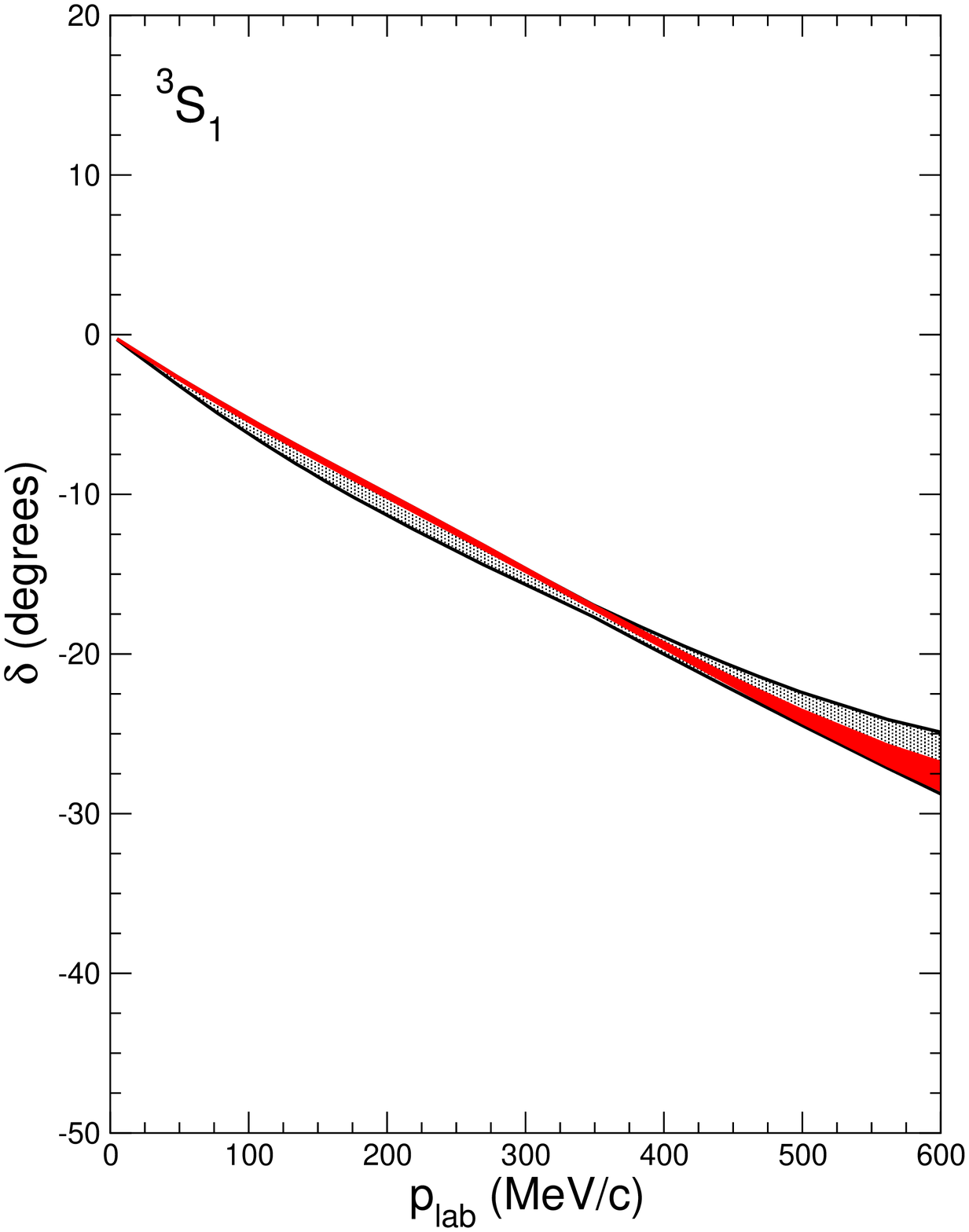}
\vspace*{-0.8cm}
\caption{Uncertainty estimate for the $YN$ interaction in the $\Si^+ p$ 
channel, employing the method suggested in Ref.~\cite{Epelbaum:2015vj}.  
Same description of the curves as in Fig.~\ref{fig:uln}. 
%\vspace*{-0.7cm}
}
\label{fig:usn}
\end{figure*}

In Figs.~\ref{fig:uln} and \ref{fig:usn}, we show our uncertainty estimates
for the cross sections and the $S$-wave phase shifts for $\La p$ 
and $\Si^+p$ following the procedure proposed in Ref.~\cite{Epelbaum:2015vj}.
Certainly, for addressing the question of convergence thoroughly,
orders beyond N$^2$LO 
are needed. Higher orders are also required to avoid that accidentally
close-by results lead to an underestimation of the uncertainty.
%%%%
For the $YN$ interaction, any uncertainty estimate is
difficult since the data are not sufficient to unambiguously 
determine all LECs. For example, recall that the strength of the 
$\La N$ interaction in the $^1S_0$ partial wave was fixed ``by hand''
and not based on actual $\La p$ scattering data. 
For this reason, there is definitely some bias in the quantification of the uncertainty
of phase shifts in individual partial waves.
Nonetheless, we want to emphasize that the estimated
uncertainty appears sensible and also plausible.
In particular, it encases the variations due to the regulator
dependence and, thus, is consistent with the expectation
that cutoff variations provide a lower bound for the
theoretical uncertainty \cite{Epelbaum:2015vj}.
For details of the method and a thorough discussion of
the underlying concept, we refer the reader to \cite{LENPIC:2015qsz}.
We should add that in case of the chiral $NN$ interaction more sophisticated
tools like a Bayesian approach~\cite{Furnstahl:2015rha} have been 
applied, too.

%%%%%%%%%%%%%%%%%%%%%%%%%%%%%%%%%%%%%%%%%%%%%%%%%%
\section{Summary and outlook} 

In the present work, we have established a hyperon-nucleon potential for the 
strangeness $S=-1$ sector ($\Lambda N$, $\Sigma N$) up to next-to-next-to-leading
order in the chiral expansion. SU(3) flavor symmetry is imposed for constructing the 
interaction, however, the explicit SU(3) symmetry breaking by the physical masses 
of the pseudoscalar mesons ($\pi$, $K$, $\eta$) and in the leading-order 
contact terms is taken into account. 
A novel regularization scheme, the so-called semilocal momentum-space 
regularization, has been employed which has been already successfully
applied in studies of the nucleon-nucleon interaction within chiral
effective field theory up to high orders \cite{Reinert:2018ip}. 

An excellent description of the low-energy $\Lambda p$, $\Sigma^- p$ and
$\Sigma^+ p$ scattering cross sections could be achieved with a $\chi^2$ of 
$15$-$16$ for the commonly considered $36$ data points \cite{Haidenbauer:2013oca}. 
At low energies, the results are also very close to those of our earlier $YN$
interactions NLO13 \cite{Haidenbauer:2013oca} and NLO19 \cite{Haidenbauer:2019boi}, 
that are based on a different regularization scheme. 
New measurements of angular distributions for the $\Sigma N$ channels
from J-PARC \cite{J-PARCE40:2021qxa,J-PARCE40:2021bgw,J-PARCE40:2022nvq} have 
been analyzed in an attempt to determine the strength of the contact interactions
in the $P$-waves. Although those data can be fairly well described, considering 
the experimental uncertainties and the fact that the pertinent momenta 
$p_{\rm lab} \gtrsim 450$~MeV are close to the limit of applicability 
of the N$^2$LO interaction, they are not included in the total $\chi^2$.

Separation energies for the hypertriton have been presented. These are not 
``true'' predictions of the theory, because we required the $^3_\La$H to be 
bound as additional constraint to fix the spin dependence of the $\Lambda N$
interaction. Anyway, the obtained values of $120$-$170$~keV are 
well within the range of the presently existing experimental evidence
\cite{Adam:2019phl,ALICE:2022rib,Eckert:2022srr,Mainz:webpage}. Compared to NLO13 and NLO19,
the new interaction seems to be more attractive. This also shows up 
in the results for $^4_\Lambda$He which are closer to the experimental 
values.  
A simple uncertainty estimate for the chiral expansion \cite{Epelbaum:2015vj}, 
performed for a
selected set of $YN$ observables, exhibits a similar pattern as has been found
for the $NN$ interaction. Certainly, at the level of N$^2$LO one can not
expect to see fully converged results, in contrast to the $NN$ sector 
where the calculation have progressed up to N$^4$LO (and beyond) \cite{Reinert:2018ip}. 

As a next step, one should explore the $YN$ potential in calculations of 
light $\La$-hypernuclei within, e.g., the no-core shell model 
(feasible up to $A\approx 10$). Of course, for that chiral ($\La NN$, $\Si NN$) 
three-body forces should be included, which arise 
at N$^2$LO in the chiral expansion \cite{Petschauer:2015elq}. 
Moreover, a possible charge-symmetry breaking in the $\La p$ and $\La n$
interactions should be introduced. Such a CSB component has been found to
be essential for understanding the level splittings in the 
$^4_\La$H-$^4_\La$He mirror nuclei \cite{Gazda:2016ir,Haidenbauer:2021wld}.
For example, an earlier study by us, based on the NLO13 and NLO19 
interactions, suggests that $\Delta a_s = a^{\La p}_s-a^{\La n}_s 
\approx 0.62 \pm 0.08$~fm for $^1S_0$ and 
$\Delta a_t  \approx -0.10\pm 0.02$~fm for $^3S_1$
\cite{Haidenbauer:2021wld}. 
Clearly, the reproduction of the large CSB effect in the $^1S_0$ partial wave 
requires a noticeable modification of the present $\La N$ interaction.
In any case, one has to keep in mind that the actual CSB splittings
for $^4_\La$H-$^4_\La$He are not yet that well settled experimentally,
cf. Refs.~\cite{Achenbach:2016xul,Mainz:webpage,STAR:2022zrf}. 
Finally,  
a more elaborate effort to determine the strength of the contact terms in 
the $P$-waves should be done in the future when $\La p$ angular distributions 
from the J-PARC E86 experiment have become available \cite{Miwa:2022coz}. 

%%%%%%%%%%%%%%%%%%%%%%%%%%%%%%%%%%%%%%%%%%%%%%%%%%%
\section*{Acknowledgements}
We thank Stefan Petschauer for his collaboration in the early stage of
this work. 
This project is part of the ERC Advanced Grant ``EXOTIC'' supported 
the European Research Council (ERC) under the European Union's Horizon 2020 
research and innovation programme (grant agreement No. 101018170).
This work is further supported in part by the DFG and the NSFC through
funds provided to the Sino-German CRC 110 ``Symmetries and
the Emergence of Structure in QCD'' (DFG grant. no. TRR~110),
and the VolkswagenStiftung (grant no. 93562).
The work of UGM was supported in part by The Chinese Academy
of Sciences (CAS) President's International Fellowship Initiative (PIFI)
(grant no.~2018DM0034). We also acknowledge support of the THEIA net-working activity 
of the Strong 2020 Project. The numerical calculations were performed on JURECA
and the JURECA-Booster of the J\"ulich Supercomputing Centre, J\"ulich, Germany.

%%%%%%%%%%%%%%%%%%%%%%%%%%%%%%%%%%%%%%%%%%%%%%%%%%
%\eject        

\appendix 
\vglue 0.5cm
\section{Semilocal momentum-space baryon-baryon potential at NLO and \texorpdfstring{N$^2$LO}{N2LO}}
\label{sec:smspot}

In order to implement the local cutoff in the two-meson contributions we 
follow Ref.~\cite{Reinert:2018ip} and write the corresponding 
potentials in terms of their spectral representation:
\begin{eqnarray}
V(q) &=& \frac{2}{\pi} \int_{2M_P}^\infty \, \mu\, d \mu
\frac {\rho (\mu)} {\mu^2 + q^2}~, \nonumber\\ 
\rho (\mu) &=& {\rm\, Im} V(q= 0^+ - {\rm i}\,\mu) \ ,  
\label{Eq:Sp1}
\end{eqnarray}
with $q$ the momentum transfer \(q=\left|\bf{p^{\,\prime}} - \bf{p}\,\right|\) 
and $M_P$ the mass of the exchanged meson. The regularized potential is then given by 
\begin{equation}
V(q) = e^{-\frac{q^2}{2 \Lambda^2}}
\frac{2}{\pi} \int_{2M_P}^\infty \, \mu\, d \mu
\frac {\rho (\mu)} {\mu^2 + q^2} \,
e^{-\frac{\mu^2}{2 \Lambda^2}}~.
\label{Eq:Sp2} 
\end{equation}

\subsection{Contributions at NLO}

\begin{figure*}[t]
 \centering 
 \includegraphics[width=\feynwidth]{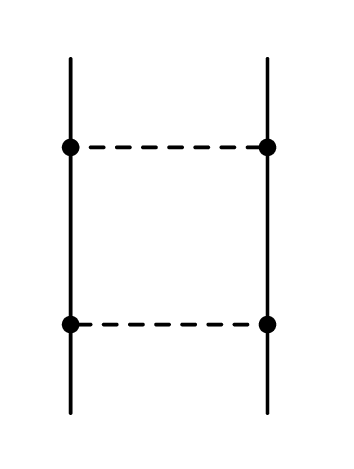}
 \includegraphics[width=\feynwidth]{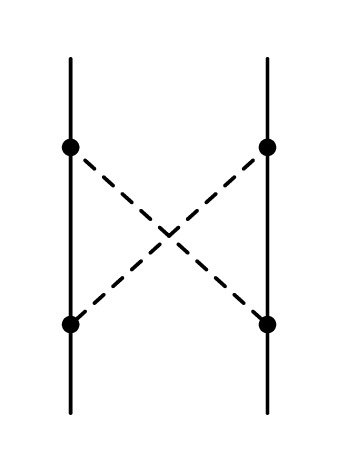}
 \includegraphics[width=\feynwidth]{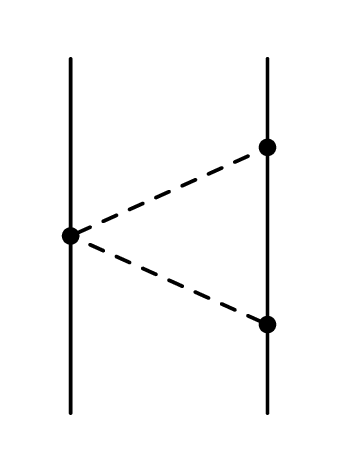}
 \includegraphics[width=\feynwidth]{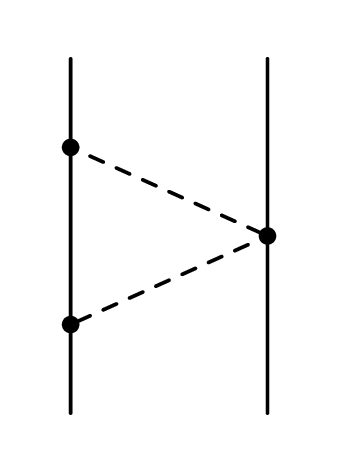}
 \includegraphics[width=\feynwidth]{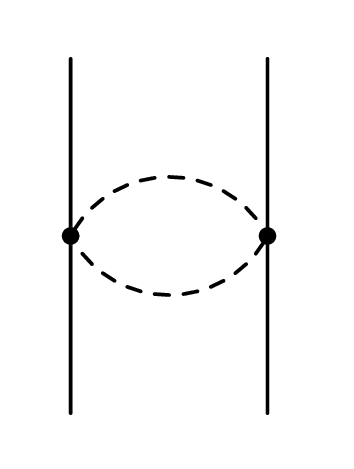}
 \caption{Relevant diagrams at next-to-leading order. Solid and dashed lines 
 denote octet baryons and pseudoscalar mesons, respectively. 
 From left to right: planar box, crossed box, left triangle, right triangle, football
 diagram. Note that from the planar box, only the irreducible part contributes to the potential.}
 \label{fig:NLO2}
\end{figure*}
Diagrams representing the contributions at NLO (chiral order $\nu=2$) 
are shown in Fig.~\ref{fig:NLO2}.
At NLO one obtains a central potential (\(V_\mathrm C\)),
a spin-spin potential $(V_S)$
and a tensor-type potential $(V_T)$ \cite{Haidenbauer:2013oca}, so that
$V^{(2)} = V^{(2)}_\mathrm C + 
\mbox{\boldmath $\sigma$}_1\cdot\mbox{\boldmath $\sigma$}_2\, V^{(2)}_S +
\mbox{\boldmath $\sigma$}_1\cdot{\bf q} \, \mbox{\boldmath $\sigma$}_2\cdot{\bf q}\,
V^{(2)}_T$. We provide here explicit expressions of the irreducible potentials 
for two-pion exchange \cite{Reinert:2018ip,Petschauer:priv}. Clearly, those formulae 
are also valid for $\eta\eta$ and/or $KK$ ($K\bar K$) exchange. General expressions
of the spectral functions for non-identical meson masses are given in \ref{sec:spectrfunc} below. 
\begin{widetext}
\begin{equation}
V^{(2)}_{C,S}(q) = \frac{2q^4}{\pi} \int_{2M_\pi}^\infty \, d \mu
\frac {\rho_{C,S} (\mu)} {\mu^3\, (\mu^2 + q^2)}, \quad 
V^{(2)}_{T}(q) = -\frac{2q^2}{\pi} \int_{2M_\pi}^\infty \, d \mu
\frac {\rho_{T} (\mu)} {\mu\, (\mu^2 + q^2)}.
\label{Eq:Sp3}
\end{equation}
\end{widetext}
The contributions (spectral functions) of the individual diagrams are: \\

\noindent {Planar box (pb)}
\begin{eqnarray}
\label{eq:TPE_17}
\rho^{pb}_C(\mu) &=& -\frac{N}{3072\pi f_0^4 \sqrt{\mu^2-4M_\pi^2}\,\mu} \cr 
& & \times (-23\mu^4 + 112 \mu^2 M_\pi^2 - 128 M_\pi^4), \\
\nonumber\\
\rho^{pb}_T(\mu) &=& \frac{\rho^{pb}_S(\mu)}{\mu^2} = \frac{N \sqrt{\mu^2-4M_\pi^2}}{256\pi f_0^4\,\mu}.
\end{eqnarray}

\noindent {Crossed box (xb)}
\begin{eqnarray}
\rho^{xb}_C(\mu) &=& -\rho^{pb}_C(\mu), \nonumber \\
\rho^{xb}_S(\mu) &=& \phantom{-} \rho^{pb}_S(\mu), \nonumber \\
\rho^{xb}_T(\mu) &=& \phantom{-} \rho^{pb}_T(\mu). 
\label{eq:TPE_18}
\end{eqnarray}

\noindent {Triangle diagrams (tr)}
\begin{eqnarray}
\label{eq:TPE_19}
\rho_C(\mu) &=& -\frac{N\, \sqrt{\mu^2-4M_\pi^2} }{3072 \pi f_0^4\,\mu}
(5\mu^2 - 8 M_\pi^2). \\
\end{eqnarray}

\noindent {Football diagram (fb)}
\begin{eqnarray}
\label{eq:TPE_20}
\rho_C(\mu) &=& -\frac{N\, (\mu^2-4M_\pi^2)^{3/2} }{6144 \pi f_0^4\,\mu }.
\end{eqnarray}

The quantities
$N$ are an appropriate product of coupling constants and isospin factors:
\begin{eqnarray}
N^{pb,xb}&=&f_{B_1B_{il}M_1}f_{B_{il}B_3M_2} \cr & & \times f_{B_2B_{ir}M_2}f_{B_{ir}B_4M_1}
(2f_0)^4\,{\mathcal I}_{B_1B_2\to B_3B_4} \ , \nonumber \\
N^{tr}&=&f_{B_1B_{i}M_1}f_{B_iB_3M_2}(2f_0)^2\,
{\mathcal I}_{B_1B_2\to B_3B_4} \ , \nonumber \\
N^{fb}&=&{\mathcal I}_{B_1B_2\to B_3B_4} \ . 
\end{eqnarray}
The isospin factors are summarized in Table~\ref{tab:iso}
whereas the coupling constants are specified in Eqs.~(\ref{eq:SU3}).
$B_{il}$ and $B_{ir}$ denote the (left and right) baryons in the intermediate 
state. Note that the relations (\ref{eq:TPE_18}) concern only the $\mu$
dependence, but not the factors $N^{pb}$ and $N^{xb}$! 
In case of the $NN$ system the expressions for the spectral functions
(and the potential) can be reduced to those given in 
Ref.~\cite{Reinert:2018ip} by simply representing the pertinent isospin 
coefficients in Table~\ref{tab:iso} in operator form: 
$-2\, \fet \tau_1 \cdot \fet \tau_2 + 3$ for the planar box,
$ 2\, \fet \tau_1 \cdot \fet \tau_2 + 3$ for the crossed box,
$-4\, \fet \tau_1 \cdot \fet \tau_2$ for the triangle diagrams, 
and $8\, \fet \tau_1 \cdot \fet \tau_2$ for the football diagram.
Then, since $V_{C}^{xb} = - V_{C}^{pb}$, see Eq.~(\ref{eq:TPE_18}), 
the central component of the spectral function (potential) is 
proportional to $\fet \tau_1 \cdot \fet \tau_2$ (denoted by $\eta_C$ and $W_C$,
respectively, in Ref.~\cite{Reinert:2018ip}), while for the spin- and tensor 
components the contributions from planar and crossed box add up and the isospin
dependence drops out ($\rho_{S,T}$ and $V_{S,T}$ in Ref.~\cite{Reinert:2018ip}).

\begin{table*}
\caption{Isospin factors ${\mathcal I}$ for the NLO diagrams. 
The baryons in the intermediate state of the planar box, crossed box,
and the triangle diagrams are indicated to the left of the factors. 
}
\begin{center}
\renewcommand{\arraystretch}{1.20}
\begin{tabular}[t]{|c||cc|cc|cc|cc|cc|}
\hline \hline
transition & \multicolumn{2}{c|}{planar}& \multicolumn{2}{c|}{crossed}& 
\multicolumn{2}{c|}{triangle}& \multicolumn{2}{c|}{triangle}& \multicolumn{2}{c|}{football}  \\
(isospin)  & \multicolumn{2}{c|}{box}& \multicolumn{2}{c|}{box}& 
\multicolumn{2}{c|}{left}& \multicolumn{2}{c|}{right}& \multicolumn{2}{c|}{diagram}  \\
\hline \hline
$N N\to N N$ & & & & & & & & & &\\
($I=0$) & $N N$         & $ 9$ & $NN$ & $-3$ & $N$ & $12$ & $N$ & $12$ & $ $ & $-24$ \\
($I=1$) & $N N$         & $ 1$ & $NN$ & $5$  & $N$ & $-4$ & $N$ & $-4$ & $ $ & $  8$ \\
\hline
$\Sigma N\to \Sigma N$  & & & & & & & & & &\\
($I=1/2$) & $\Sigma N$  & $4$ & $\Sigma N$  & $ 0$ & $N$ & $16$ & $\Sigma$ & $ 4$ & $  $ & $-32$ \\
          & $\Lambda N$ & $3$ & $\Lambda N$ & $-1$ & $ $ & $  $ & $\Lambda$& $ 4$ & $  $ & $  $  \\
($I=3/2$) & $\Sigma N$  & $1$ & $\Sigma N $ & $ 3$ & $N$ & $-8$ & $\Sigma$ & $-2$ & $  $ & $ 16$ \\
          & $\Lambda N$ & $0$ & $\Lambda N$ & $ 2$ & $ $ & $  $ & $\Lambda$& $-2$ & $  $ & $  $  \\
\hline
$\Lambda N\to \Sigma N$ & & & & & & & & & &\\
($I=1/2$) & $\Sigma N$  & $ 2\sqrt{3}$ & $\Sigma N$  & $-2\sqrt{3}$ & $N$ & $0$ & $\Sigma$ & $4\sqrt{3}$ & $ $ & $0$ \\
\hline
$\Lambda N\to \Lambda N$ & & & & & & & & & &\\
($I=1/2$) & $\Sigma N$  & $3$ & $\Sigma N $ & $ 3$ & $N$ & $0$ & $\Sigma $ & $0$ & $  $ & $ 0$ \\
\hline \hline
\end{tabular}
\label{tab:iso}
\end{center}
\end{table*}

\subsection{Contributions at N$^2$LO}

\begin{figure*}[t]
 \centering 
 \includegraphics[width=\feynwidthbig]{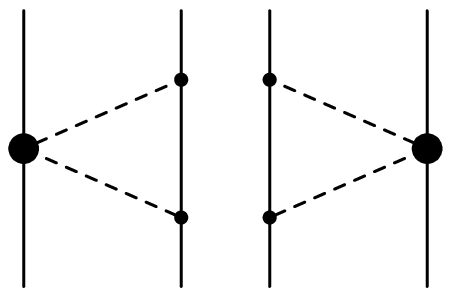}{\quad}
 \includegraphics[width=\feynwidthbig]{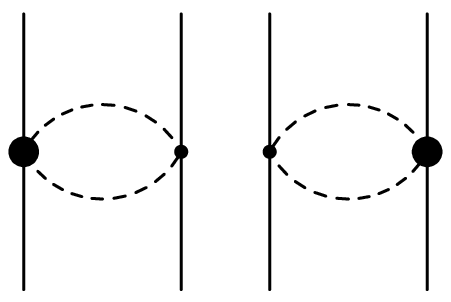}
 \caption{Relevant diagrams at next-to-next-to-leading order. Solid and dashed lines 
 denote octet baryons and pseudoscalar mesons, respectively. 
Triangle (left) and football (right) diagram.} \label{fig:NLO3}
\end{figure*}

Diagrams that arise at N$^2$LO are shown in Fig.~\ref{fig:NLO3}. 
It should be noted, however, that only the triangle diagrams
contributes. There is no contribution from the football diagram because
of parity conservation. The potential consists again of 
central, spin-spin, and tensor components, $V^{(3)}_{C,S,T}$, and 
those components can be evaluated from representations analogous to
Eq.~(\ref{Eq:Sp3}). The
spectral functions in question are given by \cite{Petschauer:priv} 
\begin{eqnarray}
\label{eq:SPE_17}
\rho_C(\mu) &=& \frac{N_1}{512\mu f^4_0} (\mu^2-2M_\pi^2) + \frac{N_2}{256\mu f^4_0} (\mu^2-2M_\pi^2)^2,\cr && \\
%\end{eqnarray}
%
%\begin{eqnarray}
\label{eq:SPE_18}
\rho_T(\mu) &=& \frac{\rho_S(\mu)}{\mu^2} = -\frac{N_3}{512\mu f^4_0} 
(\mu^2 - 4 M_\pi^2).
\end{eqnarray}
 
 The coefficients $N_i$ ($i=1,2,3$) are combinations of the coupling constants at
 the involved $BBM$ vertices and of elements of the sub-leading (${\cal O} (q^2)$)
 meson-baryon Lagrangian \cite{Frink:2006hx,Oller:2006yh}, 
in particular of the meson-baryon LECs $b_D$, $b_F, b_0$, $b_1$-$b_4$, 
and $d_1$-$d_3$, see Sect.~IV in Ref.~\cite{Petschauer:2015elq} for details
and/or Sect.~4.3 in \cite{Petschauer:2020urh}.
The concrete relations are as follows:

\noindent for $NN$
\begin{eqnarray}
N_1 &=& \phantom{-}96\, c_1 \, g_A^2\, M^2_\pi, \nonumber \\
N_2 &=& \phantom{-}12\, c_3 \, g_A^2, \nonumber \\
N_3 &=& -4\, c_4 \, g_A^2 \, \fet\tau_1\cdot\fet\tau_2,
\label{NN:N}
\end{eqnarray}
with 
$c_1 = (2b_0+b_D+b_F)/2$, $c_3 = b_1+b_2+b_3+2b_4$, $c_4 = 4(d_1+d_2)$ \cite{Frink:2004ic},
where the $c_i$ are the conventional LECs used in the nucleonic sector. 

\vskip 0.5cm 
\noindent for $\Sigma N$
\begin{eqnarray}
N_1 &=& \left[48\, c^\Sigma_1 \  g_A^2  + 32\, c_1 \  (2\alpha g_A)^2 \right. \cr & & \left. + 16\, c_1 \  \frac{4}{3}((1-\alpha) g_A)^2\right] M^2_\pi, \nonumber \\
N_2 &=& 4 c^\Sigma_3 \  g_A^2 + 4 c_3 \ (2 \alpha g_A)^2 + 2 c_3 \  \frac{4}{3}((1-\alpha) g_A)^2, \nonumber \\
N_3 &=& - \left(4 d^\Sigma \  g_A^2 + \right. \cr & & \left. c_4 \ (2 \alpha g_A)^2 + c_4 \ \frac{4}{3}((1-\alpha) g_A)^2\right) \, 
\fet T_1\cdot\fet\tau_2, \cr & & 
\label{SN:N}
\end{eqnarray}
with $c^\Sigma_1 = b_0+b_D$, $c^\Sigma_3 =4b_1+2b_2+3b_4$, and $d^\Sigma = 4d_2+d_3$ 
and
$\langle \fet T_1\cdot\fet\tau_2  \rangle =-2,\, 1$ for isospin $I=1/2,\,3/2$.

\noindent for $\Lambda N$
\begin{eqnarray}
N_1 &=& \left[ 16\, c^\Lambda_1 \  g_A^2 + 48\, c_1 \  \frac{4}{3}((1-\alpha) g_A)^2\right] M^2_\pi, \nonumber \\
N_2 &=&  4\, c^\Lambda_3 \  g_A^2 + 6\, c_3 \  \frac{4}{3}((1-\alpha) g_A)^2, 
\nonumber \\
N_3 &=& 0, 
\label{LN:N}
\end{eqnarray}
with $c^\Lambda_1 = 3b_0+b_D$, $c^\Lambda_3 = 2b_2+3b_4$.

\vskip 0.5cm 
\noindent for $\Lambda N \to \Sigma N$
\begin{eqnarray}
N_1 &=& 0, \nonumber \\
N_2 &=& 0, \nonumber \\
N_3 &=&   16 d_1 \ g_A^2 + 2 \sqrt{3}\, c_4 \ \frac{4}{\sqrt{3}} \alpha (1-\alpha) g_A^2.
\label{SLN:N}
\end{eqnarray}
Note that in Eqs.~(\ref{SN:N}) - (\ref{SLN:N}) we have re-expressed the 
$\Si\Si \pi$ and $\Si\La\pi$ coupling constants in terms of the SU(3)
relations given in Eq.~(\ref{eq:SU3}), i.e. 
$f_{\Sigma\Sigma\pi} = 2\alpha f_{NN\pi}$ and 
$f_{\Sigma\Lambda\pi} = ({2}/{\sqrt{3}})(1-\alpha) f_{NN\pi}$ 
with $f_{NN\pi}= g_A / (2 f_\pi)$.

In our calculation we take the $\pi N$ LECs, i.e. $c_1$-$c_4$, 
from  Refs.~\cite{Hoferichter:2015tha,Hoferichter:2015hva}, 
obtained from matching the chiral expansion of the pion-nucleon scattering 
amplitude to the solution of the Roy-Steiner equations.
Specifically we use the values employed in the SMS $NN$ potential up to N$^2$LO. 
Fixing the values for the other LECs, without direct experimental evidence which 
can be used as constraint, is, however, difficult, and to some extent arbitrary. 
Here we try to find the best possible set instead of insisting on an 
intrinsically consistent selection. Since theoretical studies of the
baryon masses yield, in general, $b_1$,...,$b_4$ values that imply a
$c_3$ very far away from the results obtained from $\pi N$ scattering 
we consider the values from decuplet saturation as the most realistic
choice. Accordingly, we take the values for the $b$'s and $d$'s 
(i.e. $b_1$-$b_4$, $d_1$-$d_3$) for the $\pi \Si$ and $\pi \La$
vertices from Ref.~\cite{Petschauer:2017gd}.
Since $b_D$, $b_F$, $b_0$ are zero in this case, we use here values
from Ref.~\cite{Bernard:1995dp}, fixed in a study of the baryon mass 
splittings and the $\pi N$ sigma term. Anyway exploratory calculations 
indicated that the $YN$ results are fairly insensitive to the specific 
values adopted for the LECs $b_D$, $b_F$, $b_0$. 
The actual values used are (all in units of GeV$^{-1}$):
$c_1 = -0.74$, $c_3 = -3.61$, $c_4 = -2.44$  \cite{Hoferichter:2015tha,Hoferichter:2015hva},
$b_D = 0.066$, $b_F = -2.13$, $b_0 = -0.517$ \cite{Bernard:1995dp},
$b_1 = 0.59$, $b_2 = 0.76$, $b_3 = -1.01$, $b_4 = -1.51$,
$d_1 = 0.25$, $d_2 = 0.08$, $d_3 = -0.50$
\cite{Petschauer:2017gd}.

\subsection{Subtractions in the spectral integrals} 
As in case of the LO term and following the procedure 
in the $NN$ interaction \cite{Reinert:2018ip} we perform
subtractions according to Eqs. (42) and (44) of that reference
in the spectral integrals for the NLO and N$^2$LO potentials
so that the final form of those contributions read 
\begin{widetext}
\be
\label{eq:NLO}
V_{C}^{(2,3)} (q) &=& e^{-\frac{q^2}{2 \Lambda^2}} \;
\frac{2}{\pi} \int_{2 M_\pi}^\infty \, \frac{d \mu}{\mu^3}
\rho_{C}^{(2,3)} (\mu)\, \bigg(\frac{q^4}
{\mu^2 + q^2} + C_{C, 1}^2 (\mu ) +
C_{C, 2}^2 (\mu ) \, q^2\bigg)
\, e^{- \frac{\mu^2}{2
    \Lambda^2}} \,,\nn
V_{S}^{(2,3)} (q) &=& e^{-\frac{q^2}{2 \Lambda^2}} \;
\frac{2}{\pi} \int_{2 M_\pi}^\infty \, \frac{d \mu}{\mu^3} \,
\rho_{S}^{(2,3)} (\mu)\, \bigg( \frac{q^4}{q^2 + \mu^2} + C_{S, 1}^2 (\mu ) +
C_{S, 2}^2 (\mu ) \, q^2\bigg) \, e^{- \frac{\mu^2}{2
    \Lambda^2}} \,, \nn
V_{T}^{(2,3)} (q) &=& - e^{-\frac{q^2}{2 \Lambda^2}} \;
\frac{2}{\pi} \int_{2 M_\pi}^\infty \, \frac{d \mu}{\mu^3} \,
\rho_{S}^{(2,3)} (\mu)\, \bigg(
\frac{q^2}
{\mu^2 + q^2} + C_{T}^1 (\mu) \bigg)\, e^{- \frac{\mu^2}{2
    \Lambda^2}} \,,
\ee
\end{widetext}
The functions $C_i^{2} (\mu )$ and $C_T^{1} (\mu)$ appearing in the 
(single- and double-)subtracted spectral integrals have the form
\cite{Reinert:2018ip}: 
\be
C_{C,1}^{2}(\mu) &=&\Big[ 2 \Lambda \mu^2 \left(2 \Lambda ^4-4 \Lambda ^2 \mu ^2-
   \mu ^4 \right)  \cr 
   & & + \sqrt{2 \pi } \mu ^5 e^{\frac{\mu ^2}{2 \Lambda ^2}} \left(5 \Lambda ^2+\mu ^2\right)
   \text{erfc}\left(\frac{\mu }{\sqrt{2} \Lambda }\right)\Big]\cr 
   & & \quad /(4 \Lambda ^5)
\,, \nn
C_{C,2}^{2}(\mu) &=&- \Big[ 2 \Lambda \left( 6 \Lambda ^6-2 \Lambda ^2 \mu ^4-
   \mu ^6 \right) \cr 
   & & +\sqrt{2 \pi } \mu ^5 e^{\frac{\mu ^2}{2 \Lambda ^2}} \left(3 \Lambda ^2+\mu ^2\right)
   \text{erfc}\left(\frac{\mu }{\sqrt{2} \Lambda }\right)\Big] \cr 
   && /({12 \Lambda ^7})\,, \nn
C_{S,1}^{2}(\mu) &=&\Big[ 2 \Lambda \mu^2 \left( 2 \Lambda ^4-4 \Lambda ^2 \mu ^2-
   \mu ^4 \right) \cr & & 
   + \sqrt{2 \pi } \mu ^5 e^{\frac{\mu ^2}{2 \Lambda ^2}} \left(5 \Lambda ^2+\mu ^2\right)
   \text{erfc}\left(\frac{\mu }{\sqrt{2} \Lambda }\right)\Big] \cr && \quad / ({6 \Lambda ^5})\,, \nn
C_{S,2}^{2}(\mu) &=& -\Big[ 2 \Lambda \left( 15 \Lambda ^6- \Lambda ^4 \mu ^2-3 \Lambda ^2
   \mu ^4-2  \mu ^6 \right) \cr 
   & & +\sqrt{2 \pi } \mu ^5 e^{\frac{\mu ^2}{2 \Lambda ^2}} \left(5 \Lambda ^2+2 \mu ^2\right)
   \text{erfc}\left(\frac{\mu }{\sqrt{2} \Lambda }\right)\Big] \cr 
   & & \quad /({30 \Lambda
   ^7}) \, , \nn
C_{T}^{1}(\mu) &=&-\Big[ 2 \Lambda \left( 15 \Lambda ^6-3 \Lambda ^4 \mu ^2+ \Lambda ^2 \mu ^4-
  \mu ^6 \right) \cr 
  & & + \sqrt{2 \pi } \mu ^7 e^{\frac{\mu ^2}{2 \Lambda ^2}} \text{erfc}\left(\frac{\mu }{\sqrt{2} \Lambda
   }\right)\Big] \cr 
   & & \quad / ({30 \Lambda ^7})\, \ .
\ee

%%%%%%%%%%%%%%%%%%%%%%%%%%%%%%%%%%%%%%%%%%%%%%%%%%%%%%%%%%%
\section{Spectral functions for unequal meson masses}
\label{sec:spectrfunc}

For completeness we provide here expressions for the spectral functions
when the masses of the mesons are different. Those can be used to 
evaluate the contributions from exchanges of $\pi K$, $\eta K$, etc.,
which arise formally in SU(3) chiral EFT at NLO and N$^2$LO.
However, as already 
emphasized in the main text, given the present choice of the cutoff in
the local regulator of $\Lambda = 500-600$~MeV, those contributions 
are strongly suppressed and, therefore, omitted in the present study.
Denoting the meson masses by $M_1$ and $M_2$ the spectral functions are
as follows: 

\noindent for NLO
\be
\rho^{pb}_C (\mu) &=&   \frac{- \displaystyle \frac{N } 
{3072 \pi f^4_0 }}{\sqrt{\left[\mu^2 - (M_1 + M_2)^2\right]
\left[\mu^2 - (M_1 - M_2)^2\right]}} \cr 
&\times& 
\bigg[-23\mu^4+\frac{(M^2_1 - M^2_2)^4}{\mu^4} 
+ 56 \mu^2 (M^2_1 + M^2_2) \cr 
& & + 8 \frac{(M^2_1 + M^2_2)(M^2_1 - M^2_2)^2}{\mu^2}
 \nonumber \\
&-& 2(21M^4_1 + 22 M^2_1M^2_2 + 21 M^4_2) \bigg] \\
\rho^{pb}_T(\mu) &=& \frac{\rho^{pb}_S(\mu)}{\mu^2} \cr 
& & = \frac{N \sqrt{\left[\mu^2-(M_1 + M_2)^2\right]
\left[\mu^2-(M_1 - M_2)^2\right]}}{256\pi\mu^2 f^4_0} \cr & & 
\ee
For crossed-box diagrams the relations given in Eq.~(\ref{eq:TPE_18}) apply. 
\be
\rho^{tr}_C(\mu) &=& -\frac{N \sqrt{\left[\mu^2-(M_1 + M_2)^2\right]
\left[\mu^2-(M_1 - M_2)^2\right]}}{3072\pi\mu^4 f^4_0} \nonumber \\
&\times&\left[5\mu^4-4\mu^2(M^2_1+M^2_2)-(M^2_1-M^2_2)^2\right] 
\ee

\be
\rho^{fb}_C(\mu) &=& \frac{N \left[\mu^2-(M_1 + M_2)^2\right]^{\frac{3}{2}}
\left[\mu^2-(M_1 - M_2)^2\right]^{\frac{3}{2}} }{6144\pi\mu^4 f^4_0} \cr & & 
\ee

\noindent for N$^2$LO
\be
\rho^{tr}_C(\mu) &=& \frac{N_1}{512\mu f^4_0} (\mu^2 - M^2_1-M^2_2) \cr 
& & + \frac{N_2}{256\mu f^4_0} (\mu^2 - M^2_1-M^2_2)^2
\ee

\be
\rho^{tr}_T(\mu) &=& \frac{\rho^{tr}_S(\mu)}{\mu^2} \cr 
& = & 
-\frac{N_3}{512\mu^3 f^4_0} \left[\mu^2 - (M_1+M_2)^2\right] \cr 
& & \qquad  \qquad \times \left[ \mu^2 - (M_1-M_2)^2\right] \cr & & 
\ee

%%%%%%%%%%%%%%%%%%%%%%%%%%%%%%%
\section{Tables with LECs} 
\label{sec:lecs}

The $YN$ LECs employed in the present study are summarized in
Tables~\ref{tab:F1} and \ref{tab:F2}. With those LECs the contribution
of the contact terms to the potentials in the various $YN$ channels
can be calculated, based on Eqs.~(\ref{VC0}) to (\ref{VC}).
With regard to the $P$-waves SU(3) symmetry is preserved so that the 
potentials follow from the appropriate SU(3) combination as specified in 
Table~\ref{tab:SU3}. In case of the $^1S_0$ and $^3S_1$-$^3D_1$
partial waves, leading-order SU(3) breaking terms have been considered in the 
fitting procedure, in line with the power counting \cite{Petschauer:2013uua}.
Here, we list the LECs in the isospin basis for the $\La N$ and $\Si N$
channels and the $\La N \leftrightarrow \Si N$ transition (Table~\ref{tab:F1}).
Since for the $^3S_1$-$^3D_1$ partial wave SU(3) symmetry implies 
that $V_{\La N \to \La N}= V_{\Si N \to \Si N \, (I=1/2)}=(C^{8_a}+C^{10^*})/2$,
cf. Table~\ref{tab:SU3}, one can directly read off the amount of symmetry 
breaking in the contribution of the contact potential from the values in
Table~\ref{tab:F1}. In general, it is small or even zero.

\begin{table*}
\caption{The $YN$ contact terms for the $^1S_0$ and $^3S_1$-$^3D_1$
partial waves for various cut--offs. The values of the $\tilde C$'s are in
$10^4$ ${\rm GeV}^{-2}$ the ones of the $C$'s in $10^4$ ${\rm GeV}^{-4}$;
the values of $\Lambda$ in MeV. 
}
\renewcommand{\arraystretch}{1.2}
\label{tab:F1}
\vspace{0.2cm}
\centering
\begin{tabular}{|c|c||rrr||rrr|}
\hline
\multicolumn{2}{|c||}{}& 
\multicolumn{3}{c||}{SMS NLO} & \multicolumn{3}{c|}{SMS N$^2$LO} \\
\multicolumn{2}{|c||}{$\Lambda$} & $500$ & $550$ & $600$& $500$& $550$& $600$  \\
\hline
$\La N\to\La N$&$\tilde C_{^1S_0}$  &  $-$0.02935 &  $-$0.00329 &   0.14237 &   0.00494 &   0.07219 &   0.08299 \\ 
&$ C_{^1S_0}$  &   0.63280 &   0.61297 &   0.79287 &   0.26538 &   0.37189 &   0.09995 \\ 
$\La N\to\Si N$&$\tilde C_{^1S_0}$  &  $-$0.03286 &  $-$0.03023 &  $-$0.11525 &  $-$0.02415 &  $-$0.06843 &  $-$0.07698 \\ 
&$ C_{^1S_0}$  &  $-$0.29427 &  $-$0.26766 &  $-$0.33429 &  $-$0.11513 &  $-$0.17396 &  $-$0.02095 \\ 
$\Si N\to\Si N\, (1/2)$&$\tilde C_{^1S_0}$  &   0.11221 &   0.12683 &   0.14184 &   0.09486 &   0.17822 &   0.29980 \\ 
&$\ C_{^1S_0}$  &  $-$0.15191 &  $-$0.10078 &  $-$0.09857 &  $-$0.04162 &  $-$0.09201 &  0.04409 \\ 
$\Si N\to\Si N\, (3/2)$&$\tilde C_{^1S_0}$  &  $-$0.01620 &   0.02679 &   0.17627 &   0.00309 &   0.06730 &   0.07276 \\ 
&$ C_{^1S_0}$  &   0.73089 &   0.70219 &   0.90430 &   0.30375 &   0.42988 &   0.10693 \\ 
 \hline
$\La N\to\La N$&$\tilde C_{^3S_1}$  &   0.09667 &   0.10212 &   0.14003 &   0.16132 &   0.18609 &   0.21782 \\ 
&$ C_{^3S_1}$  &   0.72758 &   0.56012 &   0.54597 &   0.39114 &   0.34187 &   0.16242 \\ 
$\La N\to\Si N$&$\tilde C_{^3S_1}$  &   0.15685 &   0.18472 &   0.19931 &   0.17541 &   0.16851 &   0.18866 \\ 
&$ C_{^3S_1}$  &   0.72892 &   0.27346 &  $-$0.05722 &   0.58414 &   0.49310 &   0.15400 \\ 
$\Si N\to\Si N\, (1/2)$&$\tilde C_{^3S_1}$  &   0.09667 &   0.10212 &   0.14003 &   0.18681 &   0.18877 &   0.21267 \\ 
&$ C_{^3S_1}$  &   0.72758 &   0.56012 &   0.54597 &   0.39114 &   0.34187 &   0.16242 \\ 
$\Si N\to\Si N\, (3/2)$&$\tilde C_{^3S_1}$  &   0.05032 &   0.06086 &   0.06355 &   0.15319 &   0.11209 &   0.13002 \\ 
&$ C_{^3S_1}$  &   0.08219 &   0.08044 &   0.03758 &   0.40259 &  $-$0.06077 &  $-$0.10236 \\ 
 \hline
$\La N\to\La N$&$ C_{^3SD_1}$  &   0.08863 &   0.09803 &   0.08863 &   0.34868 &   0.13053 &   0.12134 \\ 
$\La N\to\Si N$&$ C_{^3SD_1}$  &   0.31634 &   0.32118 &   0.31634 &   0.52449 &   0.32367 &   0.34512 \\ 
$\Si N\to\Si N\, (1/2)$&$ C_{^3SD_1}$  &   0.08863 &   0.09803 &   0.08863 &   0.34868 &   0.13053 &   0.12134 \\ 
$\Si N\to\Si N\, (3/2)$&$ C_{^3SD_1}$  &   0.20000 &   0.24793 &   0.24793 &   0.21463 &   0.21463 &   0.18000 \\
\hline
\end{tabular}
\renewcommand{\arraystretch}{1.0}
\end{table*}
\begin{table*}
\caption{The $YN$ contact terms for the $P$-waves for various cut--offs.
The values of the LECs are in
$10^4$ ${\rm GeV}^{-4}$; the values of $\Lambda$ in MeV.
The superscripts $a$ and $b$ denote the two variants introduced in 
Sect.~\ref{sec:Spp}. 
}
\renewcommand{\arraystretch}{1.6}
\label{tab:F2}
\vspace{0.2cm}
\centering
\begin{tabular}{|c||rrr||rrrr|}
\hline
& 
\multicolumn{3}{c||}{SMS NLO} & \multicolumn{4}{c|}{SMS N$^2$LO} \\
{$\Lambda$} & $500$ & $550$ & $600$& $500$& $550^a$& $550^b$ & $600$  \\
\hline
\hline
 $C^{27}_{^3P_0}$  & 0.17477 & 0.22196 & 0.26500 &  0.44332 &  0.45000 &0.62505& 0.61226  \\ 
 $C^{8_s}_{^3P_0}$ & 1.65980 & 2.75900 & 2.06930 &  2.39600 &  0.82218 &1.60990& 2.42460 \\ 
 \hline
 $C^{10}_{^1P_1}$  & 2.41220 & 1.62550 & 0.82692 &  2.41640 &  1.55430 &1.73930& 3.14090 \\ 
 $C^{10^*}_{^1P_1}$& 0.00000 & 0.00000 & 0.00000 &  0.32168 &  0.20699 &$-$0.09449& 0.03500 \\ 
 $C^{8_a}_{^1P_1}$ & 0.06119 &$-$0.10810 &$-$0.14370 &  0.14853 &  0.19339 &0.20985& 0.32815 \\ 
 \hline
 $C^{27}_{^3P_1}$  & 0.22993 & 0.19541 & 0.18500 &  0.32000 &  0.48651 &0.65850& 0.58177 \\ 
 $C^{8_s}_{^3P_1}$ & 0.39891 & 0.07447 & 0.10685 &  0.49239 &  0.51190 &0.52342& 0.79248 \\ 
 \hline
 $C^{27}_{^3P_2}$  &$-$0.29185 &$-$0.25446 &$-$0.22000 & $-$0.08937 & $-$0.10000 &$-$0.01692&$-$0.01802 \\ 
 $C^{8_s}_{^3P_2}$ & 1.06570 &$-$0.06844 &$-$0.20845 &  2.96720 &  3.30420 &3.32790& 2.96270 \\
\hline
\end{tabular}
\renewcommand{\arraystretch}{1.0}
\end{table*}

\bibliographystyle{unsrturl}

\bibliography{bib/literature}

\end{document}